\documentclass[pra,a4paper,twocolumn,superscriptaddress,longbibliography]{revtex4-2}
\usepackage[utf8]{inputenc}
\usepackage{amssymb}
\usepackage{amsmath}
\usepackage{amsfonts}
\usepackage{amsthm}
\usepackage{amsbsy}
\usepackage{bm}
\usepackage{bbold}

\usepackage{graphicx}
\usepackage{xcolor}
\usepackage{multirow}
\usepackage{float}
\usepackage{comment}
\usepackage{ulem}
\usepackage{relsize}
\usepackage{cmap}
\usepackage{physics}
\usepackage[colorlinks]{hyperref}

\DeclareMathOperator{\tTr}{tr}

\DeclareMathOperator{\im}{Im}
\DeclareMathOperator{\re}{Re}

\DeclareMathOperator{\arctanh}{arctanh}

\begin{document}
	
	\title{Reaction-diffusive dynamics of number-conserving dissipative quantum state preparation}
	
	\author{P. A. Nosov}
	
	\affiliation{Stanford Institute for Theoretical Physics, Stanford University, Stanford, California 94305, USA}
		
	\author{D. S. Shapiro}
	
\affiliation{\hbox{Institute for Quantum Materials and Technologies, Karlsruhe Institute of Technology, 76021 Karlsruhe, Germany}}
 
	\affiliation{\hbox{National University of Science and Technology MISiS, 119049 Moscow, Russia}}

	\author{M. Goldstein}
	
	\affiliation{\hbox{Raymond and Beverly Sackler School of Physics and Astronomy, Tel Aviv University, Tel Aviv 6997801, Israel}}

       \author{I. S. Burmistrov}
	
	\affiliation{\hbox{L.~D.~Landau Institute for Theoretical Physics, acad.\ Semenova av.\ 1-a, 142432 Chernogolovka, Russia}}
	\affiliation{Laboratory for Condensed Matter Physics, HSE University, 101000 Moscow, Russia}

	\date{\today, v.1.16} 
	
	\begin{abstract}
The use of dissipation for the controlled creation of nontrivial quantum many-body correlated states is of much fundamental and practical interest.
What is the result of imposing number conservation, which, in closed system, gives rise to diffusive spreading?
We investigate this question for a paradigmatic model of a two-band system, with dissipative dynamics aiming to empty one band and to populate the other, which had been introduced before for the dissipative stabilization of topological states.
Going beyond the mean-field treatment of the dissipative dynamics, we demonstrate the emergence of a diffusive regime for the particle and hole density modes at intermediate length- and time-scales, which, interestingly, can only be excited in nonlinear response to external fields. We also identify processes that limit the diffusive behavior of this mode at the longest length- and time-scales. Strikingly, we find that these processes lead to a reaction-diffusion dynamics governed by the Fisher-Kolmogorov-Petrovsky-Piskunov equation, making the designed dark state unstable towards a state with a finite particle and hole density.
	\end{abstract}

	\maketitle
	%%%%%%%%%%%%%%%%%%%%%%%%%%%%%%%%%%%%

\section{Introduction}

Recently the driven-dissipative dynamics of open quantum many-body systems has become an arena of active research~\cite{Diehl2016,LeHur2018,Skinner2019,Rudner2020,Kamenev2022,fava2023nonlinear,poboiko2023}. Such open quantum systems have non-thermal stationary states with nontrivial non-equilibrium transient dynamics. This paves the way towards the possibility of nonequilibrium phases of matter~\cite{Lechner2013,Piazza2014,Keeling2014,Altman2015,Kollath2016} and of phase transitions between them~\cite{DallaTorre2010,Sieberer2013,Raftery2014,Li2018,Li2019,Roy2020,Garratt2021} in different universality classes as compared with their equilibrium counterparts.

A central role is played by driven dissipative dynamics described by the Gorini-Kossakowski-Sudarshan-Lindblad (GKSL) master equation~\cite{Diehl2016}. The driving and dissipation often act to suppress quantum coherence. However, it has been realized that, by proper tuning of the coupling between the system and the reservoirs, one may tailor-design an intricate dark state, and thus obtain a many-body steady state which inherits its nontrivial properties from the dissipation rather than from the internal Hamiltonian dynamics~\cite{Plenio1999,Diehl2008,Kraus2008,Verstraete2009,Weimer2010,Otterbach2014,Lang2015,Zhou2017}.
Particular effort was centered around the dissipative creation of topological states~\cite{Diehl2011,Bardyn2012,Bardyn2013,Koenig2014,Budich2015,Iemini2016,Gong2017,Goldstein2019,Shavit2020,Tonielli2020,Yoshida2020,Gau2020a,Gau2020b,Bandyopadhyay2020,Santos2020,Altland2021,Beck21,Nava2023}.

An important role in quantum many body dynamics is played by symmetries and the ensuing conservation laws.
In closed systems, diffusive dynamics may then be induced by the combination of conservation laws and disorder averaging, or, alternatively, by dephasing caused by a coupling to a bath \cite{diffusion2005,Castro-Alvaredo2016,Dhar2019,Jin2022} or by the degrees of freedom of the system itself~\cite{Kamenev2009}.
But in dissipative state preparation one tries to avoid disorder, and to implement bath couplings which enforce rather than suppress coherence. How would the dynamics look then? While most studies considered particle-number changing dynamics, a recent work~\cite{Tonielli2020} introduced a paradigmatic two-band model, with particle-conserving Lindblad operators which empty one band and fill the other. Mean-field analysis shows that the system converges to the desired dark state at a rate independent of the system size, and leads to quantized response functions if the bands are also topologically-nontrivial (as opposed to the particle-nonconserving case~\cite{Shavit2020}). What happens beyond mean field? This is the problem addressed by this work.

To provide a complete characterization of the dynamics, we employ a two-pronged strategy.
In one direction, we investigate the question of diffusion by introducing the beyond-mean-field vertex corrections into the Lindblad Keldysh diagramatics, motivated by their importance in disordered electronic systems~\cite{Kamenev2009}.
Using this tool we demonstrate the existence of a diffusive regime for the particle and hole density modes at an intermediate range of length- and time scales. Interestingly, in contrast to the textbook examples for Hamiltonian dynamics, this diffusive density mode cannot be induced by the linear response to a scalar potential. However, we demonstrate that the diffusive mode can be excited in the nonlinear response, cf.~Eq.~\eqref{eq:main:res}.

In the second direction we address the limitations (on long length- and time-scales) to the diffusive behavior of the particle-hole density mode due to transitions between the bands. In addition to the expected recombination (annihilation between particles in the emptied band and holes in the filled band), which are second order in the deviation of the densities from the steady state, we surprisingly find first order contributions which actually create particles and holes, resulting in an instability of the desired dark state.
The diffusion and transitions between the band combine to give rise to reaction-diffusion dynamics 
governed by the iconic Fisher-Kolmogorov-Petrovsky-Piskunov (FKPP) equation, cf.~Eq.~\eqref{eq:main:res:FKPP}. We expect it to arise generically in number-conserving state preparation.

The outline of the paper is as follows. In Sec.~\ref{Sec:Model} we specify the model of dissipative dynamics, then translate it to Keldysh form in Sec.~\ref{Sec:Keldysh}. Next, in Sec.~\ref{Sec:SelfEnergy} we study the structure of the dark state within the mean-field approximation, that is, a self-consistent solution for the self-energy. The linear response of the density to an external scalar potential is analyzed in Sec.~\ref{Sec:Linear}, followed by nonlinear response in Sec.~\ref{Sec:Nonlinear}.
The recombination rate for the particles due to dissipative dynamics is computed in Sec.~\ref{Sec:Rec}, while in Sec.~\ref{Sec:Pumping} the rate of change of the number of particles due to transitions of particles from lower band to  the upper band is estimated. We end the paper with discussions and conclusions in Secs.~\ref{Sec:Discussion} and \ref{Sec:Conclusions}, respectively. Some details of calculations are presented in the Appendices.

%%%%%%%%%%%%%%%%%%%
\begin{figure*}[ht]
\centerline{\includegraphics[width=0.3\textwidth]{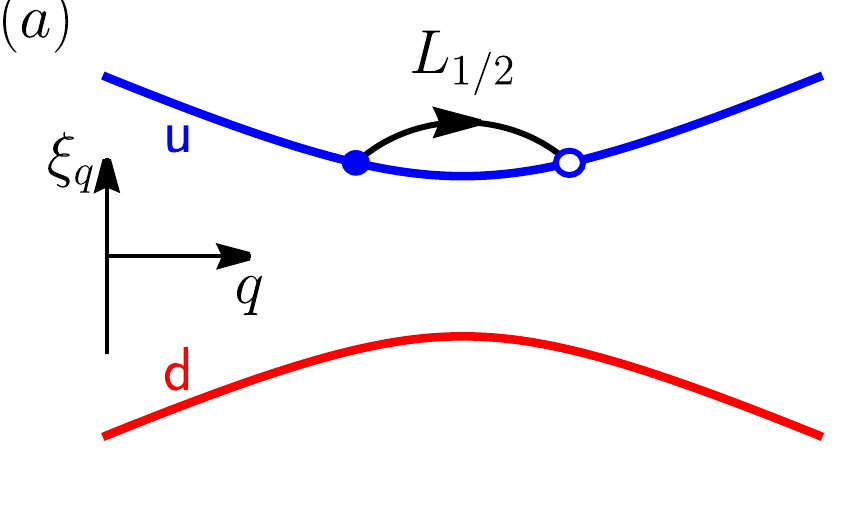}
\qquad \includegraphics[width=0.3\textwidth]{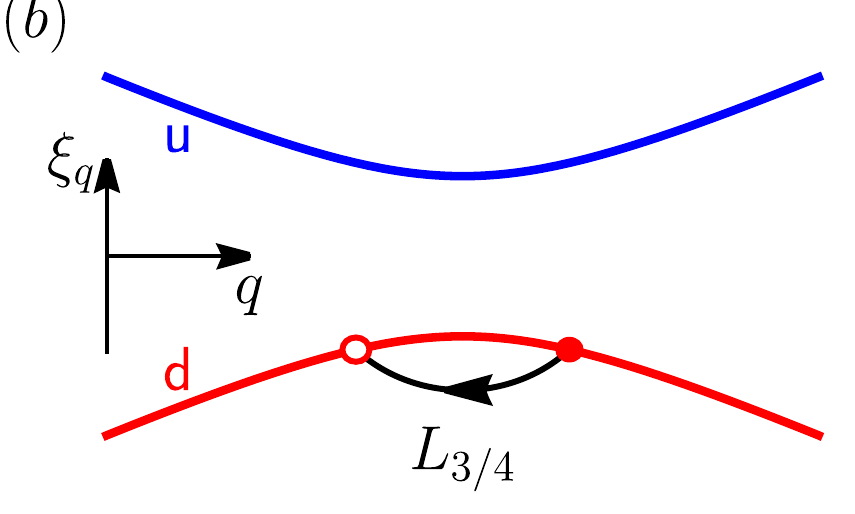}
\qquad \includegraphics[width=0.3\textwidth]{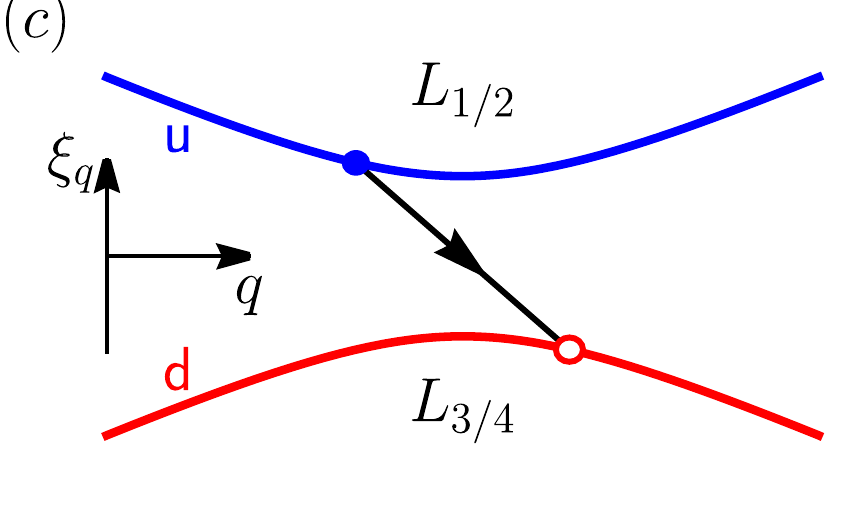}
}
\caption{Sketch of the action of the jump operators. $L_{1/2}$ redistributes a particle in the upper band (panel (a)) or transfers it to the lower band (panel (c)). $L_{3/4}$ redistributes a particle in the lower band (panel (b)) or creates a particle in the lower band by annihilating it in the upper band (panel (c)).}
\label{fig:Jump}
\end{figure*}
%%%%%%%%%%%%%%%%%%%

\section{Model\label{Sec:Model}}

We consider the dissipative model of Ref.~\cite{Tonielli2020}, whose time evolution is governed by the GKSL equation for the density matrix:
\begin{equation}
\frac{d\rho}{dt} = \int\limits_{\bm{x}} 
\Bigl ( i [\rho,H_{\rm 0}] +\sum_{\alpha=1}^4 \gamma_\alpha 
\bigl (2 L_\alpha \rho L_\alpha^\dag - \{L_\alpha^\dag L_\alpha, \rho\}\bigr )\Bigr ),
\label{eq:GKSL}
\end{equation}
where $\int_{\bm{x}}{\equiv} \int d^d\bm{x}$. The unitary part of the evolution is described by the two-band Hamiltonian $\mathcal{H}{=}\int_{\bm{x}} H_0{=} \int_{\bm{q}} \Psi^\dagger_{\bm{q}} H_0({\bm{q}})\Psi_{\bm{q}}$, parameterized as
\begin{equation}\label{eq:Hamiltonian}
H_0(\bm{q})=\bm{d}_{\bm{q}} {\cdot} \bm{\sigma}\;,\quad\bm{d}_{\bm{q}}=\{2m q_x,2m q_y,q^2-m^2\}\;,
\end{equation}
where $\Psi_{\bm{q}}{=}\{\psi_{1,\bm{q}},\psi_{2,\bm{q}}\}$, and $\sigma_{0,x,y,z}$ are the standard Pauli matrices acting in a spin space (`$1$' and `$2$'  indexes). We also introduced shorthand notations for momentum integrals, $\int_{\bm{q}} {\equiv} \int d^d\bm{q}/(2\pi)^d$. Below we shall focus on the cases of $d{=}1$ and $d{=}2$ dimensions. In the former case the $y$ component of momentum vanishes, whereas in the latter case, the Hamiltonian in Eq.~\eqref{eq:Hamiltonian} describes a two-dimensional Chern insulator with a Chern number $\theta{=}{-}1$ for any finite $m^2$. In the eigenbasis, the Hamiltonian density $H_0({\bm{q}})$ in Eq.~\eqref{eq:Hamiltonian} becomes $|\bm{d}_{\bm{q}}| \sigma_z$, where $|\bm{d}_{\bm{q}}|{=}q^2{+}m^2$ is the spectrum.

The dissipative part of dynamics in \eqref{eq:GKSL} is defined through the jump operators $L_\alpha$
\begin{equation}
L_{1/2}  = \psi^\dagger_{1/2}(\bm{x})  l_{\textsf{u}}(\bm{x}) ,  \quad 
L_{3/4} = \psi_{1/2}(\bm{x}) l^\dagger_{\textsf{d}}(\bm{x})\;, 
\label{eq:Jump:Operators1:def}
\end{equation} 
where the operators $l{=}\{l_{\textsf{u}},l_{\textsf{d}}\}$ bring the Hamiltonian to a diagonal form $\mathcal{H}{=}\int_{\bm{q}} l_{\bm{q}}^\dagger \sigma_z l_{\bm{q}}$, and could be expressed in terms of the original operators $\psi$ as
\begin{equation}\label{eq:l_def_op}
   l_{\bm{q}}= \sqrt{d_{\bm{q}}}U^\dagger_{\bm{q}}\Psi_{\bm{q}}\;,\quad U_{\bm{q}} = \frac{q_x\sigma_0-i q_y \sigma_z-i m \sigma_y }{\sqrt{d_q}}\;.
\end{equation}
Here $d_q{=}|\bm{d}_{\bm{q}}|{=}q^2{+}m^2$, the unitary matrix $U_{\bm{q}}$ diagonalizes $H_0(\bm{q})$, and the extra factor $\sqrt{d_{\bm{q}}}$ in the definition of $l_{\bm{q}}$ is introduced to ensure the locality of the operators $l_{\textsf{u}/\textsf{d}}(\bm{x})$ in the coordinate space (i.e., $l(\bm{x})$ contains only the first spatial derivatives of $\Psi(\bm{x})$). The labels $\textsf{u}/\textsf{d}$ are referring to the `\textit{up}' and `\textit{down}' bands in the eigenbasis of $H_0(\bm{q})$. The jump operators $L_\alpha$ not only transfer a particle from the upper band to the lower one but also move a particle  within the bands, see Fig. \ref{fig:Jump}. In real space, the jump operators $L_\alpha$ may transfer a particle to a distance ${\lesssim} 1/m$. We note in passing that some possible ways to experimentally engineer such local jump operators conserving the total number of particles are discussed in Refs.~\cite{Diehl2011,Bardyn2012,Bardyn2013,Iemini2016,exprealization2}.

We are going to focus on the half-filled configuration, corresponding to the ground state $|D\rangle $ of $\mathcal{H}$ with a fully occupied `down' band and an empty `up' band (we will be referring to this state as the \textit{dark state}). As one can easily check, the density matrix $\rho_D = |D\rangle \langle D|$ is a steady state of the dynamics, Eq.~\eqref{eq:GKSL}, as it obeys $L_\alpha |D\rangle =0$. In fact, the Lindblad operators $L_\alpha$ are designed so as to stabilize this state, even in the absence of the Hamiltonian $\mathcal{H}$.

One of the crucial aspects of the model outlined above is that the {\it total} number of particles is conserved. Indeed, let us define the operator of the total number of particles as $\hat{N}{=}\int_{\bm{x}} [\psi^\dag_{1}(\bm{x}) \psi_{1} (\bm{x}){+} \psi^\dag_{2}(\bm{x}) \psi_{2} (\bm{x})]$. Using the following commutation relations,
\begin{equation}
\begin{split}
[\psi_{1/2}^\dag(\bm{x}), \hat N] & = - \psi_{1/2}^\dag(\bm{x}),  \,\, [\psi_{1/2}(\bm{x}), \hat N]  = \psi_{1/2}(\bm{x}) , \\
[l_{{\textsf{u}}/{\textsf{d}}}^\dag(\bm{x}), \hat N]  & = - l_{{\textsf{u}}/{\textsf{d}}}^\dag(\bm{x}),  \,\, [l_{{\textsf{u}}/{\textsf{d}}}(\bm{x}), \hat N]  = l_{{\textsf{u}}/{\textsf{d}}}(\bm{x}) ,
\end{split} 
\end{equation}
one can check that $[L_\alpha^\dag,\hat{N}]{=}0$. This implies that the dynamics of the density matrix described  by Eq. \eqref{eq:GKSL} conserves the total number of particles, $d\tTr (\hat N\rho)/dt {=}0$. As we shall see below, the conservation of the number of particles in dissipative dynamics has important consequences, reminiscent of the case of unitary evolution. However, we stress that the number of particles in the upper (lower) band is not separately conserved, and thus, the response of the associated density modes to external perturbations is expected to decay at sufficiently long times. 

\section{Keldysh field theory approach \label{Sec:Keldysh}}

In order to study the long time/distance dynamics governed by Eq.~\eqref{eq:GKSL}, we re-formulate this equation as the action on the Keldysh contour (see Ref. \cite{Diehl2016} for details). The corresponding Keldysh partition function reads as
\begin{equation}
Z= \int \mathcal{D}[\overline{\Psi},\Psi]\,  e^{i S_{\rm 0}[\overline{\Psi},\Psi] + i S_{\rm L}[\overline{\Psi},\Psi]},
\label{eq:K:action}
\end{equation}
which is expressed in terms of spin $1/2$ fermionic annihilation and creation fields on the `$+$' and `$-$' parts of the Keldysh contour, $\Psi{=}\{\psi_{1,+},\psi_{2,+},\psi_{1,-},\psi_{2,-}\}^T$ and $\overline{\Psi}{=}\{\overline{\psi}_{1,+},\overline{\psi}_{2,+},\overline{\psi}_{1,-},\overline{\psi}_{2,-}\}$ [we emphasize that these fields must be distinguished from their counterparts used in the operator formalism of Eq.~\eqref{eq:GKSL}]. The part of the Keldysh action corresponding to the unitary (Hamiltonian) dynamics of the density matrix is given by
\begin{gather}
S_{\rm 0}  {=} \int\limits_{\bm{q},t} \overline{\Psi}_{\bm{q}}(t) \bigl(  i\sigma_0 \partial_t {-} H_{\rm 0}(\bm{q}) \bigr )\tau_z \Psi_{\bm{q}}(t).
\end{gather}
Here $H_0(\bm{q})$ is defined in Eq.~\eqref{eq:Hamiltonian}, $\tau_{0,x,y,z}$ are the standard Pauli matrices acting in Keldysh space (`$\pm$' indices), and $\int_t {\equiv} \int_{-\infty}^\infty dt$ stands for the integration in the time domain.

The dissipative part of the Keldysh action \eqref{eq:K:action} reads in the coordinate representation 
\begin{gather}
S_{\rm L} {=} {-} {i} \int\limits_{\bm{x},t} \sum_{\alpha=1}^4 \gamma_\alpha \Bigl [
2 L_{\alpha,+}(t_-) \overline{L}_{\alpha,-}(t) - \overline{L}_{\alpha,+}(t)L_{\alpha,+}(t_-) 
\notag \\
- \overline{L}_{\alpha,-}(t) L_{\alpha,-}(t_+)
\Bigr ]\;.
\end{gather}
The times $t_\pm{=}t{\pm}\delta$ with $\delta{=}0^+$ take into account the specific regularization of equal time terms that is of crucial importance for correct causality of the Keldysh action (see Ref. \cite{Diehl2016} for details). 

In order to define $L_{\alpha,\pm}$ and $\overline{L}_{\alpha,\pm}$ corresponding to the jump operators $L_{\alpha}$ and $L_{\alpha}^\dagger$ of Eq.~\eqref{eq:Jump:Operators1:def} on the Keldysh contour, it is convenient to introduce two additional sets of fermionic fields. First we define the fields $C{=}\{c_{\textsf{u},+},c_{\textsf{d},+},c_{\textsf{u},-},c_{\textsf{d},-}\}^T$ and $\overline{C}{=}\{\overline{c}_{\textsf{u},+},\overline{c}_{\textsf{d},+},\overline{c}_{\textsf{u},-},\overline{c}_{\textsf{d},-}\}$, which form the eigenbasis (`\textit{up}' and `\textit{down}' states) of the Hamiltonian $H_0$, 
\begin{equation}
S_{\rm 0} =  \int_{\bm{q},t} \bar{C}_{\bm{q}}(t) \bigl(  i\sigma_0 \partial_t - \xi_q \sigma_z \bigr )\tau_z C_{\bm{q}}(t)\;. 
\label{eq:S0:ud}
\end{equation}
Here we introduce a parameter $\xi_q {\equiv} d_q {=} |\bm{d}_{\bm{q}}|{=}q^2{+}m^2$, which is the same as $d_q$, but will allow us to distinguish the contributions from unitary and dissipative parts of the dynamics. In particular, it allows us to consider the case in which Hamiltonian dynamics is absent, by setting $\xi_q$ to zero. We note that the two sets of fermionic fields, $\Psi$ and $C$, are related by a canonical transformation,
\begin{equation}\label{eq:Psi_to_C}
\Psi_{\bm{q}} = U_{\bm{q}} C_{\bm{q}} , \qquad \overline{\Psi}_{\bm{q}} = \overline{C}_{\bm{q}}  U_{\bm{q}}^\dag ,
\end{equation} 
where the matrix $U_{\bm{q}}$ was introduced in Eq.~\eqref{eq:l_def_op}.
Second, we introduce the fermionic fields $\ell = \{l_{\textsf{u},+},l_{\textsf{d},+},l_{\textsf{u},-},l_{\textsf{d},-}\}^T$
and $\overline{\ell} {=}\{\overline{l}_{\textsf{u},+},\overline{l}_{\textsf{d},+},\overline{l}_{\textsf{u},-},\overline{l}_{\textsf{d},-}\}$ that are related with the fields $C$ and $\overline{C}$ as  (cf. Eq.~\eqref{eq:l_def_op})
\begin{equation}
\ell_{\bm{q}}= \sqrt{d_q} C_{\bm{q}}, \qquad \overline{\ell}_{\bm{q}} = \sqrt{d_q}\, \overline{C}_{\bm{q}} .
\end{equation}
Then the jump operators $L_{\alpha}$ and $L_{\alpha}^\dagger$ can be written in the Keldysh theory by (we suppress the Keldysh indices) 
\begin{equation}
\begin{split}
L_{1/2} & = \overline{\psi}_{1/2}(\bm{x})  l_{\textsf{u}}(\bm{x}) ,  \quad 
L_{3/4} = \psi_{1/2}(\bm{x}) \bar{l}_{\textsf{d}}(\bm{x}), \\
\overline{L}_{1/2} & = \overline{l}_{\textsf{u}}(\bm{x}) \psi_{1/2}(\bm{x}) , \quad 
\overline{L}_{3/4} = l_{\textsf{d}}(\bm{x}) \overline{\psi}_{1/2}(\bm{x}) .
\end{split}
\label{eq:Jump:Operators:def}
\end{equation} 
Expressing  $\psi$, $\overline{\psi}$ and $\ell$, $\overline{\ell}$ in terms of $c$ and $\overline{c}$, we  represent the jump operators in the following form\footnote{For $L_{3/4}$ the correct ordering of operators is maintained by proper choice of infinitesimally-separated time arguments for the creation and annihilation operators in Eq.~\eqref{eq:SL:ud} below.}
\begin{equation}
L_\alpha = \int\limits_{\bm{qp}} e^{i(\bm{p}-\bm{q})\bm{x}}\, \overline{c}_{\bm{p}} \mathcal{L}_{\bm{pq}}^{(\alpha)} c_{\bm{q}}, \, 
\overline{L}_\alpha = \int\limits_{\bm{qp}} e^{i(\bm{p}-\bm{q})\bm{x}}\, \overline{c}_{\bm{p}} \overline{\mathcal{L}}_{\bm{pq}}^{(\alpha)} c_{\bm{q}} ,
\end{equation}
Here $\mathcal{L}^{(\alpha)}$ and $\overline{\mathcal{L}}^{(\alpha)}$ are $2{\times} 2$ matrices in the space of up/down states, which are constructed as
\begin{equation}
\begin{split}
[\mathcal{L}^{(1/2)}_{\bm{pq}}]_{\textsf{ab}}& =\sqrt{d_{q}} [U_{\bm{p}}^\dag]_{\textsf{a},1/2}\delta_{\textsf{bu}}, \\ [\mathcal{L}^{(3/4)}_{\bm{pq}}]_{\textsf{ab}} & = -\sqrt{d_{p}} [U_{\bm{q}}]_{1/2,\textsf{b}}\delta_{\textsf{ad}} ,
\end{split}
\label{eq:A:matrices:0}
\end{equation}  
and $\overline{\mathcal{L}}_{\bm{pq}}^{(\alpha)}{=} [\mathcal{L}_{\bm{qp}}^{(\alpha)}]^\dag$.
For convenience, we write the matrices $\mathcal{L}^{(\alpha)}$ explicitly
\begin{equation}
\begin{split}
\mathcal{L}^{(1)}_{\bm{pq}}{=}& \sqrt{\frac{d_q}{d_p}}
\begin{pmatrix}
p_x {+} i p_y & 0 \\
-m & 0
\end{pmatrix} ,
\quad
\mathcal{L}^{(2)}_{\bm{pq}}{=}\sqrt{\frac{d_q}{d_p}} 
\begin{pmatrix}
m & 0 \\
p_x {-} i p_y & 0
\end{pmatrix} , \\ 
\mathcal{L}^{(3)}_{\bm{pq}}{=}& \sqrt{\frac{d_p}{d_q}} 
\begin{pmatrix}
0 & 0 \\
- q_x {+} i q_y & m
\end{pmatrix} ,
\quad
\mathcal{L}^{(4)}_{\bm{pq}}{=}{-}\sqrt{\frac{d_p}{d_q}}
\begin{pmatrix}
0 & 0 \\
m & q_x {+} i q_y
\end{pmatrix} .
\end{split}
\label{eq:A:matrices}
\end{equation}

Now we can write the dissipative part of the action in terms of the $C$ and $\overline{C}$ fields explicitly,
\begin{widetext}
\begin{gather}
S_{\rm L} {=} {-} i (2\pi)^d  \!\!  \int\limits_{\bm{p_{j}},t}\!
\delta(\bm{p_1}{-}\bm{p_2}{+}\bm{p_3}{-}\bm{p_4})
\sum_{\alpha=1}^4 \gamma_\alpha
\Bigl [2 \overline{c}_{\bm{p_1},-}(t)\overline{\mathcal{L}}^{(\alpha)}_{\bm{p_1p_2}} c_{\bm{p_2},-}(t{+}\epsilon_\alpha)
\overline{c}_{\bm{p_3},+}(t{-}\delta)\mathcal{L}^{(\alpha)}_{\bm{p_3p_4}}c_{\bm{p_4},+}(t{-}\delta{-}\epsilon_\alpha)
- \overline{c}_{\bm{p_1},+}(t{+}\epsilon_\alpha)\notag \\ 
\times
\overline{\mathcal{L}}^{(\alpha)}_{\bm{p_1p_2}} c_{\bm{p_2},+}(t)
\overline{c}_{\bm{p_3},+}(t{-}\delta)\mathcal{L}^{(\alpha)}_{\bm{p_3p_4}}c_{\bm{p_4},+}(t{-}\delta{-}\epsilon_\alpha)
-\overline{c}_{\bm{p_1},-}(t{-}\epsilon_\alpha)\bar{\mathcal{L}}^{(\alpha)}_{\bm{p_1p_2}} c_{\bm{p_2},-}(t)
\overline{c}_{\bm{p_3},-}(t{+}\delta)\mathcal{L}^{(\alpha)}_{\bm{p_3p_4}}c_{\bm{p_4},-}(t{+}\delta{+}\epsilon_\alpha)
\Bigr ] .
\label{eq:SL:ud}
\end{gather}
\end{widetext}
Here we indicate regularization at coinciding times explicitly by $\epsilon_{1,2}{=}0^+$ and $\epsilon_{3,4}{=}0^-$. We note that, as a consequence of the conservation of the total number of particles, the Keldysh action $S_{\rm 0}{+}S_{\rm L}$ (\eqref{eq:S0:ud} and \eqref{eq:SL:ud}) has a strong symmetry being invariant under global $\mathrm{U}(1){\times} \mathrm{U}(1)$ transformations,
$\overline{c}_{\sigma}{\to} e^{-i\alpha_\sigma} \overline{c}_{\sigma}$ and $c_{\sigma}{\to} e^{ i\alpha_\sigma} c_{\sigma}$, where $\sigma{=}\pm$. At the same time, translation invariance is only a weak symmetry in this model \cite{Diehl2016,Buca2012,Albert2014}, meaning that the Keldysh action $S_{\rm L}$ is invariant only under translations acting simultaneously on the forward and backward branches of the Keldysh contour. As a consequence, the jump operators $L_\alpha$ can lead to momentum relaxation, as depicted in Fig.~\ref{fig:Jump}.

It is worth noting that the interaction in Eq.~\eqref{eq:SL:ud} could be formally decoupled by introducing some auxiliary bosonic fields which represent bath degrees of freedom. The latter, however, can be shown to be far from thermal equilibrium, as they would violate the principle of detailed balance. In particular, the correlation functions of such auxiliary bath fields disobey the fluctuation-dissipation theorem reflecting the non-Hermitian nature of the jump operators $L_\alpha$.

\section{Self-consistent solution for the dark state\label{Sec:SelfEnergy}}

We now demonstrate how the dark state appears within the Keldysh field theory, see Eqs.~\eqref{eq:S0:ud} and \eqref{eq:SL:ud}.

\subsection{Definitions}

We are interested in the single-particle Green function for the theory \eqref{eq:S0:ud} and \eqref{eq:SL:ud}. The structure of the exact Green functions in the Keldysh space has the standard form~\cite{Kamenev2009}: 
\begin{equation}
\begin{split}
\langle c_{\bm{q},+}(t)\overline{c}_{\bm{q},-}(t^\prime)\rangle = i\mathcal{G}_{\bm{q}}^<(t,t^\prime), \\
\langle c_{\bm{q},-}(t)\overline{c}_{\bm{q},+}(t^\prime)\rangle = i\mathcal{G}_{\bm{q}}^>(t,t^\prime) ,
\end{split}
\end{equation}
and 
\begin{equation}
\langle c_{\bm{q},\pm}(t)\overline{c}_{\bm{q},\pm}(t^\prime)\rangle {=} i\mathcal{G}_{\bm{q}}^{{T}/{\tilde{T}}}(t,t^\prime) {=} i \begin{cases}
\mathcal{G}_{\bm{q}}^{{>}/{<}}(t,t^\prime), & t{>}t^\prime ,\\
\mathcal{G}_{\bm{q}}^{{<}/{>}}(t,t^\prime), & t{<}t^\prime .
\end{cases}
\end{equation}
We note that the Green functions are $2{\times}2$ matrices in the `up/down' space. 
The standard relation $\mathcal{G}^T{+}\mathcal{G}^{\tilde{T}}{=}\mathcal{G}^>{+}\mathcal{G}^<$ holds to preserve causality.

Using the action $S_{\rm 0}$, Eq. \eqref{eq:S0:ud}, we find that the bare Green functions are diagonal in the `up/down' space,
\begin{equation}
\begin{split}
i G_{\textsf{u}/\textsf{d},\bm{q}}^>(t,t^\prime) & = (1- n_{\textsf{u}/\textsf{d},\bm{q}}) e^{\mp i \xi_q (t-t^\prime)}, \\ 
i G_{\textsf{u}/\textsf{d},\bm{q}}^<(t,t^\prime) & = - n_{\textsf{u}/\textsf{d},\bm{q}} e^{\mp i \xi_q (t-t^\prime)} .
\end{split}
\end{equation} 
Here $n_{\textsf{u},q}$ ($n_{\textsf{d},q}$) denotes the momentum distribution function for the particles in the upper (lower) band.

As usual, in order to proceed further it is convenient to make the Keldysh rotation from $c_{\pm}$ to classical and quantum components $c_{\textsf{cl}/\textsf{q}}$ \cite{Larkin1986,Kamenev2009},
\begin{equation}
\begin{split}
\begin{pmatrix}
c_{\textsf{cl}} \\
c_{\textsf{q}} \\
\end{pmatrix} 
& = \frac{1}{\sqrt{2}} 
\begin{pmatrix}
1 & 1 \\
1 & -1 
\end{pmatrix}
\begin{pmatrix}
c_{+} \\
c_{-} 
\end{pmatrix}  , \\
\begin{pmatrix}
\overline{c}_{\textsf{cl}}  &
\overline{c}_{\textsf{q}} 
\end{pmatrix} 
& = \begin{pmatrix}
\overline{c}_{+} &
\overline{c}_{-} 
\end{pmatrix}\frac{1}{\sqrt{2}} 
\begin{pmatrix}
1 & 1 \\
-1 & 1 
\end{pmatrix} .
\end{split}
\end{equation}
After the Keldysh rotation, the Green function acquires the familiar form
\begin{equation}
-i \langle c_{\bm{q}}(t) \overline{c}_{\bm{q}}(t^\prime)\rangle  =
 \mathcal{G}_{\bm{q}}(t,t^\prime) = \begin{pmatrix}
 \mathcal{G}^R_{\bm{q}}(t,t^\prime) & \mathcal{G}^K_{\bm{q}}(t,t^\prime) \\
 0 &  \mathcal{G}^A_{\bm{q}}(t,t^\prime) 
 \end{pmatrix} .
\end{equation}
The bare retarded and advanced Green functions are given by (in the energy representation)
\begin{equation}
G^{R/A}_{\bm{q}}(\varepsilon) = \frac{1}{\varepsilon-\xi_q \sigma_z \pm i0^+} .
\end{equation}
The exact and bare Green functions are related to each other via the Dyson equation,
\begin{equation}
[\mathcal{G}_{\bm{q}}^{R/A}]^{-1}=[G_{\bm{q}}^{R/A}]^{-1} - \Sigma_{\bm{q}}^{R/A} ,
\label{eq:Dyson:RA}
\end{equation}
where $\Sigma_{\bm{q}}^{R/A}$ stands for the self-energy.
The Keldysh component of the Green function is given by
\begin{equation}
\mathcal{G}_{\bm{q}}^{K}(\varepsilon)=\mathcal{G}_{\bm{q}}^{R}(\varepsilon) \Sigma^K_{\bm{q}}(\varepsilon) \mathcal{G}_{\bm{q}}^{A}(\varepsilon). 
\label{eq:Green:K:2}
\end{equation}

% %%%%%%%%%%%%%%%%%%%
% \begin{figure}[t]
% \centerline{\includegraphics[width=0.5\columnwidth]{fig2a_v2.pdf}
% \quad
% \includegraphics[width=0.5\columnwidth]{fig2b_v2.pdf}
% }
% \caption{Self-energy diagrams of the Fock- and Hartree-type  in the self-consistent Born approximation. The solid blue line is the full Green function. The dashed orange semicircle indicates the parameter $\gamma_\alpha$ controlling dissipation-induced interaction. The orange filled circles correspond to operators $\overline{\mathcal{L}}^{(\alpha)}$, while the empty circles denote operators $\mathcal{L}^{(\alpha)}$. There are similar diagrams with the operators $\overline{\mathcal{L}}^{(\alpha)}$ and $\mathcal{L}^{(\alpha)}$ interchanged.}
% \label{fig:1}
% \end{figure}
% %%%%%%%%%%%%%%%%%%%
%%%%%%%%%%%%%%%%%%%
\begin{figure}[t]
\centerline{\includegraphics[width=0.95\columnwidth]{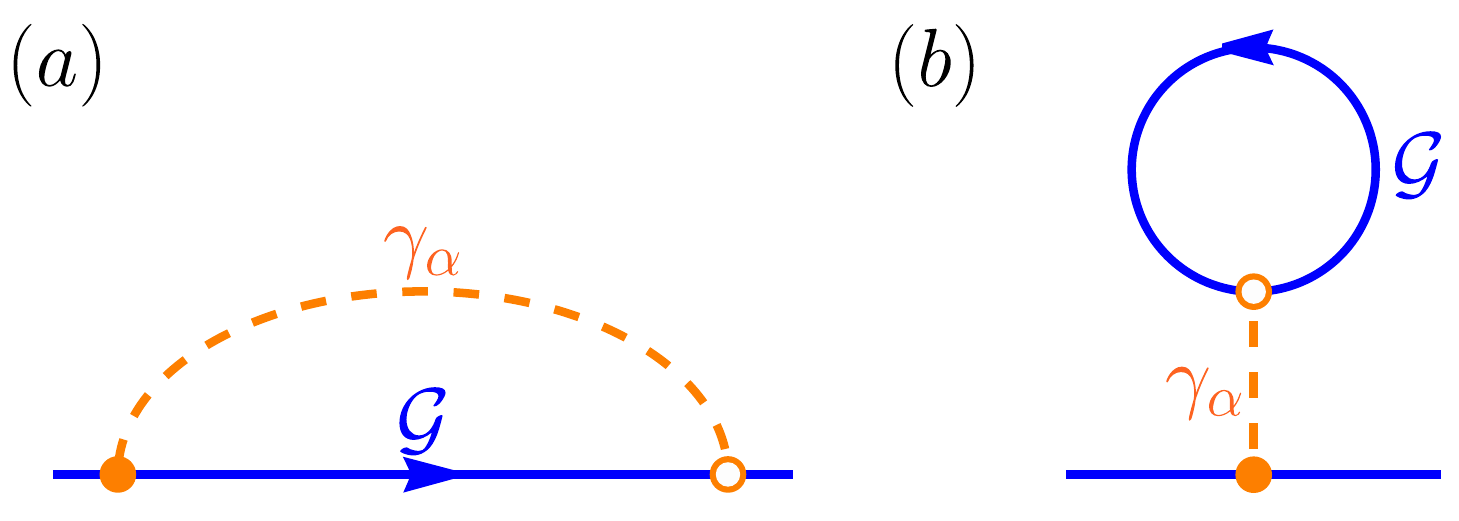}
}
\caption{Self-energy diagrams of the Fock- and Hartree-type  in the self-consistent Born approximation. The solid blue line is the full Green function. The dashed orange semicircle indicates the parameter $\gamma_\alpha$ controlling dissipation-induced interaction. The orange filled circles correspond to operators $\overline{\mathcal{L}}^{(\alpha)}$, while the empty circles denote operators $\mathcal{L}^{(\alpha)}$. There are similar diagrams with the operators $\overline{\mathcal{L}}^{(\alpha)}$ and $\mathcal{L}^{(\alpha)}$ interchanged.}
\label{fig:1}
\end{figure}
%%%%%%%%%%%%%%%%%%%

\subsection{Self-consistent equation for the self-energy}

The self-energy is induced by the dissipative part of the action, Eq. \eqref{eq:SL:ud}.
In order to compute it we consider the simplest (Born-type) diagram shown in Fig. \ref{fig:1}. We apply a self-consistent scheme, assuming that the internal line is the exact Green function (we will refer to this approximation as the ``self-consistent Born approximation'', SCBA, in analogy with the theory of disordered systems~\cite{Kamenev2009}). Then we find the following expressions for the retarded/advanced self-energy,
\begin{align}
\Sigma_q^{R/A} = &  
 \!\sum_{\alpha=1}^2 \!\gamma_\alpha\! \int\limits_{\bm{p}} \! \Bigl [\mathcal{L}^{(\alpha)}_{\bm{qq}} \tr  \bar{\mathcal{L}}^{(\alpha)}_{\bm{pp}} \mathcal{G}_{\bm{p}}^<(t,t)
-
\bar{\mathcal{L}}^{(\alpha)}_{\bm{qq}} \tr \mathcal{L}^{(\alpha)}_{\bm{pp}} \mathcal{G}_{\bm{p}}^<(t,t) \Bigr ]
\notag \\ 
+ & \!\sum_{\alpha=3}^4 \!\gamma_\alpha\! \int\limits_{\bm{p}}\! \Bigl [ \mathcal{L}^{(\alpha)}_{\bm{qq}} \tr  \bar{\mathcal{L}}^{(\alpha)}_{\bm{pp}} \mathcal{G}_{\bm{p}}^>(t,t)- 
\bar{\mathcal{L}}^{(\alpha)}_{\bm{qq}} \tr \mathcal{L}^{(\alpha)}_{\bm{pp}} \mathcal{G}_{\bm{p}}^>(t,t)
\Bigr ] 
\notag \\
\mp & \!\sum_{\alpha=1}^4 \!\gamma_\alpha\! \int\limits_{\bm{p}}\! \Bigl [
\mathcal{L}^{(\alpha)}_{\bm{qp}}\mathcal{G}_{\bm{p}}^<(t,t)\bar{\mathcal{L}}^{(\alpha)}_{\bm{pq}}
-\bar{\mathcal{L}}^{(\alpha)}_{\bm{qp}}\mathcal{G}_{\bm{p}}^>(t,t) \mathcal{L}^{(\alpha)}_{\bm{pq}}
\Bigr ] .
\label{eq:self-energy:gen:RA}
\end{align}
We stress that the Keldysh structure of terms produced by the jump operators $L_{1/2}$ and $L_{3/4}$ are different.
The Keldysh component of the self-energy becomes
\begin{equation}
\Sigma_q^K = 
2 \sum\limits_{\alpha=1}^4\gamma_\alpha\int\limits_{\bm{p}} \Bigl [
 \mathcal{L}^{(\alpha)}_{\bm{qp}}\mathcal{G}_{\bm{p}}^<(t,t)\bar{\mathcal{L}}^{(\alpha)}_{\bm{pq}}
+\bar{\mathcal{L}}^{(\alpha)}_{\bm{qp}}\mathcal{G}_{\bm{p}}^>(t,t) \mathcal{L}^{(\alpha)}_{\bm{pq}}
\Bigr ]\;.
\label{eq:self-energy:gen:K}
\end{equation}  
We note that the self-energy in the steady state is independent of time $t$, as expected. Taking into account the relation $\mathcal{G}^{>/<}_{\bm{q}}{=}(\mathcal{G}^{K}_{\bm{q}}{\pm}\mathcal{G}^{R}_{\bm{q}}{\mp}\mathcal{G}^{A}_{\bm{q}})/2$, Eqs. \eqref{eq:self-energy:gen:RA} and \eqref{eq:self-energy:gen:K} are self-consistent equations for the self-energies. 

In order to solve Eqs. \eqref{eq:self-energy:gen:RA} and \eqref{eq:self-energy:gen:K} we parametrize the exact Green functions at equal times as follows,
\begin{equation}
\mathcal{G}^<_{\bm{p}}(t,t)  = i \begin{pmatrix}
n_{\textsf{u},\bm{p}} & \eta_{\bm{p}}\\
\eta_{\bm{p}}^* & n_{\textsf{d},\bm{p}}
\end{pmatrix},  \quad
\mathcal{G}^>_{\bm{p}}(t,t)  = - i + \mathcal{G}^<_{\bm{p}}(t,t),
\label{eq:exactG:anzats}
\end{equation} 
and, consequently, 
\begin{equation}
\mathcal{G}^K_{\bm{p}}(t,t) = - i \mathcal{F}_{\bm{p}}, \quad \mathcal{F}_{\bm{p}}= \begin{pmatrix}
1-2n_{\textsf{u},\bm{p}} & -2\eta_{\bm{p}}\\
-2\eta_{\bm{p}}^* & 1-2n_{\textsf{d},\bm{p}} 
\end{pmatrix} .
\label{eq:exactG:K:anzats}
\end{equation}
Here $n_{\textsf{u},\bm{p}}$ and $n_{\textsf{d},\bm{p}}$ denotes the momentum distribution of particles in the `up' and `down' states, respectively. 
The off-diagonal parameter $\eta_{\bm{p}}$ describes possible correlations between particles in the upper and lower bands.

Using the ansatz \eqref{eq:exactG:anzats} and the expressions \eqref{eq:A:matrices}, we straightforwardly find
 a somewhat lengthy expression for the self-energy in the self-consistent Born approximation (see Fig.~\ref{fig:1}),
\begin{gather}
\Sigma^K_{\bm{q}} {=} 2 i\! \int\limits_{\bm{p}}\! \Biggl\{  \frac{d_p}{d_q}
\begin{pmatrix}
\gamma_{[1,3]} q^2{+}\gamma_{[2,4]} m^2 & (\gamma_{[2,4]}{-}\gamma_{[1,3]})  m q_+ \\
(\gamma_{[2,4]}{-}\gamma_{[1,3]})  m q_-
& \gamma_{[1,3]} m^2{+}\gamma_{[2,4]}q^2
\end{pmatrix}
\notag \\
{-}  \frac{d_q}{d_p}  \begin{pmatrix}
 \Gamma^{(12,>)}_{\bm{p}}
 & \!\! 0 \\
0 & \!\!{-}\Gamma^{(34,<)}_{\bm{p}} 
\end{pmatrix}
\notag\\
{-} 2  m \re (p_-\eta_{\bm{p}}) \frac{d_q}{d_p} \begin{pmatrix}
 \gamma_{1,2}^{(-)} 
 & 0 \\
0 & \gamma_{4,3}^{(-)}
\end{pmatrix}
\Biggr \} .
\label{eq:Sigma:K:1}
\end{gather}
Here for the sake of a brevity we introduced short-hand notations,
$\gamma_{[\alpha,\beta]}{=}\gamma_\alpha n_{\textsf{u},\bm{p}} {-}\gamma_\beta  (1-n_{\textsf{d},\bm{p}})$, $\gamma_{\alpha,\beta}^{(\pm)}{=} \gamma_\alpha \pm\gamma_\beta$, 
and
$\Gamma^{(\alpha\beta,<)}_{\bm{p}}  {=}\gamma_\alpha [p^2 n_{\textsf{u},\bm{p}}{+}m^2 n_{\textsf{d},\bm{p}}] 
 {+} \gamma_\beta [p^2 n_{\textsf{d},\bm{p}} {+} m^2 n_{\textsf{u},\bm{p}}]$.
The quantity $\Gamma^{(\alpha\beta,>)}_{\bm{p}}$ is obtained from $\Gamma^{(\alpha\beta,<)}_{\bm{p}}$ by replacing $n_{\textsf{u}/\textsf{d},\bm{p}}$ with $1{-}n_{\textsf{u}/\textsf{d},\bm{p}}$.

The retarded component of the self-energy reads
\begin{widetext}
\begin{align}
\Sigma^R_{\bm{q}} & {=}  
 i \! \int\limits_{\bm{p}}\!
\begin{pmatrix}
2i \gamma_1 [\bm{p}\times\bm{q}] n_{\textsf{u},\bm{p}}{+}2 i m \im [(\gamma_1 q_-{-}\gamma_2 p_-)\eta_{\bm{p}}] & 
(\gamma_{\{1,4\}} p_+{-}\gamma_{\{2,3\}} q_+) m 
{-}
(\gamma_{1,4}^{(+)}m^2{+}\gamma_{2,3}^{(+)}q_+p_-)\eta_{\bm{p}}  \\
(\gamma_{\{2,3\}} q_-{-}\gamma_{\{1,4\}} p_-) m  
{+}(\gamma_{1,4}^{(+)}m^2{+}\gamma_{2,3}^{(+)}q_-p_+)\eta_{\bm{p}}^* & 
-2i \gamma_4 [\bm{p}\times\bm{q}] (1-n_{\textsf{d},\bm{p}}){-} 2 i m \im [(\gamma_4 q_-{-}\gamma_3 p_-)\eta_{\bm{p}}]
\end{pmatrix}
\notag\\
 {-} & i \!\int\limits_{\bm{p}}\! \Biggl \{  \frac{d_p}{d_q}
\begin{pmatrix}
\gamma_{\{1,3\}} q^2{+}\gamma_{\{2,4\}} m^2 & (\gamma_{\{4,2\}}{-}\gamma_{\{3,1\}})  m q_+ \\
(\gamma_{\{4,2\}}{-}\gamma_{\{3,1\}})  m q_-
& \gamma_{\{1,3\}} m^2{+}\gamma_{\{2,4\}}q^2
\end{pmatrix}
{+} \frac{d_q}{d_p}\left [  \begin{pmatrix}
\Gamma^{(12,>)}_{\bm{p}} & \!\!0 \\
0 &  \!\!\Gamma^{(34,<)}_{\bm{p}} 
\end{pmatrix} 
{+}2 m \re (p_-\eta_{\bm{p}})  \begin{pmatrix}
  \gamma_{1,2}^{(-)}
 & 0 \\
0 &  \gamma_{4,3}^{(-)}
\end{pmatrix}
\right ]
\Biggr \}
,
\label{eq:Sigma:RA:1}
\end{align}
\end{widetext}
where $\gamma_{\{\alpha,\beta\}}{=}\gamma_\alpha n_{\textsf{u},\bm{p}} {+}\gamma_\beta  (1-n_{\textsf{d},\bm{p}})$. 
The expression for the advanced self-energy $\Sigma_{\bm{q}}^A$ can be obtained from Eq. \eqref{eq:Sigma:RA:1} by hermitian conjugation.  

The self-consistent Green functions can be then written as follows
% \begin{equation}
% [\underline{\mathcal{G}}^{R/A}_{\bm{q}}(\varepsilon)]^{-1}=\varepsilon-\xi_q \sigma_z -  \Sigma_{\bm{q}}^{R/A} , 
% \end{equation}
\begin{gather}
[\underline{\mathcal{G}}^{R/A}_{\bm{q}}(\varepsilon)]^{-1}=\varepsilon-\xi_q \sigma_z -  \Sigma_{\bm{q}}^{R/A} , \notag \\
 \underline{\mathcal{G}}^{K}_{\bm{q}}(\varepsilon)=\underline{\mathcal{G}}^{R}_{\bm{q}}(\varepsilon)
\Sigma_{\bm{q}}^{K} \underline{\mathcal{G}}^{A}_{\bm{q}}(\varepsilon) .
\end{gather}
% whereas the Keldysh Green function reads as
% \begin{equation}
%      \underline{\mathcal{G}}^{K}_{\bm{q}}(\varepsilon)=\underline{\mathcal{G}}^{R}_{\bm{q}}(\varepsilon)
% \Sigma_{\bm{q}}^{K} \underline{\mathcal{G}}^{A}_{\bm{q}}(\varepsilon).
% \end{equation}
Taking into account the parametrization \eqref{eq:exactG:K:anzats}, the self-consistent equation for $n_{\textsf{u}/\textsf{d},\bm{q}}$ and $\eta_{\bm{q}}$ becomes
\begin{equation}
\underline{\mathcal{G}}^{K}_{\bm{q}}(t,t)=\int \frac{d\varepsilon}{2\pi} \,\underline{\mathcal{G}}^{R}_{\bm{q}}(\varepsilon)
\Sigma_{\bm{q}}^{K} \underline{\mathcal{G}}^{A}_{\bm{q}}(\varepsilon).
\label{eq:self:Born}
\end{equation}

\subsection{The dark state solution}

Using Eqs.~\eqref{eq:Sigma:K:1} and \eqref{eq:Sigma:RA:1}, one can check  that 
the self-consistent equations \eqref{eq:self-energy:gen:RA} and \eqref{eq:self-energy:gen:K} have the solution 
\begin{equation}
n_{\textsf{u},\bm{p}}=\eta_{\bm{p}}=0 , \qquad n_{\textsf{d},\bm{p}}= 1,
\label{eq:steady:self-consistent Born approximation}
\end{equation}
corresponding to the dark state. In the self-consistent Born approximation the retarded/advanced Green functions become
\begin{equation}
\underline{\mathcal{G}}^{R/A}_{\bm{q}}(\varepsilon) = \begin{pmatrix}
\frac{1}{\varepsilon-\xi_q \pm i \gamma_{\textsf{u}} n d_q} & 0 \\
0 & \frac{1}{\varepsilon+\xi_q \pm i \gamma_{\textsf{d}} n d_q}  
\end{pmatrix} ,
\label{eq:GF:self-consistent Born approximation:RA}
\end{equation}
where we introduced 
\begin{equation}
\gamma_{\textsf{u}/\textsf{d}} = \int\limits_{\bm{p}} \frac{\gamma_{1/4} p^2+\gamma_{2/3}m^2}{n(p^2+m^2)}, \qquad n=\int\limits_{\bm{p}} 1 .
\end{equation}
We note that the total particle density $n$ is determined by the ultra-violet scale (lattice spacing). The dimensionless quantity $\gamma_{\textsf{u}/\textsf{d}} n$ determines the decay rate for single-particle excitations in the upper and lower bands, respectively. In what follows, we take for simplicity $\gamma_{\alpha}{=}\gamma$ for all $\alpha$, so that $\gamma_{\textsf{u}/\textsf{d}}{=}\gamma$. We note though that our main results do not depend on this simplification. In Appendix \ref{App:AppendixSCeq} we demonstrate that the dark state, cf.\ Eq.~\eqref{eq:steady:self-consistent Born approximation}, is the only solution for the Green function in the self-consistent Born approximation.

It is worth emphasizing that the same single-particle decay rate as in Eq.~\eqref{eq:GF:self-consistent Born approximation:RA} can be formally reproduced by means of the following replacement: $L_{1/2}{\rightarrow} l_{\sf{u}},\; L_{3/4}{\rightarrow} l_{\sf{d}}^\dagger,\; \gamma {\rightarrow} \gamma n$, which was initially suggested in Ref.~\cite{Tonielli2020}  as some sort of ``mean-field" (MF) approach to the present problem. However, the underlying physical mechanism for such a decay rate is different: in the MF model, the decay rate corresponds to the fact that particles can escape the system completely, and has nothing to do with transitions between the upper and lower bands (in fact, the MF model does not couple them at all). On the other hand, in the full model, the SCBA single-particle decay rate is purely elastic, and corresponds to momentum relaxation within the upper (lower) band, analogously to the single-particle decay rate induced by disorder in closed systems~\cite{Kamenev2009}.  

The Keldysh component of the Green function for the dark state can be expressed in terms of retarded and advanced Green functions as
\begin{equation}
\underline{\mathcal{G}}^{K}_{\bm{q}}(\varepsilon)= \underline{\mathcal{G}}^{R}_{\bm{q}}(\varepsilon)\sigma_z - \sigma_z \underline{\mathcal{G}}^{A}_{\bm{q}}(\varepsilon) .
\label{eq:GF:self-consistent Born approximation:K}
\end{equation}

In Appendix \ref{App:AppendixSelf-Cons} we compute the self-energy to the second-order in $\gamma$, thus going beyond the self-consistent Born approximation. This analysis demonstrates that the dark state solution \eqref{eq:steady:self-consistent Born approximation} exists beyond the self-consistent Born approximation as expected, since, as noted above, the Lindbladian dynamics \eqref{eq:GKSL} was designed so as to have $\rho_D = |D\rangle \langle D|$ as a steady state.
We further find that corrections to the self-consistent Born approximation are controlled by the small parameter $m^d/n{\ll} 1$.

In the next section, we consider relaxation dynamics induced by perturbing the dark state with external fields. As we shall see below, the total numbers of particles in the upper and lower bands (denoted as $N_{\sf{u}}$ and $N_{\sf{d}}$, respectively) are both approximately conserved within the self-consistent Born approximation at the {\it linear} order in the density deviations from the dark state, precluding exponential decay of $N_{\sf{u}}$. However, at the quadratic order in the density deviations, recombination of particles in the upper band and holes in the lower band emerges (see Sec.~\ref{Sec:Rec}), favoring eventual slow decay of $N_{\sf{u}}$ at late times. This behavior is strongly opposed to the ``mean-field" predictions, which imply exponentially fast decay of $N_{\sf{u}}$ {\it irrespective} of the density of holes in the lower band. In addition, the effect of corrections beyond the SCBA on the time evolution of $N_{\sf{u}}$ is discussed in Sec.~\ref{Sec:Pumping}.

% We note that the self-consistent Born approximation solution for the Green functions corresponds to the approximation in which Lindblad operators, cf. Eq. \eqref{eq:Jump:Operators:def}, are simplified as  $L_{1/2}{\to} \ell_{\textsf{u}}$ and $L_{3/4}{\to} \bar{\ell}_{\textsf{d}}$ \cite{Tonielli2020}, while $\gamma$ is substituted by $\bar{\gamma}{=}\gamma n$. In this approximation, the numbers of particles in the upper and lower bands are conserved separately. Below we shall study the implications of this approximate conservation law.  

\section{Linear response\label{Sec:Linear}}

Let us now consider the response of the density of the particles in the `up' and `down' states to an external scalar potential $\phi(\bm{x},t)$ coupled to the density of  original fermions $\psi$. In the basis of the `up' and `down' states such a scalar potential transforms into a matrix in the ${\sf{u}}/{\sf{d}}$ space which is non-local in coordinate space,
\begin{equation}
\begin{split}
\Phi(\bm{x},\bm{x^\prime},t) = 
\int\limits_{\bm{pq}\omega} \Phi_{\bm{p},\bm{q};\omega}\,  e^{-i\bm{p}(\bm{x}-\bm{x^\prime})-i\bm{q}(\bm{x}+\bm{x^\prime})/2-i\omega t},  \\
\Phi_{\bm{p},\bm{q};\omega} = U_{\bm{p_+}}U^{\dag}_{\bm{p_-}} \phi_{\bm{q},\omega}, \, \phi_{\bm{q},\omega}=
\int\limits_{\bm{x}t} \phi(\bm{x},t) e^{i\bm{q}\bm{x}+i \omega t} ,
\end{split}
\label{eq:ext:pot:1}
\end{equation} 
where $\bm{p_\pm} {=} \bm{p}{\pm} \bm{q}/2$, and $\int_\omega {\equiv} \int_{-\infty}^\infty d\omega/(2\pi)$ stands for the integration in the frequency domain. We assume that it has only the classical component, i.e., that it is the same for the `+' and `-' parts of the Keldysh contour, $\Phi_{\pm}(\bm{x},\bm{x^\prime};t){=}\Phi(\bm{x},\bm{x^\prime};t)$. Under such a perturbation 
the bare retarded and advanced Green functions get modified 
\begin{equation}
[G^{R/A}]^{-1}\to [G^{R/A}]^{-1} - \Phi. 
\end{equation}
In turn, the presence of $\Phi$ affects the self-energies and the exact Green functions.

%%%%%%%%%%%%%%%%%%%
\begin{figure*}[t]
\centerline{\includegraphics[width=0.46\columnwidth]{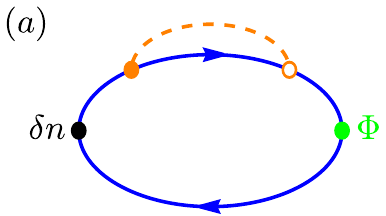}\hspace{0.05\columnwidth}\includegraphics[width=0.46\columnwidth]{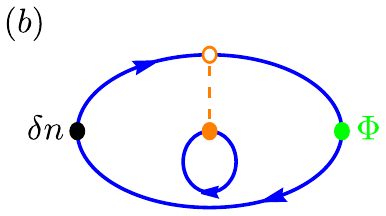}
\hspace{0.05\columnwidth}\includegraphics[width=0.46\columnwidth]{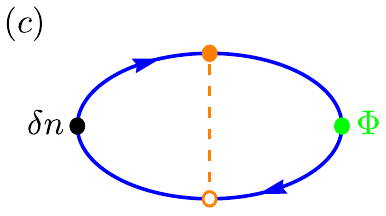}
\hspace{0.05\columnwidth}\includegraphics[width=0.46\columnwidth]{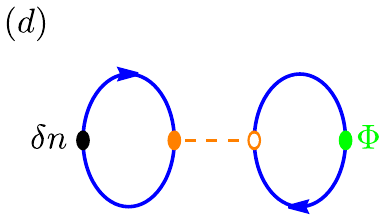}
}
\caption{Diagrams for the linear response. (a), (b) Polarization bubbles with Green functions in the self-consistent Born approximation. (c), (d) Polarization bubbles with vertex corrections. $\delta n$ stands for either of $\delta u$, $\delta d$, $\delta \eta$.}
\label{fig:2}
\end{figure*}
%%%%%%%%%%%%%%%%%%%

We note that due to the presence of $\Phi$, the Green functions become dependent on both spatial coordinates and both times. For example, Eq. \eqref{eq:Green:K:2} now reads
\begin{gather}
\mathcal{G}^K(\bm{x},t;\bm{x^\prime},t^\prime) = \int\limits_{\bm{y},\bm{y^\prime},t_1,t_2} \mathcal{G}^R(\bm{x},t;\bm{y},t_1) \Sigma^K(\bm{y},t_1;\bm{y^\prime},t_2) 
\notag \\
\times \mathcal{G}^A(\bm{y^\prime},t_2;\bm{x^\prime},t^\prime) ,
\end{gather}
or $\mathcal{G}^K {=} \mathcal{G}^R {\circ} \Sigma^K {\circ} \mathcal{G}^A$
in the short-hand notation which we shall use below.

In this and the following section we will employ a trick which allows to account for vertex corrections without having to sum over an infinite series of diagrams.
The change of the retarded and advanced Green functions due to the external potential $\Phi$ (to the linear order) is given as, cf. Eq. \eqref{eq:Dyson:RA},  
\begin{equation}
\delta  \mathcal{G}^{R/A} = \mathcal{G}^{R/A} \circ \Phi \circ \mathcal{G}^{R/A} + \mathcal{G}^{R/A} \circ \delta \Sigma^{R/A}\circ \mathcal{G}^{R/A} .
\label{eq:deltaGRA}
\end{equation} 
The shift of the Keldysh Green function due to the external potential reads 
\begin{gather}
\delta \mathcal{G}^K {=} \mathcal{G}^R \circ \delta \Sigma^K \circ \mathcal{G}^A {+} \delta \mathcal{G}^R \circ \Sigma^K \circ \mathcal{G}^A{+}\mathcal{G}^R \circ \Sigma^K \circ \delta \mathcal{G}^A \notag \\
{=} \mathcal{G}^{R}\circ \Phi \circ \mathcal{G}^{K} {+} \mathcal{G}^{K}\circ \Phi \circ \mathcal{G}^{A}
{+}\mathcal{G}^{R} \circ \delta \Sigma^{K} \circ \mathcal{G}^{A} \notag \\
{+} \mathcal{G}^{R} \circ \delta \Sigma^{R} \circ \mathcal{G}^{K}
{+} \mathcal{G}^{K} \circ \delta \Sigma^{A} \circ \mathcal{G}^{A} .
\label{eq:deltaGK:K}
\end{gather} 
Here we used the relation \eqref{eq:deltaGRA} in the last line. The variation of the self-energy can be read from Eqs. \eqref{eq:self-energy:gen:RA} and \eqref{eq:self-energy:gen:K}. First, one needs to rewrite them in the coordinate representation. Second, one needs to take into account that $\delta \mathcal{G}^{R/A}(t,t){=}0$. This follows from Eq. \eqref{eq:deltaGRA} since, as we shall see below, the variation of the self-energy is non-zero at coinciding times only. Therefore, we can write $\delta\mathcal{G}^{>/<}(x, t;x^\prime, t)=\delta\mathcal{G}^K(x, t;x^\prime, t)/2$.
Thus we find the following change of the Keldysh self-energy
\begin{gather}
\delta \Sigma^{K}(\bm{x},\!t;\!\bm{x^\prime},\!t^\prime){=} \gamma \int\limits_{\bm{y},\bm{y^\prime},\bm{z}}\!\! \Bigl[ \!\mathcal{L}^{(\alpha)}(\bm{x}{-}\bm{z},\bm{y}{-}\bm{z})\delta \mathcal{G}^K(\bm{y},\!t;\!\bm{y^\prime},\!t)\notag \\
\times \bar{\mathcal{L}}^{(\alpha)}(\bm{y^\prime}{-}\bm{z},\bm{x^\prime}{-}\bm{z})
 + (\mathcal{L} \leftrightarrow \bar{\mathcal{L}}) 
 \Bigr ] \delta(t-t^\prime) .
 \label{eq:dSigmaK:35}
\end{gather}
Here and afterwards the summation over repeated index $\alpha$ is assumed. We introduced the coordinate representation for the matrices $\mathcal{L}^{(\alpha)}$ and $\bar{\mathcal{L}}^{(\alpha)}$:
\begin{equation}
\mathcal{L}^{(\alpha)}(\bm{x},\bm{y}) {=} \!\int\limits_{\bm{qp}} e^{i\bm{p}\bm{y}{-}i\bm{q}\bm{x}} \mathcal{L}^{(\alpha)}_{\bm{qp}} \;,
\end{equation}
and similarly for $\bar{\mathcal{L}}^{(\alpha)}(\bm{x},\bm{y})$.
The variations of the retarded/advanced self-energies are given by
\begin{gather}
\delta \Sigma^{R/A}
(\bm{x},t;\bm{x^\prime},t^\prime)
{=} -\frac{\gamma}{2} \int\limits_{\bm{y},\bm{y^\prime},\bm{z}}\!\! \Bigl[ \bar{\mathcal{L}}^{(\alpha)}(\bm{x}{-}\bm{z},\bm{x^\prime}{-}\bm{z}) 
\notag \\
\times
 \tr \mathcal{L}^{(\alpha)}(\bm{y}{-}\bm{z},\bm{y^\prime}{-}\bm{z}) \delta \mathcal{G}^K(\bm{y^\prime},t;\bm{y},t) 
 \pm \mathcal{L}^{(\alpha)}(\bm{x}{-}\bm{z},\bm{y}{-}\bm{z})
\notag \\
\times \delta \mathcal{G}^K(\bm{y},t;\bm{y^\prime},t)
 \bar{\mathcal{L}}^{(\alpha)}(\bm{y^\prime}{-}\bm{z},\bm{x^\prime}{-}\bm{z})
- (\mathcal{L} \leftrightarrow \bar{\mathcal{L}}) 
\Bigr ]\delta(t-t^\prime) .
\label{eq:dSigmaRA:35}
\end{gather}
We note that the variation of the Keldysh Green function determines the change in the density of the particles in the upper and lower bands due to the application of the external potential, 
\begin{equation}
\delta \hat{n}(\bm{x},t) = -
\frac{i}{2} 
\delta \mathcal{G}^K(\bm{x},t;\bm{x},t) .
\end{equation}

Next 
we obtain the closed-form equation for the change of the Keldysh component of the Green's function due to the presence of the external potential
\begin{widetext}
\begin{gather}
\Pi_{\bm{p},\bm{q};\omega}{=} \delta \mathcal{G}^K_{\bm{p},\bm{q};\omega} 
{-}\gamma \! \int\limits_{\bm{k},\varepsilon} \! \mathcal{G}^{R}_{\bm{p_+},\varepsilon_+}
\Bigl [
 \mathcal{L}^{(\alpha)}_{\bm{p_+k_+}}\delta \mathcal{G}^K_{\bm{k},\bm{q};\omega}\bar{\mathcal{L}}^{(\alpha)}_{\bm{k_-p_-}}{+}\bar{\mathcal{L}}^{(\alpha)}_{\bm{p_+k_+}}\delta \mathcal{G}^K_{\bm{k},\bm{q};\omega}\mathcal{L}^{(\alpha)}_{\bm{k_-p_-}} \Bigr ] 
 \mathcal{G}^{A}_{\bm{p_-},\varepsilon_-}
{+}\frac{\gamma}{2} \int\limits_{\bm{k},\varepsilon} \mathcal{G}^{R}_{\bm{p_+},\varepsilon_+}\Bigl [
\bar{\mathcal{L}}^{(\alpha)}_{\bm{p_+p_-}} \tr \mathcal{L}^{(\alpha)}_{\bm{k_- k_+}} \delta \mathcal{G}^K_{\bm{k},\bm{q};\omega}
\notag \\
{-} \mathcal{L}^{(\alpha)}_{\bm{p_+p_-}} \tr  \bar{\mathcal{L}}^{(\alpha)}_{\bm{k_- k_+}} \delta \mathcal{G}^K_{\bm{k},\bm{q};\omega} {+} 
 \mathcal{L}^{(\alpha)}_{\bm{p_+k_+}}\delta \mathcal{G}^K_{\bm{k},\bm{q};\omega}\bar{\mathcal{L}}^{(\alpha)}_{\bm{k_-p_-}} {-}\bar{\mathcal{L}}^{(\alpha)}_{\bm{p_+k_+}}\delta \mathcal{G}^K_{\bm{k},\bm{q};\omega}\mathcal{L}^{(\alpha)}_{\bm{k_-p_-}}
\Bigr ]
 \mathcal{G}^{K}_{\bm{p_-},\varepsilon_-}
{+}\frac{\gamma}{2} \int\limits_{\bm{k},\varepsilon} \mathcal{G}^{K}_{\bm{p_+},\varepsilon_+}\Bigl [
\bar{\mathcal{L}}^{(\alpha)}_{\bm{p_+p_-}} \tr \mathcal{L}^{(\alpha)}_{\bm{k_- k_+}} \delta \mathcal{G}^K_{\bm{k},\bm{q};\omega}
 \notag \\
{-} \mathcal{L}^{(\alpha)}_{\bm{p_+p_-}} \tr  \bar{\mathcal{L}}^{(\alpha)}_{\bm{k_- k_+}} \delta \mathcal{G}^K_{\bm{k},\bm{q};\omega} {-} 
 \mathcal{L}^{(\alpha)}_{\bm{p_+k_+}}\delta \mathcal{G}^K_{\bm{k},\bm{q};\omega}\bar{\mathcal{L}}^{(\alpha)}_{\bm{k_-p_-}} +\bar{\mathcal{L}}^{(\alpha)}_{\bm{p_+k_+}}\delta \mathcal{G}^K_{\bm{k},\bm{q};\omega}\mathcal{L}^{(\alpha)}_{\bm{k_-p_-}}
\Bigr ]
 \mathcal{G}^{A}_{\bm{p_-},\varepsilon_-}
.
 \label{eq:self-consistent-pol-oper:2}
\end{gather}
\end{widetext}
Here, we introduced for brevity, $\bm{p_\pm} {=} \bm{p}{\pm} \bm{q}/2$, $\bm{k_\pm} {=} \bm{k}{\pm} \bm{q}/2$ and $\varepsilon_\pm {=} \varepsilon{\pm} \omega/2$. 
Also we used the following 
representation 
for 
the spatial and temporal dependence of 
the Keldysh 
component of the 
Green's function,
\begin{equation}
\delta G^K(\bm{x},t;\bm{x^\prime},t) {=} \!
\int\limits_{\bm{pq}\omega} \! \delta G^K_{\bm{p},\bm{q};\omega} 
e^{-i\bm{p_+}\bm{x}+i \bm{p_-}\bm{x^\prime}-i\omega t} .
\end{equation}

The bare density-density bubble (retarded polarization operator) is given by 
\begin{equation}
\Pi_{\bm{p},\bm{q};\omega}= \int\limits_\varepsilon \Bigl [ {\mathcal{G}}^{R}_{\bm{p_+},\varepsilon_+} \Phi_{\bm{p},\bm{q};\omega} {\mathcal{G}}^{K}_{\bm{p_-},\varepsilon_-} +{\mathcal{G}}^{K}_{\bm{p_+},\varepsilon_+}\Phi_{\bm{p},\bm{q};\omega} {\mathcal{G}}^{A}_{\bm{p_-},\varepsilon_-}\Bigr ] .\label{eq:Pi_integral_exp}
\end{equation}
We note that the terms proportional to $\gamma$ on the right-hand side of Eq. \eqref{eq:self-consistent-pol-oper:2} describe vertex corrections (see Fig. \ref{fig:2}).

%%%%%%%%%%%%%%%%%%%
\begin{figure*}[t]
\centerline{\includegraphics[width=0.95\textwidth]{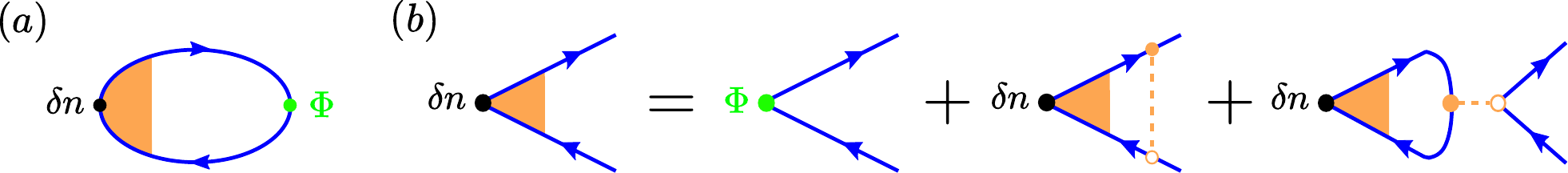}
}
\caption{Sketch of the diagrams for the density-density response corresponding to Eqs.~\eqref{eq:eq:du:1} and \eqref{eq:Diff:deltaEta}. $\delta n$ stands for either of $\delta u$, $\delta d$, $\delta \eta$. The response is given by the bubble diagram (a), with the density vertex (b) Here the blue curves correspond to the Green functions computed in the self-consistent Born approximation.}
\label{fig:3n}
\end{figure*}
%%%%%%%%%%%%%%%%%%%

Replacing the Green functions in the density-density bubble by their expressions in the self-consistent Born approximation, cf. Eqs. \eqref{eq:GF:self-consistent Born approximation:RA} and \eqref{eq:GF:self-consistent Born approximation:K}, we find
\begin{gather}
\Pi_{\bm{p},\bm{q};\omega}
= 
 \begin{pmatrix}
 0& 
 \frac{2 i \Phi^{\textsf{(ud)}}_{\bm{p},\bm{q};\omega}}{\omega-\xi_{\bm{p},\bm{q}}+i\bar{\gamma} d_{\bm{p},\bm{q}}} \\
\frac{-2 i \Phi^{\textsf{(du)}}_{\bm{p},\bm{q};\omega}}{\omega+\xi_{\bm{p},\bm{q}}+i\bar{\gamma} d_{\bm{p},\bm{q}}}& 0
 \end{pmatrix} ,
\end{gather}
where $\xi_{\bm{p},\bm{q}}{=}\xi_{\bm{p_+}}{+}\xi_{\bm{p_-}}$ and $d_{\bm{p},\bm{q}}{=}d_{\bm{p_+}}{+}d_{\bm{p_-}}$, while, as before, $\bar{\gamma}{=}\gamma n$.
We also represented $\Phi_{\bm{p},\bm{q};\omega}$ as an auxiliary $2{\times}2$ matrix in $u/d$-space,
\begin{gather}
\Phi_{\bm{p},\bm{q};\omega} = 
\begin{pmatrix}
\Phi^{\textsf{(uu)}}_{\bm{p},\bm{q};\omega} & \Phi^{\textsf{(ud)}}_{\bm{p},\bm{q};\omega}\\
\Phi^{\textsf{(du)}}_{\bm{p},\bm{q};\omega} & \Phi^{\textsf{(dd)}}_{\bm{p},\bm{q};\omega}
\end{pmatrix} .
\end{gather}
We emphasize that the matrix $\Pi_{\bm{p},\bm{q};\omega}$ has
vanishing diagonal elements, which follows from the fact that the distribution function for the dark state $\mathcal{F}_{\bm{p}}{=}\sigma_z$ (see Eq.~\eqref{eq:exactG:K:anzats}) is momentum independent. As a result, both poles of the integrand in Eq.~\eqref{eq:Pi_integral_exp} are always in the same complex half-plane, and the integral vanishes. Note that this cancellation has nothing to do with the symmetries of the Hamiltonian or with the relation between the external frequency and the spectral gap (the latter becomes important for linear response of the off-diagonal modes $\eta_{\bm{p},\bm{q},\omega}$ since they have a finite lifetime). In the case of an external scalar potential applied to the $\psi$-particles, cf. Eq. \eqref{eq:ext:pot:1}, the off-diagonal components of $\Phi_{p,q;\omega}$ read
\begin{equation}
\Phi^{\textsf{(ud)}}_{\bm{p},\bm{q};\omega} = - \frac{m(q_x+i q_y)}{\sqrt{d_{\bm{p_+}}d_{\bm{p_-}}}}\phi_{\bm{q},\omega}, \, 
\Phi^{\textsf{(du)}}_{\bm{p},\bm{q};\omega} = \frac{m(q_x-i q_y)}{\sqrt{d_{\bm{p_+}}d_{\bm{p_-}}}}\phi_{\bm{q},\omega} .
\label{eq:Phi:ud}
\end{equation} 

In order to solve Eq. \eqref{eq:self-consistent-pol-oper:2}, we use the following parametrization
 \begin{equation}
 \delta \mathcal{G}^K_{\bm{p},\bm{q};\omega}= 2 i \begin{pmatrix}
 \delta u_{\bm{p},\bm{q};\omega} & \delta \eta_{\bm{p},\bm{q};\omega} \\
 \delta \eta^*_{\bm{p},\bm{q};\omega} & \delta d_{\bm{p},\bm{q};\omega}
 \end{pmatrix} .
 \label{eq:dGK:par}
 \end{equation}
In addition, we replace all the Green functions by their expressions in the self-consistent Born approximation, cf. Eqs.~\eqref{eq:GF:self-consistent Born approximation:RA} and \eqref{eq:GF:self-consistent Born approximation:K}. We then arrive at the following equation for $\delta u_{\bm{p},\bm{q};\omega}$,
\begin{gather}
2 i   \delta u_{\bm{p},\bm{q};\omega}  {+}  \frac{4 \gamma}{\sqrt{d_{\bm{p_+}}d_{\bm{p_-}}}}\frac{d_p{-}q^2/4{+}i[\bm{p}{\times}\bm{q}]}{[\omega{+}\xi_{\bm{p_-}}{-}\xi_{\bm{p_+}} {+} i\bar{\gamma}(d_{\bm{p_-}}{+}d_{\bm{p_+}})]} \notag \\
\times \int\limits_{\bm{k}} \sqrt{d_{\bm{k_+}}d_{\bm{k_-}}} \, \delta u_{\bm{k},\bm{q};\omega}
= \Pi^{\textsf{(uu)}}_{\bm{p},\bm{q};\omega}\equiv 0 .
\label{eq:eq:du:1}
\end{gather}
Therefore, the external potential does not induce a change in the density of the `up' band, $\delta u_{\bm{p},\bm{q};\omega}{=}0$, and similarly for the `down' band, $\delta d_{\bm{p},\bm{q};\omega}{=}0$. We emphasize, however, that linear response is absent only for the densities of the eigenmodes (i.e. $\delta u_{\bm{p},\bm{q};\omega}$ and $\delta d_{\bm{p},\bm{q};\omega}$). In contrast, the response of the density of original $\psi$-fermions $\psi_1^\dagger\psi_1{+}\psi_2^\dagger\psi_2$ is non-zero because it also involves contributions from the off-diagonal mode (i.e. the matrix element $\eta_{\bm{p},\bm{q},\omega}$ in the Keldysh Green function, $\delta \mathcal{G}^K_{\bm{p},\bm{q};\omega}$, see Eq.~\eqref{eq:dGK:par}). The decay of this mode is governed by the following equation
\begin{gather}
\Bigl [ \xi_{\bm{p_+}}{+}\xi_{\bm{p_-}} {-} i\bar{\gamma}(d_{\bm{p_+}}{+}d_{\bm{p_-}}) {-} \omega\Bigr ] \delta\eta_{\bm{p},\bm{q};\omega} 
{-} i\gamma \int\limits_{\bm{k}} \delta\eta_{\bm{k},\bm{q};\omega}
 \Biggl[
\Bigl(m^2
\notag \\
{+} \bm{k_+}\bm{p_+}{+}i [\bm{k_+}{\times}\bm{p_+}]\Bigr )
\sqrt{\frac{d_{\bm{k_-}}d_{\bm{p_-}}}{d_{\bm{k_+}}d_{\bm{p_+}}}}
{+} \Bigl(m^2{+} \bm{k_-}\bm{p_-}
\notag \\
{+}i [\bm{k_-}{\times}\bm{p_-}]\Bigr )
\sqrt{\frac{d_{\bm{k_+}}d_{\bm{p_+}}}{d_{\bm{k_-}}d_{\bm{p_-}}}}
\,
\Biggr ]
{=} -\Phi^{\textsf{(ud)}}_{\bm{p},\bm{q};\omega} .
\label{eq:Diff:deltaEta}
\end{gather}
Taking the limit $q{\to} 0$, we find the following solution 
\begin{gather}
 \delta\eta_{\bm{p},\bm{q};\omega}  {\simeq} {}\frac{1}{\omega{-}2\xi_{p}{+} 2 i\bar{\gamma}d_{p}} \Biggl [\Phi^{\textsf{(ud)}}_{\bm{p},\bm{q};\omega} {-}
  \frac{\int_{\bm{k}} \frac{2i\gamma m^2 \Phi^{\textsf{(ud)}}_{\bm{k},\bm{q};\omega}}{\omega{-}2\xi_{k}{+} 2 i\bar{\gamma}d_{k}}}{1{+}\int_{\bm{k}}\frac{2i\gamma m^2}{\omega-2\xi_{k}{+} 2 i\bar{\gamma}d_{k}}}
  \Biggr ]\;.
  \label{eq:delta:eta}
\end{gather}
 We mention that the second term in the square brackets on the right-hand side of Eq.~\eqref{eq:delta:eta} describes the effect of vertex corrections. They correspond to the summation of the infinite series of ladder-type diagrams (see Fig.~\ref{fig:3n}).
 However, these vertex corrections for $\delta\eta$ disappear in the limit $\bar{\gamma}{\to} 0$. Eq.~\eqref{eq:delta:eta} 
 has clear physical meaning: In order to excite a particle from the filled `down' band to the empty `up' band, the external potential has to overcome the energy gap equal to $2\xi_p$. The decay rate of such an exciting state is finite and is given by  $1/\tau_p{=}2\bar{\gamma} d_p$. In the time domain Eq.~\eqref{eq:delta:eta} translates into exponential decay of the linear response with the rate $1/\tau{=}2\bar{\gamma} m^2$  (cf. Ref.~\cite{Tonielli2020}).

\section{Nonlinear response\label{Sec:Nonlinear}}

Since the external potential $\Phi$ does not lead to finite $\delta u$ and $\delta d$ within linear response, let us compute the second order response.
The variations of the Green function to second order in $\Phi$ can be written straightforwardly, 
\begin{align}
{\,}\hspace{-0.25cm}\delta \mathcal{G}^K & {=}  \delta \mathcal{G}^R {\circ} \Sigma^K {\circ} \mathcal{G}^A
{+} \mathcal{G}^R {\circ} \Sigma^K {\circ} \delta \mathcal{G}^A{+} \mathcal{G}^R {\circ} \delta \Sigma^K {\circ} \mathcal{G}^A 
\notag \\
{+} &  \delta \mathcal{G}^R {\circ} \Sigma^K {\circ} \delta \mathcal{G}^A {+} \delta \mathcal{G}^R {\circ} \delta\Sigma^K {\circ} \mathcal{G}^A
{+} \mathcal{G}^R {\circ} \delta \Sigma^K {\circ} \delta \mathcal{G}^A, 
\label{eq:dGK:2}
\end{align}
and
\begin{gather}
\delta \mathcal{G}^{R} {=} \mathcal{G}^{R}{\circ} \Phi {\circ} \mathcal{G}^{R}{+}
\mathcal{G}^{R} {\circ} \Phi {\circ} \mathcal{G}^{R}
 {\circ} \Phi {\circ} \mathcal{G}^{R}  {+} \mathcal{G}^{R} {\circ} \delta \Sigma^{R} {\circ} \mathcal{G}^{R} 
 \notag\\
 {+} \mathcal{G}^{R}{\circ} \Phi {\circ} \mathcal{G}^{R} {\circ} \delta \Sigma^{R} {\circ} \mathcal{G}^{R}
 {+} \mathcal{G}^{R} {\circ} \delta \Sigma^{R} {\circ} \mathcal{G}^{R} {\circ} \Phi {\circ} \mathcal{G}^{R} 
 \notag \\
 {+} \mathcal{G}^{R} {\circ} \delta \Sigma^{R} {\circ} \mathcal{G}^{R}{\circ} \delta \Sigma^{R} {\circ} \mathcal{G}^{R} .
 \label{eq:dGR:2}
\end{gather}
The expression for $\delta \mathcal{G}^{A}$ can be obtained from Eq. \eqref{eq:dGR:2} after replacing all the retarded Green functions by the corresponding advanced ones.

%%%%%%%%%%%%%%%%%%%
\begin{figure*}[t]
\centerline{\includegraphics[width=0.45\columnwidth]{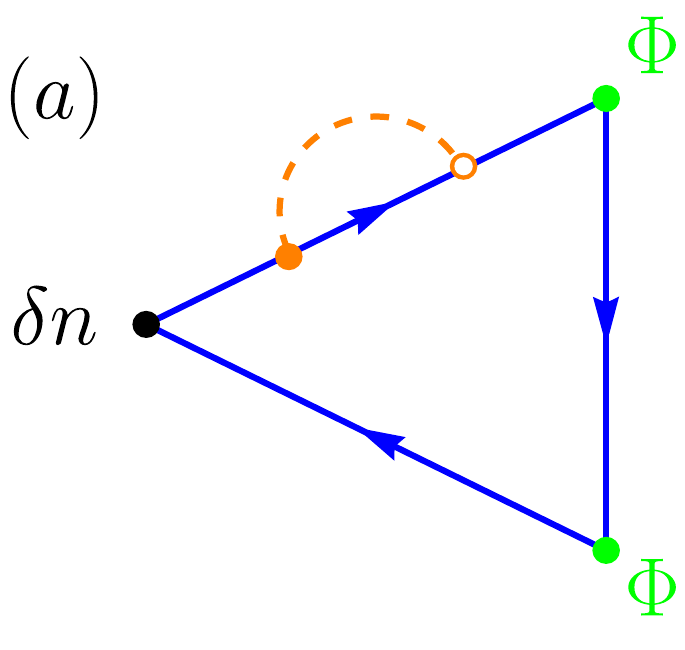}
\hspace{0.05\columnwidth}\includegraphics[width=0.45\columnwidth]{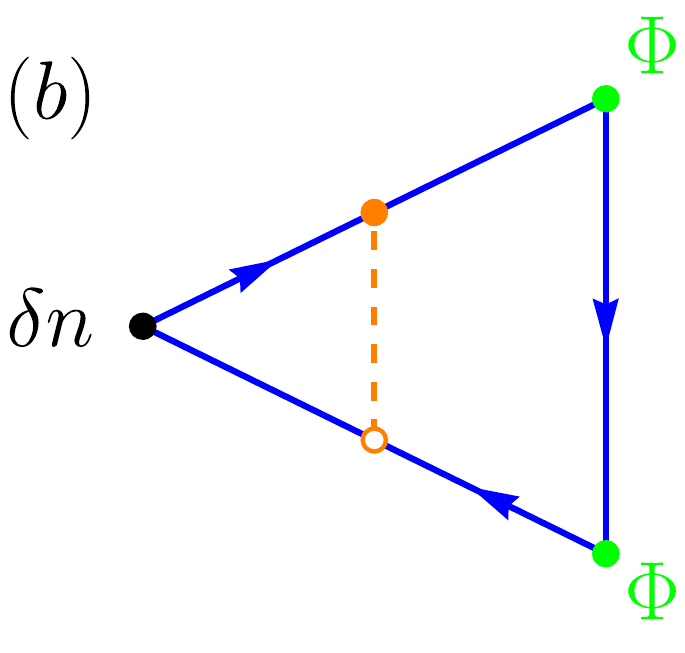}
\hspace{0.05\columnwidth}\includegraphics[width=0.45\columnwidth]{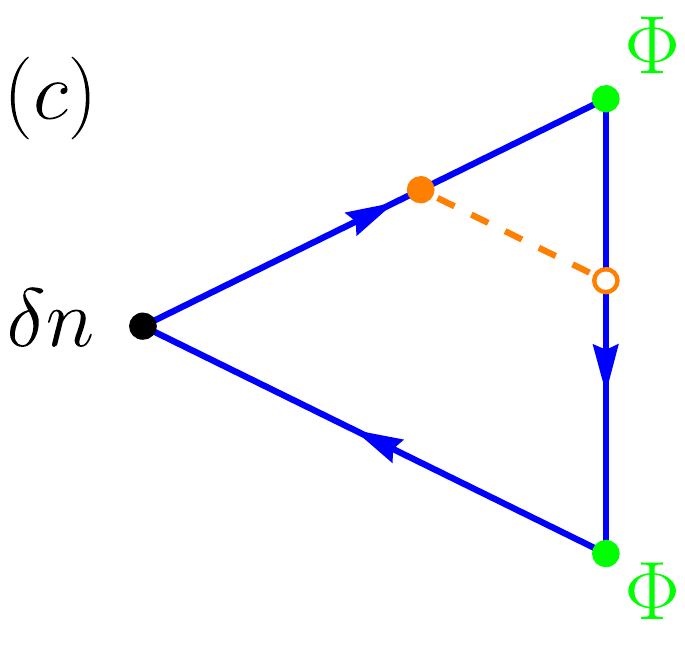}
\hspace{0.05\columnwidth}\includegraphics[width=0.45\columnwidth]{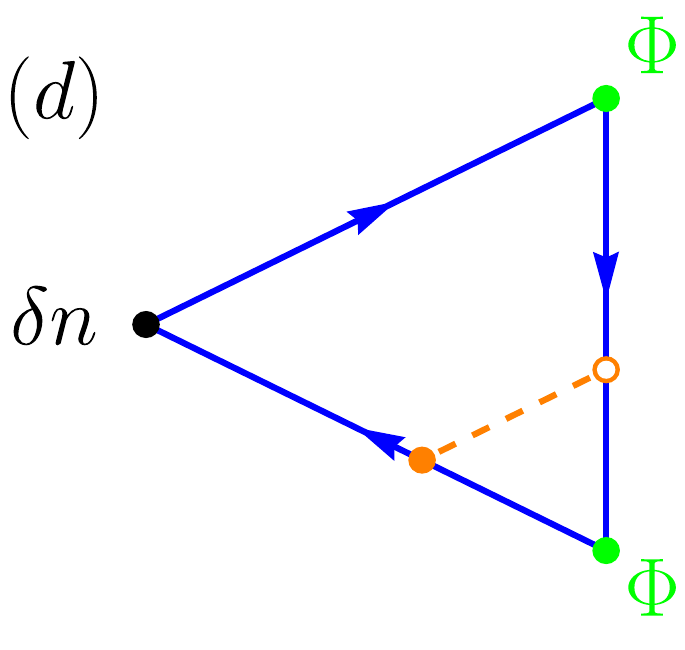}}
\caption{Diagrams for the nonlinear response. (a) The triangle diagram with Green functions in the self-consistent Born approximation. (b) The triangle diagram with the vertex correction of the density. (c),(d) The triangle diagrams with the vertex correction of the external potential. Diagrams (c) and (d) are the analogs of the diagrams in Fig. \ref{fig:2}(b) with the self-energy corrected due to the presence of the external potential.}
\label{fig:3}
\end{figure*}
%%%%%%%%%%%%%%%%%%%

Using Eq. \eqref{eq:dGK:2} and \eqref{eq:dGR:2}, we obtain the following equation for the second order contribution $\delta \mathcal{G}^K_2$ to the variation of the Keldysh Green function:
\begin{equation}
\delta \mathcal{G}^K_2 {-} \mathcal{G}^R {\circ} \delta \Sigma^K_2 {\circ} \mathcal{G}^A {-}
\mathcal{G}^{R} {\circ} \delta \Sigma^{R}_2 {\circ} \mathcal{G}^{K}
{-} \mathcal{G}^{K} {\circ} \delta \Sigma^{A}_2 {\circ} \mathcal{G}^{A}
{=} T {+} V  .
\label{eq:dGK2:eq}
\end{equation}
Here $\delta \Sigma^{R/A/K}_2$ are expressed in terms of $\delta \mathcal{G}^K_2$ in accordance with Eqs. \eqref{eq:dSigmaK:35} and \eqref{eq:dSigmaRA:35}.
The bare triangle diagram (see Fig. \ref{fig:3}) is given as,
\begin{gather}
T {=} \mathcal{G}^{R}{\circ} \Phi {\circ} \mathcal{G}^{R}
 {\circ} \Phi {\circ}\mathcal{G}^{K} 
{+} \mathcal{G}^{K}{\circ} \Phi {\circ} \mathcal{G}^{A}
 {\circ} \Phi \circ\mathcal{G}^{A} \notag \\
{+} \mathcal{G}^{R}{\circ} \Phi {\circ} \mathcal{G}^{K}
 {\circ} \Phi {\circ}\mathcal{G}^{A}  .
 \label{eq:def:T}
\end{gather}
The contribution $V$ describes vertex corrections [see Fig.~\ref{fig:3}(b)--(d).]
\begin{align}
V &{=}
\mathcal{G}^{R}{\circ} \delta \Sigma^R_1 {\circ} \mathcal{G}^{K}
 {\circ} \Phi {\circ}\mathcal{G}^{A}
 {+} \mathcal{G}^{R}{\circ} \Phi {\circ} \mathcal{G}^{K}
 {\circ} \delta \Sigma^A_1 {\circ}\mathcal{G}^{A}
 \notag \\
{+} & \mathcal{G}^{R}{\circ} \delta \Sigma^R_1 {\circ} \mathcal{G}^{K}
 {\circ} \delta \Sigma^A_1 {\circ}\mathcal{G}^{A} 
 {+} \mathcal{G}^{R}{\circ} \Phi {\circ} \mathcal{G}^{R}
 {\circ} \delta \Sigma^K_1 {\circ}\mathcal{G}^{A}
 \notag \\
 {+} & \mathcal{G}^{R}{\circ} \delta \Sigma^K_1 {\circ} \mathcal{G}^{A}
 {\circ} \Phi  {\circ}\mathcal{G}^{A}
 {+} \mathcal{G}^{R}{\circ} \delta \Sigma^K_1 {\circ} \mathcal{G}^{A}
 {\circ} \Phi  {\circ}\mathcal{G}^{A}
 \notag \\
 {+} & \mathcal{G}^{R}{\circ} \delta \Sigma^R_1 {\circ} \mathcal{G}^{R}
 {\circ} \delta \Sigma^K_1 {\circ}\mathcal{G}^{A}
 {+} \mathcal{G}^{R}{\circ} \delta \Sigma^K_1 {\circ} \mathcal{G}^{A}
 {\circ} \delta \Sigma^A_1 {\circ}\mathcal{G}^{A}
 \notag \\
 {+} & \mathcal{G}^{R}{\circ} \Phi {\circ} \mathcal{G}^{R}
 {\circ} \delta \Sigma^R_1 {\circ}\mathcal{G}^{K}
 {+} \mathcal{G}^{K}{\circ} \Phi {\circ} \mathcal{G}^{A}
 {\circ} \delta \Sigma^A_1 {\circ}\mathcal{G}^{A}
 \notag \\
 {+} &  \mathcal{G}^{R}{\circ} \delta \Sigma^R_1 {\circ} \mathcal{G}^{R}
 {\circ} \Phi {\circ}\mathcal{G}^{K}
 {+}  \mathcal{G}^{K}{\circ} \delta \Sigma^A_1 {\circ} \mathcal{G}^{A}
 {\circ} \Phi {\circ}\mathcal{G}^{A}
 \notag \\
 {+} & \mathcal{G}^{R}{\circ} \delta \Sigma^R_1{\circ} \mathcal{G}^{R}
 {\circ} \delta \Sigma^R_1 {\circ}\mathcal{G}^{K}
{+}  \mathcal{G}^{K}{\circ} \delta \Sigma^A_1 {\circ} \mathcal{G}^{A}
 {\circ} \delta \Sigma^A_1 {\circ}\mathcal{G}^{A}
 .
\end{align} 
Here $\delta \Sigma_1$ is the contribution to the self-energy due to the  correction to the $\delta \eta$ in the first order in $\Phi$. However, as we have seen in the previous section, the mode $\delta \eta$ decays at long times, $t{\gg}\tau{=}1/(2\bar{\gamma} m^2)$. Therefore, for the study of the long-time dynamics, we can safely neglect the mode $\delta \eta$ and, consequently, by all contributions involving $\delta \Sigma_1$. In other words, we can omit $V$ on the right-hand side of Eq.~\eqref{eq:dGK2:eq}. 

Using parametrization \eqref{eq:dGK:par}, we then obtain the following equation for $\delta u_{\bm{p},\bm{q};\omega}$, cf. Eq. \eqref{eq:eq:du:1},
\begin{gather}
2 i   \delta u_{\bm{p},\bm{q};\omega}  +  \frac{4 \gamma (d_p{-}q^2/4+i[\bm{p}\times\bm{q}])}{\sqrt{d_{\bm{p_-}}d_{\bm{p_+}}}[\omega+\xi_{\bm{p_-}}-\xi_{\bm{p_+}} + i\bar{\gamma}(d_{\bm{p_-}}+d_{\bm{p_+}})]} \notag \\
\times \int\limits_{\bm{k}} \sqrt{d_{\bm{k_+}}d_{\bm{k_-}}} \, \delta u_{\bm{k},\bm{q};\omega}
= T^{\textsf{(uu)}}_{\bm{p},\bm{q};\omega} .
\label{eq:eq:du:2}
\end{gather}
The $uu$-component of the triangle diagram evaluated with the Green functions computed in the self-consistent Born approximation, Eqs. \eqref{eq:GF:self-consistent Born approximation:RA}
 and \eqref{eq:GF:self-consistent Born approximation:K}, becomes
 \begin{equation}
 T_{\bm{p},\bm{q};\omega}^{\textsf{(uu)}} = - \frac{2\tilde{T}_{\bm{p},\bm{q};\omega}^{\textsf{(uu)}}}{\omega+\xi_{\bm{p_-}}-\xi_{\bm{p_+}}+ i\bar{\gamma} (d_{\bm{p_-}}+d_{\bm{p_+}})} .
 \label{eq:T:1}
 \end{equation}
Here we singled out the denominator which vanishes in the limit $q{\to}0$, $\omega{\to}0$, and $\bar{\gamma}{\to}0$; the remainder,
\begin{gather}
\tilde{T}_{\bm{p},\bm{q};\omega}^{\textsf{(uu)}} = i \int\limits_{\bm{Q},\Omega}
\frac{
\Phi^{\textsf{(ud)}}_{\bm{p_+}+\bm{Q_-}/2,\bm{-Q_-};\Omega_+} \Phi^{\textsf{(du)}}_{\bm{p_-}+\bm{Q_+}/2,\bm{Q_+};-\Omega_-}
}
{(\xi_{\bm{p_+}}+\xi_{\bm{p+Q}}- i\bar{\gamma} (d_{\bm{p_+}}+d_{\bm{p+Q}})-\Omega_+)}
\notag \\
\times 
\frac{(\xi_{\bm{p_+}}-\xi_{\bm{p_-}}- i \bar{\gamma}(d_{\bm{p_+}}+d_{\bm{p_-}}+2d_{\bm{p+Q}})-\omega)}{(\xi_{\bm{p_-}}+\xi_{\bm{p+Q}}+ i\bar{\gamma} (d_{\bm{p_-}}+d_{\bm{p+Q}})-\Omega_-)} .
\label{eq:T:2}
\end{gather}
%is finite in this limit. 
Solving Eq.~\eqref{eq:eq:du:2} we obtain
\begin{gather}
\delta u_{\bm{p},\bm{q};\omega}  =  \frac{i}{\omega+\xi_{\bm{p_-}}-\xi_{\bm{p_+}}+ i\bar{\gamma} (d_{\bm{p_-}}+d_{\bm{p_+}})}
\Biggl [
\tilde{T}_{\bm{p},\bm{q};\omega}^{\textsf{(uu)}}
\notag \\ 
 -  
\frac{i \gamma (d_p-q^2/4+i[\bm{p}\times\bm{q}])  \int_{\bm{k}} T_{\bm{k},\bm{q};\omega}^{\textsf{(uu)}}\sqrt{
\frac{d_{\bm{k_-}}d_{\bm{k_+}}}
{d_{\bm{p_-}}d_{\bm{p_+}}}
}}
{
1-2i\gamma\int_{\bm{k}} \frac{(d_k{-}q^2/4+i[\bm{k}\times\bm{q}])}
{\omega+\xi_{\bm{k_-}}-\xi_{\bm{k_+}}+ i\bar{\gamma} (d_{\bm{k_-}}+d_{\bm{k_+}})}}
\Biggr ] .
\label{eq:upq:gen:1}
\end{gather}
We emphasize that the second line of Eq.~\eqref{eq:upq:gen:1} describes the effect of vertex corrections. A striking feature of the latter is the divergence of the denominator at $q{=}\omega{=}0$.

The solution for $\delta d_{\bm{p},\bm{q};\omega}$ is obtained from Eq.~\eqref{eq:upq:gen:1} by the following steps. At first, one reverses the sign of the vector $\bm{q}$.  Secondly, one replaces $T_{\bm{k},\bm{-q};\omega}^{\textsf{(uu)}}$ by ${T}_{\bm{k},\bm{q};\omega}^{\textsf{(dd)}}$ and $\tilde{T}_{\bm{k},\bm{q};\omega}^{\textsf{(uu)}}$ by $-\tilde{T}_{\bm{k},\bm{q};\omega}^{\textsf{(dd)}}$. The expressions for ${T}_{\bm{k},\bm{q};\omega}^{\textsf{(dd)}}$ and   $\tilde{T}_{\bm{k},\bm{q};\omega}^{\textsf{(dd)}}$ are obtained from Eqs. \eqref{eq:T:1} and \eqref{eq:T:2} by interchanging the indices $u$ and $d$ and changing the signs in front of $\xi_{\bm{p_+}}$, $\xi_{\bm{p_-}}$, and $\xi_{\bm{p+Q}}$.  These relations between solutions for $\delta u_{\bm{p},\bm{q};\omega}$ and $\delta d_{\bm{p},\bm{q};\omega}$ imply that 
\begin{equation}
\delta n^{\textsf{(u)}}_{\bm{q},\omega}+ \delta n^{\textsf{(d)}}_{\bm{q},\omega}=0 ,
\label{eq:ddnn}
\end{equation}
where the Fourier transforms of the density variations of the $u$- and $d$-fermions are defined as 
\begin{equation}
\delta n^{\textsf{(u)}}_{\bm{q},\omega} = \int\limits_{\bm{p}}
\delta u_{\bm{p},\bm{q};\omega}, \quad \delta n^{\textsf{(d)}}_{\bm{q},\omega} = \int\limits_{\bm{p}}
\delta d_{\bm{p},\bm{q};\omega} .
\end{equation}
We note that Eq. \eqref{eq:ddnn} implies the conservation of  the total density.

We now concentrate on the long wavelength and low frequency regime, $|q|{\ll}\bar{\gamma} m$ and $|\omega|{\ll} \bar{\gamma} m^2$. Expanding the exact solution \eqref{eq:upq:gen:1} in powers of $q$ and $\omega$, we find the following result for the Fourier transform of the density variation of the $u$-particles,
\begin{gather}
\delta n^{\textsf{(u)}}_{\bm{q},\omega} = \int\limits_{\bm{p}}
\delta u_{\bm{p},\bm{q};\omega}
=\frac{1}{D q^2-i\omega} \int\limits_{\bm{k}} 
\tilde{T}_{\bm{k},\bm{q};\omega}^{\textsf{(uu)}}\;.
\label{eq:dn:Diff}
\end{gather}
We emphasize that it has a diffusive-pole structure that comes from the vertex correction alone.  The diffusion coefficient reads 
\begin{equation}
    D = \bar{\gamma}+\frac{2}{d\bar{\gamma}} \frac{\int_{\bm{k}}k^2/d^2_{\bm{k}}}{\int_{\bm{k}}1/d_{\bm{k}}}= \frac{1}{\bar{\gamma}}\left[1-\frac{\left(d-1\right)}{\ln\left(1+\frac{4\pi n}{m}\right)}\right]+\bar{\gamma}. 
\label{eq:D:coef}
\end{equation}
The expression \eqref{eq:D:coef} 
for the diffusion coefficient can be expressed as 
\begin{equation}
D = \frac{1}{d} \frac{\langle \bm{v}^2_{\bm{k}} \tau^2_{\bm{k}}\rangle_{\bm{k}}}{\langle \tau_{\bm{k}}\rangle_{\bm{k}}} +\bar{\gamma},
\label{eq:D:coef:2}
\end{equation}
where $\bm{v}_{\bm{k}}{=}\partial \xi_{\bm{k}}/\partial \bm{k}{=}2\bm{k}$ is the velocity. Here $\langle \dots \rangle_{\bm{k}}$ denotes averaging over the momentum $\bm{k}$. We note that the first term in Eq.~\eqref{eq:D:coef:2} originates from the interplay between the unitary (Hamiltonian) and dissipative dynamics in the GKSL equation \eqref{eq:GKSL}, whereas the second one is purely dissipative. 

In the absence of any Hamiltonian ($\mathcal{H}=0$), Eq.~\eqref{eq:D:coef:2} suggests that the diffusion coefficient remains finite (yet small), $D{=}\bar{\gamma}$. Physically, this stems from the fact that the jump operators involve derivatives, and thus allow particle transport. We verified that there are no corrections to the diffusion coefficient due to the appearance of the real part of the self-energy to the second order in $\gamma$ (see Appendix \ref{App:AppendixSelf-Cons}). However, it is quite possible that additional perturbative-in-$\gamma$ corrections to $D$ could appear from higher order vertex diagrams.

Assuming that the external potential $\phi_{\bm{q},\omega}$ varies slowly enough in both space and time, $|q|{\ll}\bar{\gamma}m$ and $|\omega|{\ll}\bar{\gamma}m^2$, we reduce Eq. \eqref{eq:dn:Diff} to the following diffusion equation 
\begin{equation}
\Bigl (\frac{\partial}{\partial t} - D\nabla^2\Bigr )\delta n^{\textsf{(u)}}(\bm{x},t)
= \frac{\bar{\gamma} \chi}{1+\bar{\gamma}^2} \bm{E}^2(\bm{x},t) 
\label{eq:main:res}
\end{equation}
where $\bm{E}= -\nabla \phi$ is the electric field induced by a scalar potential and where
\begin{equation}
\chi = 
 \int\limits_{\bm{k}} \frac{m^2}{\xi_k^3}
= \begin{cases}
3/(16 m^3), & \, d=1 , \\
1/(8\pi m^2), & \, d=2 .
\end{cases}
\end{equation}
For purely dissipative dynamics, one needs to set $D$ to zero and replace the denominator $1{+}\bar{\gamma}^2$ by $\bar{\gamma}^2$ in the diffusion equation \eqref{eq:main:res}. We note that the appearance of the electric field on the right-hand side of Eq. \eqref{eq:main:res} is not accidental, but rather guaranteed by gauge invariance \cite{Rostami2021}. Let us also note that for a static field $\bm{E}$, the right hand side of Eq.~\eqref{eq:main:res} would seem to lead to an unbounded growth of the density $\delta n^{\textsf{(u)}}$. This is countered by the recombination term discussed in the next Section.

There is a similar equation for  $\delta n^{\textsf{(d)}}(\bm{x},t)$ to ensure the conservation of the total particle density, cf. Eq. \eqref{eq:ddnn}, 
\begin{equation}
\delta n^{\textsf{(u)}}(\bm{x},t)+\delta n^{\textsf{(d)}}(\bm{x},t) = 0 .\label{eq:local_density_constraint}
\end{equation}
The range of applicability of this constraint deserves a separate comment. In this section, Eq.~\eqref{eq:local_density_constraint} is derived assuming the non-crossing approximation (corresponding to ladder diagrams depicted in Fig.~\ref{fig:3n}(b)). However, strictly speaking, the total density mode $n^{(\sf{u})}(\bm{x})+n^{(\sf{d})}(\bm{x})$ is not conserved beyond this approximation, since the jump operators $L_\alpha$ involve derivatives. As a consequence, Eq.~\eqref{eq:local_density_constraint} holds in a ``coarse-grained'' sense only, i.e. for distances greater than $1/m$ (which is much shorter than the averaged mean-free path). In other words, local deviations of the total density mode (and their coupling to the particle-hole mode) emerge only at the higher order in the spatial gradients. 

Our result \eqref{eq:main:res} demonstrates that one does not need a time-dependent potential with a characteristic frequency greater than the spectral gap $2m^2$ to induce a finite density response by the applied external field. The reason is that the density operator of the $\psi-$fermions has off-diagonal matrix elements in the $\textsf{u}/\textsf{d}-$representation. These off-diagonal components of $\Phi$, cf. Eq. \eqref{eq:Phi:ud}, allow for hybridization between the states from the upper and down bands, and thus, effectively
induce a non-zero occupation in the upper band. The corresponding density appears even in the absence of dissipation, $\gamma{=}0$, and can be estimated as (see Appendix \ref{App:ToyModel}) 
\begin{equation}
\delta n^{\textsf{(u)}}_{\textsf{0}}(\bm{x},t) \sim 
\chi \bm{E}^2(\bm{x},t) / m^2 .
\label{eq:deltaN:triv}
\end{equation}
This deviation of the density stems from the first term in Eq.~\eqref{eq:upq:gen:1}. Diagrammatically, it corresponds to the bare triangle with no ladder insertions (see Fig.~\ref{fig:3}(a)), and thus, it does not contain a diffusive pole (only the second term in Eq.~\eqref{eq:upq:gen:1} does), but it remains finite in the limit $q{=}\omega{=}0,\gamma{=}0$. Therefore, the density deviations in Eq.~\eqref{eq:main:res} should be understood as deviations from this zeroth-order density shift $\delta n^{\textsf{(u)}}_{\textsf{0}}$. We emphasize, however, that $\delta n^{\textsf{(u)}}_{\textsf{0}}$ is much smaller than the density variation obtained from the solution of Eq. \eqref{eq:main:res}.

We note that this behavior of Eq. \eqref{eq:deltaN:triv} can be understood from the toy example of a $2{\times}2$ Hamiltonian
\begin{equation}
H_{\rm TM} = \begin{pmatrix}
m^2 & \Phi \\
\Phi & - m^2 
\end{pmatrix} .
\label{eq:HTM}
\end{equation}
The Hamiltonian $H_{\rm TM}$ has two eigenstates $|{\pm} \rangle_\Phi$ with energies ${\pm} \sqrt{m^2{+}\Phi^2}$. Let us consider the case of a fully occupied lowest energy band and a fully empty upper energy band. Then the occupancy of the upper state for the Hamiltonian \eqref{eq:HTM} is nonzero once $\Phi{\ne}0$. In the limit $|\Phi|{\ll} m^2$ this occupancy is proportional to $\Phi^2/m^4$. This well-known result is fully analogous to Eq. \eqref{eq:deltaN:triv}.	

\section{Recombination\label{Sec:Rec}}

The above analysis of the linear and nonlinear response of the particle density to an external scalar potential has been restricted to the linear order in $\delta n^{\textsf{(u)}}$. Terms of the second order in the density variation describe the recombination of the particles from the `up' and `down' bands. According to the Lindblad dynamics, cf. Eq. \eqref{eq:GKSL}, such processes appear already to the first order in $\gamma$, but only in the \textit{quadratic} order in the deviation of the density from the steady state. Thus, their analysis requires appropriate modifications of the \textit{linear} integral equation~\eqref{eq:self-consistent-pol-oper:2}.

There is, however, a slightly more convenient way to account for recombination. One can make use of the following exact equation governing the time decay of the total number of particles in the `up' band,
\begin{gather}
\frac{d N_{\textsf{u}}}{dt} = \gamma \sum_\alpha \int\limits_{\bm{x}} \Tr \rho \Bigl \{
[L_\alpha^\dag, \hat N_{\textsf{u}}] L_\alpha - L_\alpha^\dag [L_\alpha, \hat N_{\textsf{u}}]
\Bigr \} . 
\end{gather}
Here we introduced the operator of the total number of particles in the upper band, $\hat N_{\textsf{u}} {=}\int_{\bm{x}} c^\dag_{\textsf{u}}(\bm{x}) c_{\textsf{u}}(\bm{x})$ (the definition of the operators $c_{\sf{u/d}}$ is analogous to the definition of the fields $c_{\sf{u/d}}$ in the Keldysh formulation, see Eq.~\eqref{eq:Psi_to_C}). Employing the following commutation relations
\begin{gather}
[l_{{\textsf{u}}}^\dag(\bm{x}), \hat N_u] = - l_{{\textsf{u}}}^\dag(\bm{x}), \qquad [l_{{\textsf{u}}}(\bm{x}), \hat N_{\textsf{u}}] = l_{{\textsf{u}}}(\bm{x}),\notag \\
[l_{{\textsf{d}}}^\dag(\bm{x}), \hat N_u] = [l_{{\textsf{d}}}(\bm{x}), \hat N_{\textsf{u}}] = 0 , \\ 
[\psi_{1/2}^\dag(\bm{x}), \hat N_{\textsf{u}}] = - \psi_{1/2;{\textsf{u}}}^\dag(\bm{x}), \, [\psi_{1/2}(\bm{x}), \hat N_{\textsf{u}}] = \psi_{1/2;{\textsf{u}}}(\bm{x}) ,\notag
\end{gather}
we obtain the exact equation
\begin{gather}
 \frac{d N_{\textsf{u}}}{dt} = - \gamma \sum_{\beta=1,2} 
\int\limits_{\bm{x}} \Tr \rho \Bigl \{ 
l_{\textsf{u}}^\dag [\psi_{\beta,{\textsf{d}}}\psi^\dag_{\beta}+\psi_{\beta}\psi^\dag_{\beta,{\textsf{d}}}] l_{\textsf{u}} \notag \\+ l_{\textsf{d}} [\psi^\dag_{\beta}\psi_{\beta,{\textsf{u}}}+\psi^\dag_{\beta,{\textsf{u}}}\psi_{\beta}] l_{\textsf{d}}^\dag
\Bigr \}\; . \label{eq:N_u_main}
\end{gather}
Here the subscript $\textsf{u}$ ($\textsf{d}$) in $\psi_{1/2;{\textsf{u}}}$ ($\psi_{1/2;{\textsf{d}}}$) denotes the part of $\psi_{1/2}$ that involves $c_{\textsf{u}}$ ($c_{\textsf{d}}$) operators only.

The right hand side of Eq.~\eqref{eq:N_u_main} can be computed by means of the Keldysh path integral theory described in the previous sections. To the lowest order in fluctuations the averages over four fermionic fields could be performed at the Gaussian level with the help of the Wick's, theorem but allowing for non-linear order in the deviations from the dark state. We then find
\begin{gather}
\frac{d N_{\textsf{u}}}{dt} \simeq - 2 \gamma \sum_{\beta=1,2} 
\int\limits_{\bm{x}} \Bigl [ \langle  l_{\textsf{u}}^\dag l_{\textsf{u}} \rangle \langle \psi_{\beta,{\textsf{d}}}\psi^\dag_{\beta,{\textsf{d}}}\rangle
+\langle l_{\textsf{d}} l_{\textsf{d}}^\dag\rangle \langle \psi^\dag_{\beta,{\textsf{u}}}\psi_{\beta,{\textsf{u}}}\rangle
\Bigr ] .
\label{app:2:dNu:1}
\end{gather}
Here we omit the averages which are off-diagonal in $u/d-$space, e.g. $\langle \psi_{\beta,{\textsf{u}}}\psi^\dag_{\beta,{\textsf{d}}}\rangle$, since such averages are proportional to $\delta \eta$. Indeed, as we have shown in Sec. \ref{Sec:Linear}, $\delta \eta$ decays on short time scales of the order of $\tau$. 
We emphasize that at the level of the self-consistent Born approximation, each of the averages involved in Eq. \eqref{app:2:dNu:1} vanishes. Going beyond the self-consistent Born approximation, we write
\begin{gather}
\langle  l_{\textsf{u}}^\dag(\bm{x},t) l_{\textsf{u}}(\bm{x},t) \rangle 
= \int\limits_{\bm{p},\bm{q};\omega} \sqrt{d_{\bm{p_+}}d_{\bm{p_-}}}  \delta u_{\bm{p},\bm{q};\omega}  e^{-i \bm{q} \bm{x}-i\omega t} ,\notag \\
\langle  l_{\textsf{d}}(\bm{x},t) l_{\textsf{d}}^\dag(\bm{x},t) \rangle 
= - \int\limits_{\bm{p},\bm{q};\omega} \sqrt{d_{\bm{p_+}}d_{\bm{p_-}}}  \delta d_{\bm{p},\bm{q};\omega}  e^{-i \bm{q} \bm{x}-i\omega t} ,\notag 
\\
\langle \psi^\dag_{\beta,{\textsf{u}}}(\bm{x},t)\psi_{\beta,{\textsf{u}}}(\bm{x},t)\rangle  = \!\!\! \int\limits_{\bm{p},\bm{q};\omega}\!\!\!\frac{d_{\bm{p}}{-}i [\bm{q}{\times}\bm{p}]}{\sqrt{d_{\bm{p_+}}d_{\bm{p_-}}} } \delta u_{\bm{p},\bm{q};\omega}  e^{-i \bm{q} \bm{x}-i\omega t} , \notag \end{gather}
\begin{gather}
\langle \psi_{\beta,{\textsf{d}}}(\bm{x},t)\psi^\dag_{\beta,{\textsf{d}}}(\bm{x},t)\rangle  
= \!{-}\!\!\!\!\int\limits_{\bm{p},\bm{q};\omega}\!\!\frac{d_{\bm{p}}{+}i [\bm{q}{\times}\bm{p}]}{\sqrt{d_{\bm{p_+}}d_{\bm{p_-}}} } \delta d_{\bm{p},\bm{q};\omega}  e^{-i \bm{q} \bm{x}-i\omega t} .
\label{eq:Wick:Aver}
 \end{gather}
 Here summation over the repeated index, $\beta{=}1,2$, is assumed. 
Substituting 
the averages \eqref{eq:Wick:Aver} into Eq. \eqref{app:2:dNu:1}, we find 
 \begin{gather}
\frac{dN_{\textsf{u}}}{dt} \simeq   4 \gamma \int\limits_{\bm{x}} \delta n^{\textsf{(u)}}(\bm{x},t) \int\limits_{\bm{p},\bm{q};\omega} d_{\bm{p}}\, \delta d_{\bm{p},\bm{q};\omega} e^{-i\bm{q}\bm{x}-i\omega t} \notag \\
\simeq
- \frac{1}{n \tau_{\rm R}}\int\limits_{\bm{x}} [\delta n^{\textsf{(u)}}(\bm{x},t)]^2 . 
\label{eq:app:dNu:dt}
\end{gather}
Here we used  the following relations
\begin{gather}
\int_{\bm{p}}  d_{\bm{p}} \delta u_{\bm{p},\bm{q};\omega} = -\int_{\bm{p}}  d_{\bm{p}} \delta d_{\bm{p},\bm{q};\omega} \simeq \frac{n}{\int_{\bm{p}} 1/d_{\bm{p}}} \delta n^{\textsf{(u)}}_{\bm{q}\omega},
\end{gather}
that follows from Eq. \eqref{eq:eq:du:1} in the absence of the external scalar potential. The recombination rate $1/\tau_R$ is given by 
\begin{gather}
\frac{1}{\tau_{\rm R}} = \frac{4\bar{\gamma} n}{\int_{\bm{p}} 1/d_{\bm{p}}}=8 \bar{\gamma}  n \begin{cases} 
 m, & d=1 ,\\
\pi/\ln(n/m^2) , & d=2 .
\end{cases}
\label{eq:app:Rec:Rate:main}
\end{gather}
We note that the obtained result, cf. Eq.~\eqref{eq:app:dNu:dt}, suggests the existence of the additional term $-[\delta n^{\textsf{(u)}}(\bm{x},t)]^2/(n \tau_{\rm R})$ on the right hand side
of Eq.~\eqref{eq:main:res}. Noting Eq.~\eqref{eq:local_density_constraint}, this term is actually $\delta n^{\textsf{(u)}}(\bm{x},t)\delta n^{\textsf{(d)}}(\bm{x},t)/(n \tau_{\rm R})$ , which highlights its role as recombination of ${\sf{u}}$-particles and ${\sf{d}}$-holes.

%%%%%%%%%%%%%%%%%%%
\begin{figure}[t]
\centerline{\includegraphics[width=0.2\textwidth]{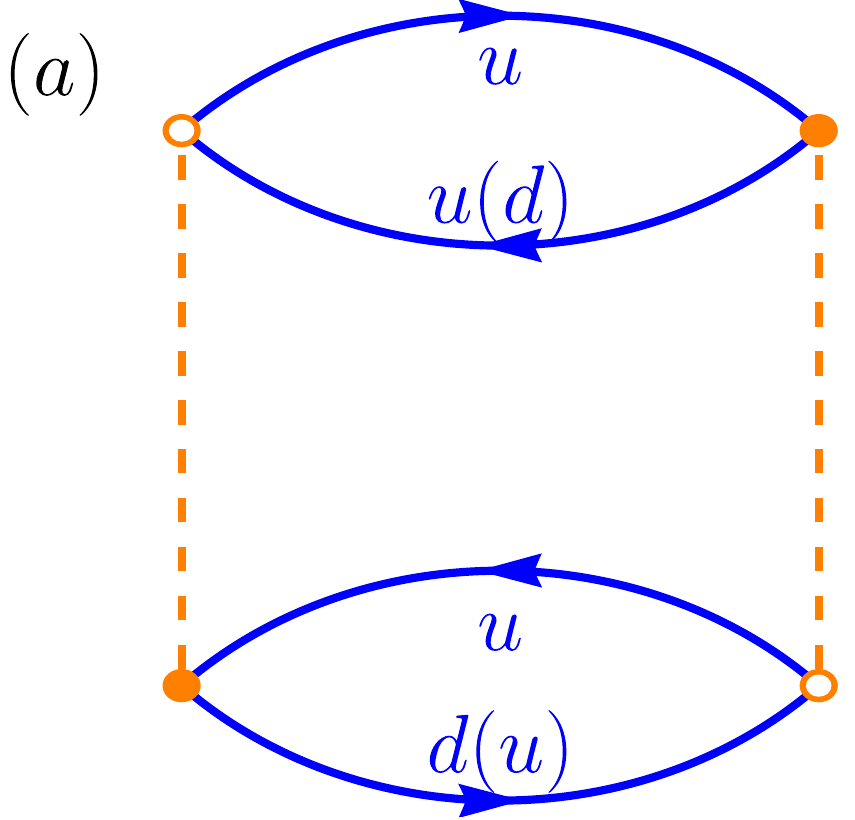}
\qquad 
\includegraphics[width=0.2\textwidth]{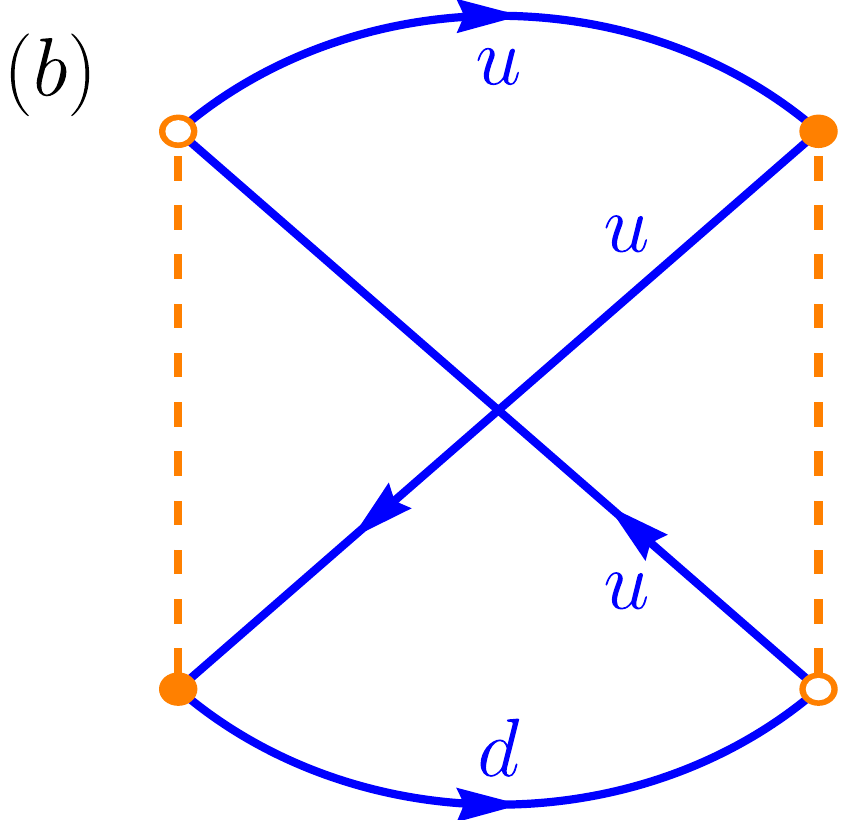}
}
\caption{Examples of diagrams for the correction to the rate of change of the number of particles in the upper band.}
\label{fig:app:22}
\end{figure}
%%%%%%%%%%%%%%%%%%%%%%%%%%

\section{Pumping of the particles into the upper band beyond the self-consistent Born approximation\label{Sec:Pumping}}

Since there is no conservation of the number of particles in `up' and `down' bands, in general, one cannot expect perfect cancellation between the self-energy and vertex corrections. Within the self-consistent Born approximation, such a cancellation does occur since self-energy and renormalized density vertex are elastic-like, i.e., they are independent of the energy. To second order in $\gamma$, beyond the self-consistent Born approximation, the self-energy becomes energy dependent (see Appendix \ref{App:AppendixSelf-Cons}). 
Similar energy dependence is expected for the vertex corrections.\footnote{The computation of the vertex correction beyond the self-consistent Born approximation is out of the scope of the present work.} In order to account for these effects, one can once more resort to Eq.~\eqref{eq:N_u_main}. This time, however, the averages on the right hand side should be computed to the second order in $\gamma$ (and linear order in the deviations of the density from the steady state), by expanding the dissipative part of the Keldysh action \eqref{eq:SL:ud} and performing irreducible contractions only. Examples of such diagrams are shown in Fig.~\ref{fig:app:22}.

We start by rewriting Eq. \eqref{eq:N_u_main} in the following way
\begin{widetext}
\begin{gather}
\frac{d N_{\textsf{u}}}{dt} {=} {-} \frac{\gamma}{4} \int\limits_{\bm{k_j}}
\delta(\bm{k_1}{-}\bm{k_2}{+}\bm{k_3}{-}\bm{k_4})
\sqrt{d_{\bm{k_1}}d_{\bm{k_4}}} \sum_{s=0,1}\sum_{a,b=\textsf{u},\textsf{d}} \Biggr \{ \Bigl [ 1- \delta_{a\textsf{u}}\delta_{b\textsf{u}}+\delta_{a\textsf{d}}\delta_{b\textsf{d}}\Bigr ]
[U_{\bm{k_2}}^\dag U_{\bm{k_3}}]_{ab} \Bigl \langle  \bigl  (\bar{c}_{\bm{k_4},u} \tau_s c_{\bm{k_1},u}\bigr ) \bigl ( c_{\bm{k_3},b} \tau_s \bar{c}_{\bm{k_2},a} \bigr )
\Bigr \rangle 
\notag \\
+ \Bigl [ 1- \delta_{a\textsf{d}}\delta_{b\textsf{d}} + \delta_{a\textsf{u}}\delta_{b\textsf{u}}\Bigr ]
[U_{\bm{k_3}}^\dag U_{\bm{k_2}}]_{ab} \Bigl \langle  \bigl  (c_{\bm{k_4},d}  \tau_s 
\bar{c}_{\bm{k_1},d} \bigr ) \bigl (\bar{c}_{\bm{k_3},a}  \tau_s  c_{\bm{k_2},b} \bigr )
\Bigr \rangle 
\Biggr \} .
\end{gather}
Here $\tau_0$ and $\tau_1$ are standard Pauli matrices acting in the Keldysh space (after Keldysh rotation). Next we perform averaging of the correlation functions in accordance with the diagrams shown in Fig. \ref{fig:app:22}. Expanding the result to the first order in deviation of the Green functions due to the presence of $\delta u_{\bm{p},\bm{q};\omega}$ and $\delta d_{\bm{p},\bm{q};\omega}$, we obtain
\begin{gather}
\frac{dN_{\textsf{u}}}{dt}=
-\frac{(2\pi)^d\gamma^2}{2} \int\limits_{t^\prime} \int\limits_{\bm{p_j}} \delta(\bm{p_1}{-}\bm{p_2}{+}\bm{p_3}{-}\bm{p_4}) d_{\bm{p_1}}\sum_{a=\textsf{u},\textsf{d}}  \Bigl [\sqrt{d_{\bm{p_2}}}
U_{\bm{p_4}}^\dag U_{\bm{p_3}} {-} \sqrt{d_{\bm{p_4}}}
U_{\bm{p_2}}^\dag U_{\bm{p_3}} \Bigr ]_{a\bar{a}}
\Bigl [ \sqrt{d_{\bm{p_2}}}
U_{\bm{p_3}}^\dag U_{\bm{p_4}} {-} \sqrt{d_{\bm{p_4}}}
U_{\bm{p_3}}^\dag U_{\bm{p_2}}\Bigr ]_{\bar{a}a}
\notag \\
\times
\Biggl\{
\underline{\mathcal{G}}^A_{a;\bm{p_2}}(t^\prime,t)\underline{\mathcal{G}}^R_{\bar{a};\bm{p_3}}(t,t^\prime)
\underline{\mathcal{G}}^A_{a;\bm{p_4}}(t^\prime,t)
[\delta \mathcal{G}^{X}_{\bm{p_1},\bm{0}}]^{{(aa)}}(t,t^\prime)
 - \underline{\mathcal{G}}^R_{a;\bm{p_2}}(t,t^\prime) \underline{\mathcal{G}}^A_{\bar{a};\bm{p_3}}(t^\prime,t)
\underline{\mathcal{G}}^R_{a;\bm{p_4}}(t,t^\prime) 
[\delta \mathcal{G}^{X}_{\bm{p_1},\bm{0}}]^{{(aa)}}(t^\prime,t)
\Biggr \} ,\end{gather}
\end{widetext}
Here we introduced $\bar{a}{=} \textsf{u} (\textsf{d})$ if $a{=} \textsf{d} (\textsf{u})$, respectively. Also we define
\begin{gather}
\notag\\
[\delta \mathcal{G}^{X}_{\bm{p_1},\bm{0}}]^{{(aa)}}(t,t^\prime) =\begin{cases}
[\delta \mathcal{G}^{<}_{\bm{p_1},\bm{0}}]^{\textsf{(uu)}}(t,t^\prime), &  a=u ,\\
-[\delta \mathcal{G}^{>}_{\bm{p_1},\bm{0}}]^{\textsf{(dd)}}(t,t^\prime), & a=d .
\end{cases}
\end{gather}

 Now using Eqs. \eqref{eq:deltaGRA}
 and \eqref{eq:deltaGK:K}, we express $\delta \mathcal{G}^{</>}_{\bm{p_1},\bm{0}}(t,t^\prime)$ in terms of $\delta u_{\bm{p},\bm{q};\omega}$ and $\delta d_{\bm{p},\bm{q};\omega}$ as follows
\begin{gather}
[\delta \mathcal{G}^{<}_{\bm{p_1},\bm{0}}]^{\textsf{(uu)}}(t,t^\prime) =
\int\limits_{\omega} \delta u_{\bm{p_1},\bm{0};\omega} \Bigr [
\underline{\mathcal{G}}^A_{u;\bm{p_1}}(t,t^\prime)  e^{-i\omega t}
\notag \\
-\underline{\mathcal{G}}^R_{u;\bm{p_1}}(t,t^\prime) e^{-i\omega t^\prime}\Bigl ] .
\end{gather}
The expression for $[\delta \mathcal{G}^{>}_{\bm{p_1},\bm{0}}]^{\textsf{(dd)}}(t,t^\prime)$ is obtained by substitution of $\textsf{u}$ and $\delta u_{\bm{p_1},\bm{0};\omega}$ by $\textsf{d}$ and $\delta d_{\bm{p_1},\bm{0};\omega}$, respectively.
Then, integrating over $t^\prime$, we find
\begin{gather}
\frac{dN_{\textsf{u}}}{dt} 
 \simeq
  2(2\pi)^d\gamma^3 n m^2  \int\limits_{\bm{p_j}} \delta(\bm{p_1}{-}\bm{p_2}{+}\bm{p_3}{-}\bm{p_4})\frac{d_{\bm{p_1}}}{d_{\bm{p_3}}} \Biggl |\sqrt{\frac{d_{\bm{p_2}}}{d_{\bm{p_4}}}}\notag \\
  \times
\bigl (p_{4x}{-}p_{3x}{+}i (p_{4y}{-}p_{3y})\bigr ) {-} 
\sqrt{\frac{d_{\bm{p_4}}}{d_{\bm{p_2}}}}
\bigl (p_{2x}{-}p_{3x}{+}i (p_{2y}{-}p_{3y})\bigr )
\Biggr |^2
\notag \\
\times 
\int\limits_\omega   \frac{e^{-i\omega t} (d_{\bm{p_1}}+d_{\bm{p_2}}{+}d_{\bm{p_3}}{+}d_{\bm{p_4}}) \delta u_{\bm{p_1},\bm{q=0};\omega}}{(\xi_{\bm{p_2}}{+}\xi_{\bm{p_3}}{+}\xi_{\bm{p_4}}{-}\xi_{\bm{p_1}})^2+\bar{\gamma}^2
(d_{\bm{p_1}}{+}d_{\bm{p_2}}{+}d_{\bm{p_3}}{+}d_{\bm{p_4}})^2}  .
\label{eq:long:int}
\end{gather}
Here we used  the relation $\delta u_{\bm{p_1},\bm{q=0};\omega}{+}\delta d_{\bm{p_1},\bm{q=0};\omega}{=}0$ and neglected $\omega$ in comparison with $1/\tau{=}2\bar{\gamma} m^2$.

%%%%%%%%%%%%%%%
\begin{figure}[t!]
\centerline{\includegraphics[width=0.6\columnwidth]{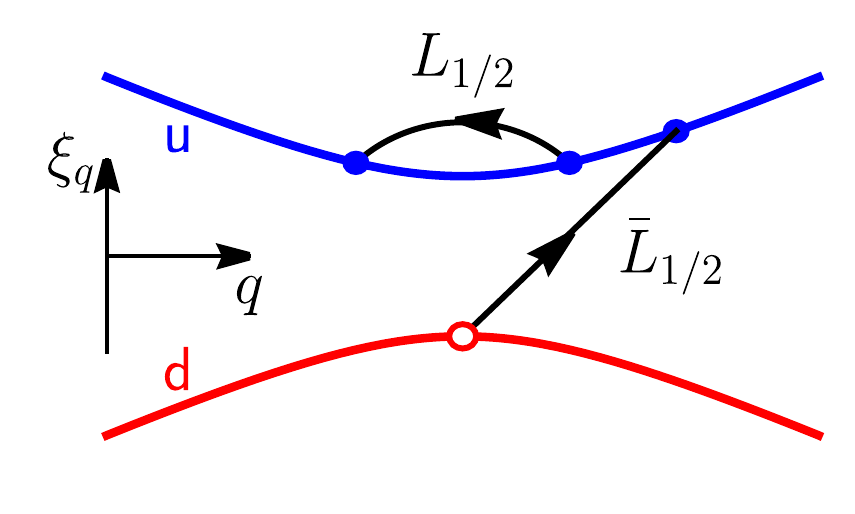}
}
\caption{An example of the process in which there is a pumping of the particles in the upper band. The initial state of a particle with a momentum $\bm{p_1}$ in the upper band and a particle with a momentum $\bm{p_3}$ in the lower band  is transformed into the state with two particles with momenta  $\bm{p_2}$ and  $\bm{p_4}$ in the upper band and a hole state with a momentum $\bm{p_3}$ in the down band.}
\label{fig:pump:1}
\end{figure}
%%%%%%%%%%%%%%%

In order to estimate the integrals over momenta in Eq.~\eqref{eq:long:int} we take into account that the integral over $\bm{p_3}$ is convergent  in $d{=}1$ while is logarithmically divergent in $d{=}2$. The integral over $\bm{p_4}$ is determined by the ultraviolet. We then find
\begin{equation}
\frac{dN_u}{dt}\simeq 
\frac{1}{\tau_{\rm p} m^2} \int\limits_{\bm{p}}\int\limits_\omega e^{-i\omega t}
 d_p \delta u_{\bm{p},\bm{q=0};\omega} .
 \label{eq:Lambda}
\end{equation}
Here the corresponding rate can be estimated as (for $d{=}1,2$)
\begin{gather}
\frac{1}{\tau_{\rm p}} \propto \bar\gamma \frac{m^{d+2}}{n} \ln^{d-1} (n/m^d)
\begin{cases}
\bar\gamma^2,  \quad & \bar\gamma \ll 1 , \\
1 , \quad &  \bar\gamma \gg 1.
\end{cases}
\label{eq:Lambda:app}
\end{gather}
We note that $\tau_{\rm p}{\gg}\tau$ due to smallness of two factors: $\bar{\gamma}{\ll}1$ and $m^d/n{\ll}1$.

We emphasize the positivity of the right hand side of Eq. \eqref{eq:Lambda}. This implies pumping of particles into the upper band, and suggests the presence of an additional term, $\delta n^{\textsf{(u)}}(\bm{x},t)/\tau_{\rm p}$, on the right hand side of Eq. \eqref{eq:main:res}. Similar terms will appear for the lower band.
Such terms destabilize the dark state. We turn to discuss their physical meaning, as well as the meaning of the previously-obtained diffusive dynamics.

\section{\textbf{Discussion of the main results: Diffusion and instability of the dark state} \label{Sec:Discussion}}

The structure of the jump operators involved in the GKSL equation, cf.  Eq. \eqref{eq:GKSL}, allows a particle to move back and forth both in momentum space and in real space, see Fig. \ref{fig:Jump}. Such motion resembles a random walk, leading to a diffusive dynamics of the particle density. In the leading approximation in $\bar{\gamma}$ the jump processes induce elastic-like mean free time $\tau$ for a single-particle excitation. In the presence of the unitary part of evolution, assuming the spectrum of the Hamiltonian is not flat, a particle has a finite velocity. Together with the mean free time this is enough to generate diffusive dynamics with the diffusion coefficient expressed in a standard way in terms of the velocity and the mean free path, cf. Eq. \eqref{eq:D:coef:2}. However, since the system is more complicated than a random walk, and, strictly speaking, the jumps are not elastic processes, the diffusive dynamics for the GKSL equation \eqref{eq:GKSL} is limited to a finite range of length and time scales. 
In other words, Eq.~\eqref{eq:main:res} cannot describe dynamics of $\delta n^{\textsf{(u)}}_{\bm{q},\omega}$ down to $q{\to}0$ and $\omega{\to}0$. 
More formally, one could anticipate this conclusion by recalling the number of particles in upper and down bands are not conserved separately by the dissipative part of the action, cf. Eq.~\eqref{eq:SL:ud}.

As we have shown in Secs. \ref{Sec:Rec} and \ref{Sec:Pumping} above, there are two competing processes that affect diffusion: recombination of particles in the upper band with holes in the lower band, and pumping of particles into the upper band, leaving behind holes into the lower band. Considering first the former process, and using the expression \eqref{eq:app:Rec:Rate:main} for the rate $1/\tau_{\rm R}$, we find that the purely diffusive kernel in Eq. \eqref{eq:main:res} is limited to frequencies and momenta $1/\tau \gg |\omega|, Dq^2 \gg \delta n^{\textsf{(u)}}/(m^d \tau)$.
Provided that the change in the density of the particles in the upper band is small, $\delta n^{\textsf{(u)}}{\ll}m^d$, there is a wide interval for diffusive dynamics (more on this below).

Let us now turn to the latter process. As expected from the absence of exact cancellation between self-energy and vertex corrections, we find a nonzero contribution to the rate of change of the total number of the particles in the upper band, $N_{\textsf{u}}$, see Eq. \eqref{eq:Lambda}. Surprisingly, this rate is positive, i.e. particles are pumped into the upper band. An example of a process that results in the growth of the number of particles in the upper band is shown in Fig.~\ref{fig:pump:1}. This process directly follows from the so-called ``non-Hermitian Hamiltonian'' part of the Lindblad equation, involving the combinations $L^\dagger_\alpha L_\alpha$. For instance, this operator acting on a pure state with the lower eigenband filled and the upper eigenband occupied by a single fermion results in a superposition of states, one of which has one hole in the lower eigenband and two fermions in the upper one. This stems from the fact that in the jump operators $L_\alpha$, one of the fermion operators is not an eigen-operator, but a linear combination of such, enabling additional mixed terms in $L^\dagger_\alpha L_\alpha$. In Appendix \ref{App:ToyModel:2}, we further illustrate this peculiar property within a toy two-band model where the dissipative dynamics also results in pumping of the population in the upper band.

Provided that $m^d{\ll}n$, the pumping time $\tau_{\rm p}$ is parametrically longer than the elastic scattering time $\tau$. Putting together all the above considerations, we find that the diffusive kernel in Eq.~\eqref{eq:main:res} is valid in a broad range, 
\begin{equation}
\frac{1}{\tau} \gg |\omega|,\; Dq^2 \gg \frac{1}{\tau_{\rm p}},\; \frac{\delta n^{\textsf{(u)}}}{m^d \tau} .
\label{eq:cond:range:2}
\end{equation}

%%%%%%%%%%%%%%%
\begin{figure}[t]
\centerline{\includegraphics[width=0.7\columnwidth]{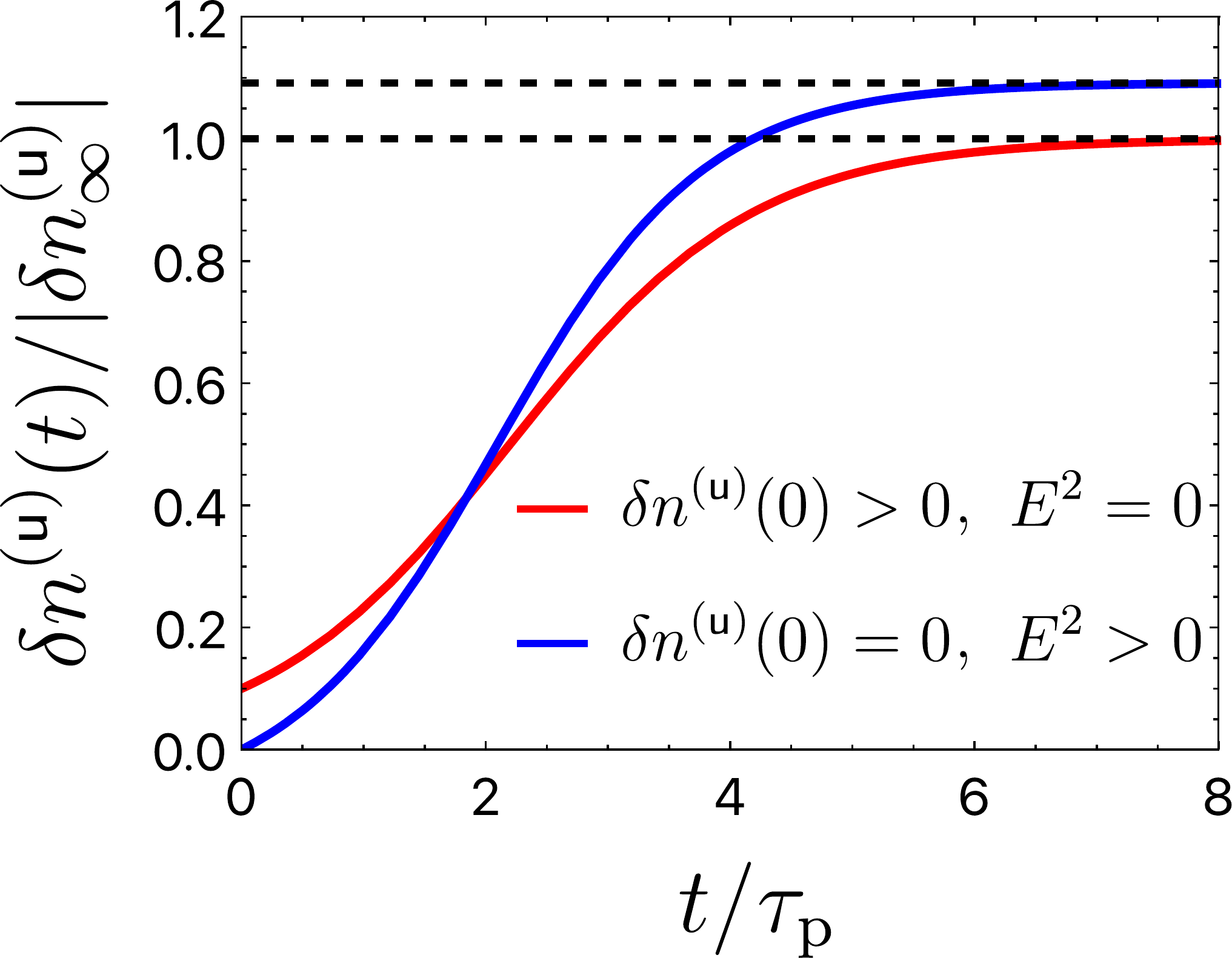}
}
\caption{The solution of Eq.~\eqref{eq:main:res:FKPP} in a spatially-homogeneous case. The red curve corresponds to the time evolution with the initial condition  $\delta n^{(\sf{u})}(x,0)/|\delta n^{(\sf{u})}_{\infty}|{=}0.1$, and in the absence of external fields. The blue curve represents the solution with $\delta n^{(\sf{u})}(x,0)/|\delta n^{(\sf{u})}_{\infty}|{=}0$, while the r.h.s. of Eq.~\eqref{eq:main:res:FKPP} is set to $0.1$ in units of $n\tau_{\rm R}/\tau_{\rm p}^2$. }
\label{fig:uniform_FKPP}
\end{figure}
 %%%%%%%%%%%%%%%

\begin{figure*}[t!]
\centerline{\includegraphics[width=1.0\textwidth]{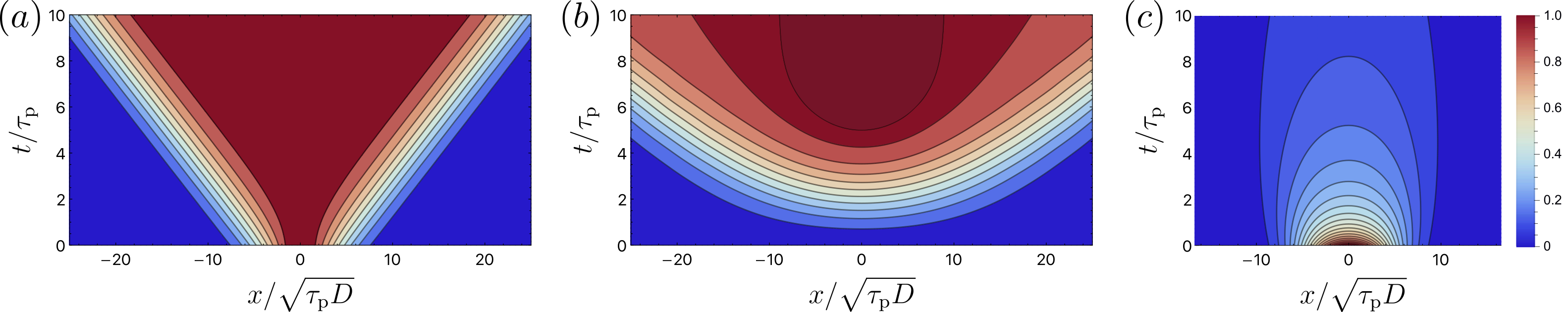}}
\caption{ Numerical solution of the reaction-diffusion equation \eqref{eq:main:res:FKPP} in the one-dimensional case on a finite interval $x/\sqrt{\tau_{\rm p}D}{\in} ({-}L,L)$ with $L=50$, and the boundary conditions $\delta n^{(\sf{u})}(\pm L,t){=}0$, (a) in the absence of external fields, and with a spatially-localized initial condition $\delta n^{(\sf{u})}(x,0){=}\exp\{{-}100(x/L\sqrt{\tau_{\rm p}D})^2\}$; (b) with vanishing initial density, $\delta n^{(\sf{u})}(x,0){=}0$, but with a local source term $0.1\exp\{{-}20(x/L\sqrt{\tau_{\rm p}D})^2\}$ in units of $n\tau_{\rm R}/\tau_{\rm p}^2$. The steady state density in the central region is slightly greater than $1$ due to the presence of the source. (c) The same as in (a), but the `pumping' term is ignored, and only recombination is included. In all cases, the density deviation $\delta n^{(\sf{u})}(x,t)$ is measured in units of $\delta n^{(\sf{u})}_{\infty}$, see Eq.~\eqref{eq:stat_density}. }
\label{fig:FKPP}
\end{figure*}

Combining the results of  Secs. \ref{Sec:Rec} and \ref{Sec:Pumping} suggests that the diffusion equation \eqref{eq:main:res} should be modified as
\begin{gather}
\Bigl (\frac{\partial}{\partial t} - D\nabla^2\Bigr )\delta n^{\textsf{(u)}}(\bm{x},t)
- \frac{\delta n^{\textsf{(u)}}(\bm{x},t)}{\tau_{\rm p}}+\frac{[\delta n^{\textsf{(u)}}(\bm{x},t)]^2}{n \tau_{\rm R}}
\notag \\
= \frac{\bar{\gamma} \chi}{1+\bar{\gamma}^2} \bm{E}^2(\bm{x},t)\;. 
\label{eq:main:res:FKPP}
\end{gather}
After defining appropriate dimensionless variables, ${\sf{t}}{=}t/\tau_{\rm p}$, $ {\sf{x}}{=}{\bm x}/\sqrt{\tau_{\rm p}D}$, ${\sf{f}}({\sf{x}},{\sf{t}}){=}(\tau_{\rm p}/\tau_{\rm R})\delta n^{({\sf{u}})}(\bm{x},t)/n$, and ${\sf{J}}({\sf{x}},{\sf{t}}){=}\bar{\gamma}\chi \tau_{\rm{p}}^2\bm{E}^2(\bm{x},t)/[n\tau_{\rm{R}}(1{+}\bar{\gamma}^2)]$,  Eq.~\eqref{eq:main:res:FKPP} attains the universal form
\begin{equation}
\Bigl (\frac{\partial}{\partial \mathsf{t}} - \tilde{\nabla}^2\Bigr ){\sf{f}}(\mathsf{x},\mathsf{t})
- {\sf{f}}(\mathsf{x},\mathsf{t})\Big(1-{\sf{f}}(\mathsf{x},\mathsf{t})\Big)
={\sf{J}}(\sf{x},\sf{t}), 
\label{eq:main:res:FKPPdim}
\end{equation}
where $\tilde{\nabla}$ corresponds to derivatives with respect to the dimensionless coordinates ${\sf{x}}$. We note that the left hand side of Eq.~\eqref{eq:main:res:FKPPdim} is nothing but the famous Fisher-Kolmogorov-Petrovsky-Piskunov (FKPP) reaction-diffusion equation \cite{Fisher1937,Kolmogorov1937}, which appears in numerous applications, including the propagation of advantageous genes and combustion fronts, the dynamics of domain walls and fluids, chemical reactions, bacterial spreading, decoherence propagation, etc.; see e.g., 
Refs.~\cite{FKPP1988,FKPP2,FKPP3,Aleiner2016,Zhou2023}. A striking hallmark of this equation is the existence of two stationary solutions ${\sf{f}}(\mathsf{x},\mathsf{t}) {=} 1$, and ${\sf{f}}(\mathsf{x},\mathsf{t}) {=} 0$, with the latter one being unstable due to a formation of a propagating wave with a constant velocity. Therefore, we arrive at the conclusion that the dark state with $\delta n_{\textsf{u}}{\equiv}n(\tau_{\rm R}/\tau_{\rm p}){\sf{f}} {=}0$ is unstable towards a new steady state with a nonzero density of particles in the upper band and holes in the lower one. The source of this instability can be either the right hand side of Eqs.~\eqref{eq:main:res:FKPP} and \eqref{eq:main:res:FKPPdim}, that is, an external %scalar potential
electric field, or a non-zero small initial $\delta n_{\textsf{u}}$, see Appendix \ref{App:ToyModel}. In the simplest, spatially-homogeneous case (i.e., assuming some uniform initial density $\delta n_{\textsf{u}}{=}n(\tau_{\rm R}/\tau_{\rm p}){\sf{f}}_0$, and a constant external field ${\sf{J}}$), Eq.~\eqref{eq:main:res:FKPPdim} can be easily solved as follows
\begin{gather}
    {\sf{f}}({\sf{t}})=\frac{1}{2} \left\{1+\sqrt{1+4 {\sf{J}}} \tanh \left[\arctanh \left(\frac{2 {\sf{f}}_0-1}{1+\sqrt{4 {\sf{J}}}}\right)\right.\right.\\\notag \left.\left.+\frac{ {\sf{t}}}{2} \sqrt{1+4 {\sf{J}}}\right]\right\}.
\end{gather}
The resulting time evolution of $\delta n^{\textsf{(u/d)}}(t)$ is depicted in Fig.~\ref{fig:uniform_FKPP}. The steady state density of the $\textsf{u}$-particles and $\textsf{d}$-holes is given by $|\delta n^{\textsf{(u/d)}}_{\infty}|{=}  n( \tau_{\rm R}/\tau_{\rm p}) (1{+}\sqrt{1{+}4{\sf{J}}})/2$, which, in the absence of the external constant source, ${\sf{J}}{=}0$, reduces to 
\begin{equation}
    |\delta n^{\textsf{(u/d)}}_{\infty}|{\sim}  n\left(\frac{\tau_{\rm R}}{\tau_{\rm p}}\right)\propto n\bar{\gamma}^2 \left(\frac{m^d}{n}\right)^2 \ln^{2(d-1)} \left(\frac{n}{m^d}\right),\label{eq:stat_density}
\end{equation}
for $\bar{\gamma}{\ll} 1$, with $d{=}1,2$ the dimensionality. Note also that $|\delta n^{\textsf{(u/d)}}_{\infty}| {\ll} m^d$. The real-space propagation of the instability, seeded by either initial local particle density $\delta n^{\textsf{(u)}}(\bm{x},0)$ or a spatially-localized external field, is depicted in Figs.~\ref{fig:FKPP}(a) and \ref{fig:FKPP}(b), respectively. After some initial diffusive relaxation, the solution assumes the form of a traveling kink with a constant velocity ${\propto} \sqrt{D/\tau_{\rm p}}$, where the exact proportionality coefficient is determined by the initial condition \cite{Fisher1937,Kolmogorov1937}. The detailed study of the resulting steady state with nonzero particle and hole densities is left for future work.

%%%%%%%%%%%%%%%%%%%
%%%%%%%%%%%%%%%%%%%

Let us finally note that, if one would fine-tune the bath couplings (beyond requiring them to vanish on the dark state), and replace the $\psi$ operators in the definitions of the Lindblad operators \eqref{eq:Jump:Operators1:def} by the corresponding $l$ operators, that is, switch $L_{1/2} {\to} l_\textsf{u/d}^\dagger l_\textsf{u}$, $L_{3/4} {\to} l_\textsf{u/d} l_\textsf{d}^\dagger$, the pumping term would not arise and only the diffusive kernel and the recombination term would remain on the left-hand side of Eq.~\eqref{eq:main:res:FKPP}, hence the dark state would be stable.
However, 
deviations of the density from the dark state would decay much more slowly than exponentially.
Indeed, an excitation composed of a single particle and a single hole localized at far-away points would need a long time (proportional to the initial distance squared) to diffuse around till they would meet and recombine. Since the initial distance could be of the order of the system size, the exponential decay rate would vanish in the thermodynamic limit.
Another way to see this is to note that, assuming spacial homogeneity, one would obtain an equation of the form 
$
(\partial /\partial t) \delta n^{\textsf{(u)}}(t) {=} {-} [\delta n^{\textsf{(u)}}(t)]^2/(n\tau_{\rm R})$,
which leads to an algebraic decay of $\delta n$ for $t{\gg} \tau_{\rm R}$. In order to emphasize the striking difference between this pure recombination dynamics and the full propagating instability shown in Fig.~\ref{fig:FKPP}(a), we present a numerical solution of Eq.~\eqref{eq:main:res:FKPPdim} in the absence of the linear `pumping' term in Fig.~\ref{fig:FKPP}(c).

\section{Conclusions\label{Sec:Conclusions}}
To summarize, we studied the emergence of the diffusive regime within the two-band dissipative quantum many-body state preparation dynamics proposed in Ref. \cite{Tonielli2020}. Although this dissipative model conserves the total number of particles only, we demonstrate the existence of a diffusive regime for the particle and hole density modes. This diffusive mode can be induced by the second-order response to a scalar potential. The diffusive regime persists up to length- and time-scales determined by the recombination processes and by pumping of particles from the `down' to the `up' band. The latter suggests an instability of the designed dark state, that is, the state with fully occupied `down' and empty `up' bands. The kinetic properties discussed above remain qualitatively the same in either one or two spatial dimensions (and even in the absence of the Hamiltonian dynamics), and as such, are not related to the Berry curvature.

These results open up many future research directions. It would be interesting to characterize the new steady state with nonzero particle and hole densities. It would also be worthwhile to study implications of the existence of a diffusive regime for nontrivial topological properties of the considered model (in particular, the effect of Berry curvature on the kinetic equation in a new steady state, similarly to \cite{SonSpivak}). In addition, one could try to derive a non-linear sigma-model-like description of the diffusive regime following the lines of Ref. \cite{Yang2022,poboiko2023}. Other potential directions include quantum absorbing phase transitions \cite{PhysRevLett.123.100604}, as well as the dynamics of systems with anyonic excitations, which cannot be created or annihilated individually even if no conservation law is imposed~\cite{Kapit2014,Umucalilar2017,Umucalilar2021,Kurilovich2022}.

\begin{acknowledgements}

We thank V.~Adler, A.~Altland, S.~Diehl, M.~Glazov, V.~Kravtsov, P. Kurilovich, V. Kurilovich, N.~O'Dea, and S.~Raghu  for useful discussions. One of us (I.S.B.) thanks A.~Lyublinskaya for collaboration on related project. The work of P.A.N. was supported in part by the US Department of Energy, Office of Basic Energy Sciences, Division of Materials Sciences and Engineering, under contract number DE-AC02-76SF00515. The work of D.S.S. was funded by DFG Grant No. MI 658/13-1. M.G. was supported by the Israel Science Foundation (ISF) and the Directorate for Defense Research and Development (DDR\&D) Grant No. 3427/21, and by the US-Israel Bi- national Science Foundation (BSF) Grant No. 2020072. The work of I.S.B. was funded by the Russian Science Foundation, grant No. 22-22-00641.
\end{acknowledgements}

\appendix
\begin{widetext}

\section{Self-consistent equation for the Green function in the Born approximation
\label{App:AppendixSCeq}}

In this Appendix, we demonstrate that, in the Born approximation, the self-consistent equations admit the only the dark state solution. We consider the case $\gamma_\alpha{=}\gamma$ for simplicity.
As one can check, a solution of Eq. \eqref{eq:self:Born} exists provided that the self-energies are related as
\begin{equation}
\Sigma_{\bm{q}}^{K}=\Sigma_{\bm{q}}^{R}\mathcal{F}_{\bm{q}}- \mathcal{F}_{\bm{q}}\Sigma_{\bm{q}}^{A} + \xi_q [\sigma_z, \mathcal{F}_{\bm{q}}] .
\label{eq:sce:maim:app:a}
\end{equation}
As one can further verify, the off-diagonal element of $\mathcal{F}_{\bm{q}}$ can be written as $\eta_{\bm{q}} = q_+ \mu_{\bm{q}}$. In what follows we assume that $n_{\textsf{u}/\textsf{d},\bm{q}}$ and $\mu_{\bm{p}}$ depend only on the length of the vector $\bm{q}$. The self-energies \eqref{eq:Sigma:K:1} and \eqref{eq:Sigma:RA:1} then become
\begin{equation}
\Sigma^K_{\bm{q}} = -2i\gamma\bigl (d_q \Delta_0 +\Delta_1+d_q n \sigma_z\bigr ) 
\end{equation}
and
\begin{equation}
\Sigma^R_{\bm{q}} = -i\gamma\Bigl [ d_q n+\delta_1+(d_q \Delta_0+i 2 m \varkappa ) \sigma_z\Bigr ] +\gamma \Re (q_+ w) \sigma_y 
+\gamma \Im (q_+w)\sigma_x .
\end{equation}
Here we introduced the following notations:
\begin{gather}
\Delta_k {=} \int\limits_q d_q^k (1{-}n_{\textsf{u},\bm{q}}{-}n_{\textsf{d},\bm{q}}), \quad 
\delta_k {=} \int\limits_q d_q^k (1{+}n_{\textsf{u},\bm{q}}{-}n_{\textsf{d},\bm{q}}) ,\notag \\
w {=} m \delta_0 {+} \int\limits_q 
q^2\mu_{\bm{q}}, \quad \varkappa = \Im \int\limits_q q^2 \mu_{\bm{q}} .
\end{gather}
Then, Eq. \eqref{eq:sce:maim:app:a} can be rewritten as the set of three equations:
\begin{equation}
\gamma w (n_{\textsf{u},\bm{q}}{-}n_{\textsf{d},\bm{q}}) {=} 2\mu_{\bm{q}} \Bigl[ \gamma (d_q n{+}\delta_1{+}i 2 m \varkappa){+}i\xi_q\Bigr ] 
\label{eq:A1} 
\end{equation}
and 
\begin{equation}
[d_q(n{\pm} \Delta_0){+}\delta_1](1{-}2n_{\textsf{u}/\textsf{d},\bm{q}})=
d_q(\Delta_0\pm n){+}\Delta_1 
{\pm} 2 q^2 \re \mu_{\bm{q}} w^* .
\label{eq:A2} 
\end{equation}
Let us impose the physical constraint that the total number of particles is fixed, $\int_q (n_{\textsf{u},\bm{q}}{+}n_{\textsf{d},\bm{q}}){=}n$, i.e.,$\Delta_0{=}0$.

First, we consider the case $\xi_q{\neq} 0$, and prove that there is no other solution independent of $\gamma$ except the dark state, cf.\ Eq.~\eqref{eq:steady:self-consistent Born approximation}. In this case, Eq.~\eqref{eq:A1}  implies that $\mu_{\bm{q}}{=}w{=}\delta_0{=}0$.  Next, we express $n_{\textsf{u}/\textsf{d},\bm{q}}$ from Eq. \eqref{eq:A2} as
\begin{equation}
1{-}n_{\textsf{u},\bm{q}}{-}n_{\textsf{d},\bm{q}}=\frac{\Delta_1}{d_q n+\delta_1}, \quad
1{+}n_{\textsf{u},\bm{q}}{-}n_{\textsf{d},\bm{q}}=\frac{\delta_1}{d_q n+\delta_1}.
\label{eq:A3}
\end{equation}
Hence we obtain that the only consistent solution is $\delta_k{=}\Delta_k{=}0$, which corresponds to the dark state: $n_{\textsf{u},\bm{q}}{=}0$ and $n_{\textsf{d},\bm{q}}{=}1$.

Second, let us consider the case $\xi_q{=}0$. Then Eq.~\eqref{eq:A1} implies that we can take $\mu_{\bm{q}}$ to be real. Consequently, $\varkappa{=}0$ and $w$ is also real. Then Eq.~\eqref{eq:A2} results in Eq.~\eqref{eq:A3}. Therefore, we find again the dark state solution. Since $\delta_0{=}0$, Eq.~\eqref{eq:A1} leads to $\mu_{\bm{q}}{=}w{=}0$.

Finally, we consider the case $\xi_q{\neq}0$, and demonstrate that the dark state is the only solution for $\bar{\gamma}{\ll} 1$. In this case, Eq. \eqref{eq:A3} is modified as follows
\begin{equation}
\begin{split}
1{-}n_{\textsf{u},\bm{q}}{-}n_{\textsf{d},\bm{q}}&=\frac{\Delta_1}{d_q n+\delta_1}, \\
1{+}n_{\textsf{u},\bm{q}}{-}n_{\textsf{d},\bm{q}}&=\frac{\delta_1-2 q^2\re (\mu_{\bm{q}} w^*)}{d_q n+\delta_1} .
\end{split}
\label{eq:A4}
\end{equation}
Hence, we find $\Delta_1{=}0$, i.e., $n_{\textsf{u},\bm{q}}{+}n_{\textsf{d},\bm{q}}{=}1$. The other unknowns, $\mu_{\bm{q}}$, $w$, $\varkappa$, and $\delta_{0,1}$, satisfy the following set of nonlinear equations
\begin{subequations}
\begin{gather}
\delta_0 = \int\limits_q \frac{\delta_1-2 q^2\re (\mu_{\bm{q}} w^*)}{d_q n+\delta_1}, 
\quad \varkappa = \int\limits_q q^2 \Im \mu_{\bm{q}},  \label{eq:A5a}
\\ \int\limits_q \frac{\delta_1^2}{d_q n+\delta_1} = -  \int\limits_q \frac{q^2d_q n\re (\mu_{\bm{q}} w^*)}{d_q n+\delta_1} , \label{eq:A5b} \\
\mu_{\bm{q}} = -  \frac{\gamma w [d_q n+ 2q^2 \re (\mu_{\bm{q}} w^*)]}{2(d_qn+\delta_1)[\gamma (d_qn+\delta_1+i 2m\varkappa)+i\xi_q]}, \label{eq:A5c} \\
w= m\delta_0 +\int\limits_q q^2 \mu_q \frac{\delta_1-2q^2\re (\mu_{\bm{q}} w^*)}{d_q n+\delta_1} . \label{eq:A5d}
\end{gather}
\end{subequations}
We note that $\delta_{0,1}$ are real and nonnegative. As follows from Eq.~\eqref{eq:A5b}, these conditions imply that $\Re (\mu_{\bm{q}} w^*){<}0$. Solving Eq.~\eqref{eq:A5c} to the lowest order in $\bar{\gamma}$, we find, indeed, $\Re (\mu_{\bm{q}} w^*){\simeq}{-}\bar{\gamma}^2 d_q n |w|^2/(2\xi_q^2){<}0$. 
However, according to Eqs.~\eqref{eq:A5a} and \eqref{eq:A5b}, it results in $\delta_{0,1}{\sim}\bar{\gamma}$. Then, as follows from Eq.~\eqref{eq:A5d}, $w{\sim}\bar{\gamma}$. But if it is so,  $\Re (\mu_{\bm{q}} w^*){\simeq} \bar{\gamma}^4$ and we can neglect it in Eqs.~\eqref{eq:A5a} and~\eqref{eq:A5b}. Then the only solution is simply $\delta_{0,1}{=}0$. Consequently, we find $\mu_{\bm{q}}{=}w{=}\varkappa{=}0$, i.e., the dark state again.

\section{Corrections to the self-consistent Born approximation\label{App:AppendixSelf-Cons}}

In this Appendix, we compute corrections to the self-energy beyond the self-consistent Born approximation. The corresponding diagrams are depicted in Fig. \ref{fig:app:1}.

%%%%%%%%%%%%%%%%%%%
\begin{figure}[t]
\centerline{\includegraphics[width=0.22\textwidth]{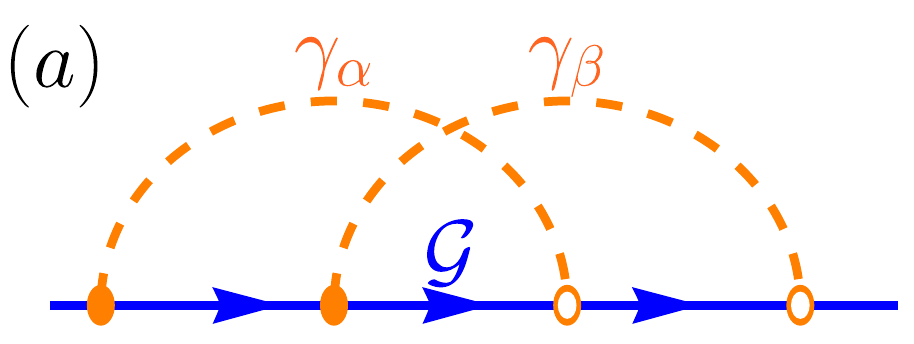}
\qquad 
\includegraphics[width=0.18\textwidth]{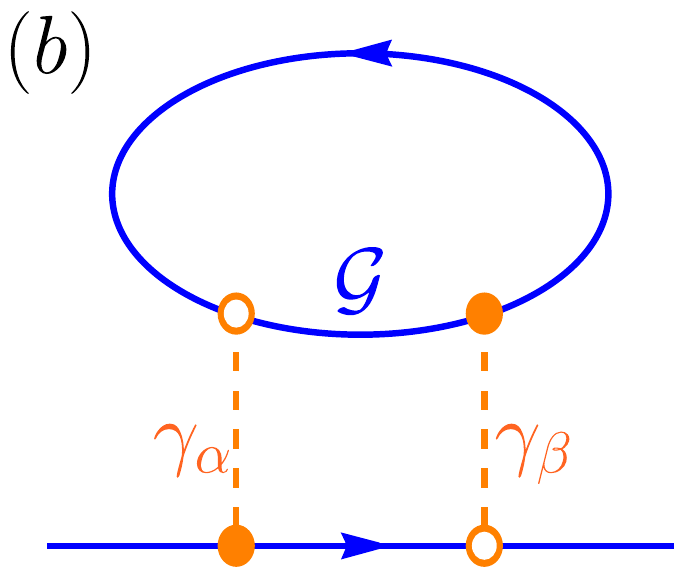}
\qquad
\includegraphics[width=0.22\textwidth]{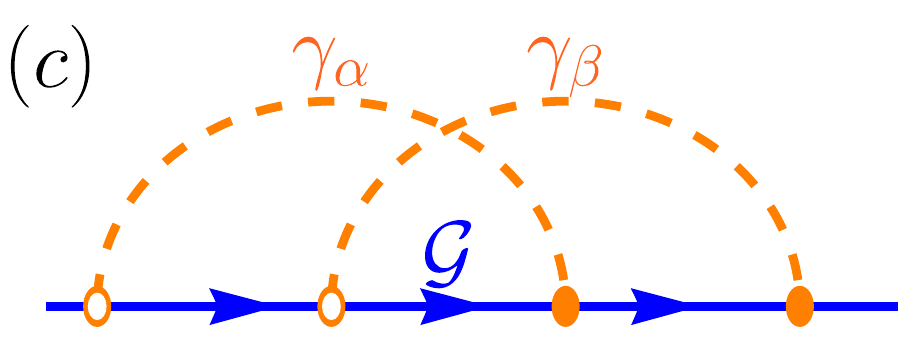}
\qquad
\includegraphics[width=0.18\textwidth]{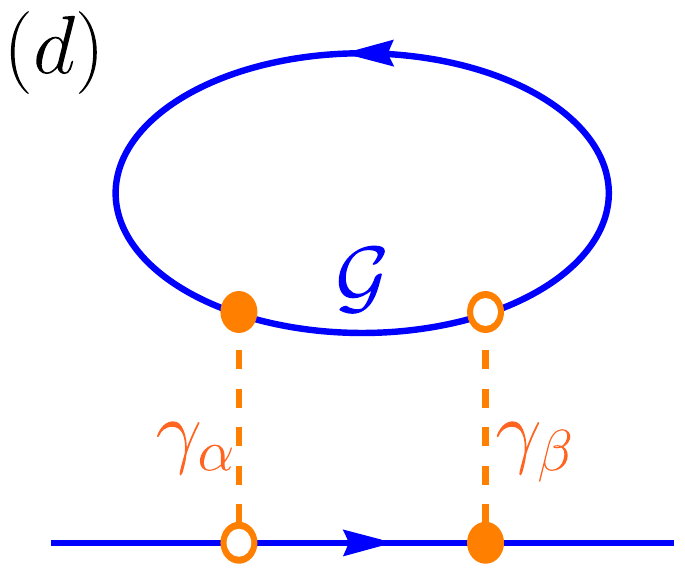}
}
\caption{Diagrams for the self-energy to second order in $\gamma$, beyond self-consistent Born approximation.}
\label{fig:app:1}
\end{figure}

%%%%%%%%%%%%%%%%%%%%%%%%%%%%%%%%%%%%%%%%%

\subsection{Diagram \ref{fig:app:1}(a)}

We start from the contributions to the self-energy to the second order in $\gamma$ shown in Fig. \ref{fig:app:1}(a). The corresponding corrections to the retarded, advanced, and Keldysh components of the self-energy read
\begin{gather}
\Sigma_{\bm{q},\varepsilon}^{a,(2),R}= \frac{\gamma^2}{4}\!\!\! \int\limits_{\bm{p},\bm{k};\omega,\Omega} \!\!\! \overline{\mathcal{L}}^{(\alpha)}_{\bm{q,p_+}}
\Biggl \{  G^R_{\bm{p_+},\varepsilon+\omega} \overline{\mathcal{L}}^{(\beta)}_{\bm{p_+},\bm{p+k}} G^K_{\bm{p+k},\varepsilon+\omega+\Omega}
 \mathcal{L}^{(\alpha)}_{\bm{p+k},\bm{k_+}} G^K_{\bm{k_+},\varepsilon+\Omega}  -G^K_{\bm{p_+},\varepsilon+\omega} \overline{\mathcal{L}}^{(\beta)}_{\bm{p_+},\bm{p+k}} G^K_{\bm{p+k},\varepsilon+\omega+\Omega}
 \mathcal{L}^{(\alpha)}_{\bm{p+k},\bm{k_+}}
 \notag \\
 \times
 G^R_{\bm{k_+},\varepsilon+\Omega} 
 + 2 G^R_{\bm{p_+},\varepsilon+\omega} \overline{\mathcal{L}}^{(\beta)}_{\bm{p_+},\bm{p+k}} G^K_{\bm{p+k},\varepsilon+\omega+\Omega}
 \mathcal{L}^{(\alpha)}_{\bm{p+k},\bm{k_+}} G^R_{\bm{k_+},\varepsilon+\Omega} 
 + G^K_{\bm{p_+},\varepsilon+\omega} \overline{\mathcal{L}}^{(\beta)}_{\bm{p_+},\bm{p+k}} G^A_{\bm{p+k},\varepsilon+\omega+\Omega}
 \mathcal{L}^{(\alpha)}_{\bm{p+k},\bm{k_+}} G^K_{\bm{k_+},\varepsilon+\Omega} 
 \notag \\
 + 2 G^K_{\bm{p_+},\varepsilon+\omega} \overline{\mathcal{L}}^{(\beta)}_{\bm{p_+},\bm{p+k}} G^A_{\bm{p+k},\varepsilon+\omega+\Omega}
 \mathcal{L}^{(\alpha)}_{\bm{p+k},\bm{k_+}} G^R_{\bm{k_+},\varepsilon+\Omega} 
 - G^R_{\bm{p_+},\varepsilon+\omega} \overline{\mathcal{L}}^{(\beta)}_{\bm{p_+},\bm{p+k}} G^A_{\bm{p+k},\varepsilon+\omega+\Omega}
 \mathcal{L}^{(\alpha)}_{\bm{p+k},\bm{k_+}} G^R_{\bm{k_+},\varepsilon+\Omega} 
\Biggr \} \mathcal{L}^{(\beta)}_{\bm{k_+},\bm{q}} ,
\end{gather}
\begin{gather}
\Sigma_{\bm{q},\varepsilon}^{a,(2),A}= -\frac{\gamma^2}{4}\!\!\! \int\limits_{\bm{p},\bm{k};\omega,\Omega} \!\!\! \overline{\mathcal{L}}^{(\alpha)}_{\bm{q},\bm{p_+}}
\Biggl \{  G^A_{\bm{p_+},\varepsilon+\omega} \overline{\mathcal{L}}^{(\beta)}_{\bm{p_+},\bm{p+k}} G^K_{\bm{p+k},\varepsilon+\omega+\Omega}
 \mathcal{L}^{(\alpha)}_{\bm{p+k},\bm{k_+}} G^K_{\bm{k_+},\varepsilon+\Omega}  -G^K_{\bm{p_+},\varepsilon+\omega} \overline{\mathcal{L}}^{(\beta)}_{\bm{p_+},\bm{p+k}} G^K_{\bm{p+k},\varepsilon+\omega+\Omega}
 \mathcal{L}^{(\alpha)}_{\bm{p+k},\bm{k_+}} 
 \notag \\
 \times
 G^A_{\bm{k_+},\varepsilon+\Omega} 
 + 2 G^A_{\bm{p_+},\varepsilon+\omega} \overline{\mathcal{L}}^{(\beta)}_{\bm{p_+},\bm{p+k}} G^K_{\bm{p+k},\varepsilon+\omega+\Omega}
 \mathcal{L}^{(\alpha)}_{\bm{p+k},\bm{k_+}} G^A_{\bm{k_+},\varepsilon+\Omega} 
 - G^K_{\bm{p_+},\varepsilon+\omega} \overline{\mathcal{L}}^{(\beta)}_{\bm{p_+},\bm{p+k}} G^R_{\bm{p+k},\varepsilon+\omega+\Omega}
 \mathcal{L}^{(\alpha)}_{\bm{p+k},\bm{k_+}} G^K_{\bm{k_+},\varepsilon+\Omega} 
 \notag \\
 + 2 G^A_{\bm{p_+},\varepsilon+\omega} \overline{\mathcal{L}}^{(\beta)}_{\bm{p_+},\bm{p+k}} G^R_{\bm{p+k},\varepsilon+\omega+\Omega}
 \mathcal{L}^{(\alpha)}_{\bm{p+k},\bm{k_+}} G^K_{\bm{k_+},\varepsilon+\Omega} 
 + G^A_{\bm{p_+},\varepsilon+\omega} \overline{\mathcal{L}}^{(\beta)}_{\bm{p_+},\bm{p+k}} G^R_{\bm{p+k},\varepsilon+\omega+\Omega}
 \mathcal{L}^{(\alpha)}_{\bm{p+k},\bm{k_+}} G^A_{\bm{k_+},\varepsilon+\Omega} 
\Biggr \} \mathcal{L}^{(\beta)}_{\bm{k_+},\bm{q}} ,
\end{gather}
and
\begin{gather}
\Sigma_{\bm{q},\varepsilon}^{a,(2),K}= \frac{\gamma^2}{4}\!\!\! \int\limits_{\bm{p},\bm{k};\omega,\Omega} \!\!\! \overline{\mathcal{L}}^{(\alpha)}_{\bm{q},\bm{p_+}}
\Biggl \{  G^A_{\bm{p_+},\varepsilon+\omega} \overline{\mathcal{L}}^{(\beta)}_{\bm{p_+},\bm{p+k}} G^K_{\bm{p+k},\varepsilon+\omega+\Omega}
 \mathcal{L}^{(\alpha)}_{\bm{p+k},\bm{k_+}} G^A_{\bm{k_+},\varepsilon+\Omega}  -2G^K_{\bm{p_+},\varepsilon+\omega} \overline{\mathcal{L}}^{(\beta)}_{\bm{p_+},\bm{p+k}} G^K_{\bm{p+k},\varepsilon+\omega+\Omega}
 \mathcal{L}^{(\alpha)}_{\bm{p+k},\bm{k_+}}
 \notag \\
 \times
 G^A_{\bm{k_+},\varepsilon+\Omega} 
 + 2 G^K_{\bm{p_+},\varepsilon+\omega} \overline{\mathcal{L}}^{(\beta)}_{\bm{p_+},\bm{p+k}} G^A_{\bm{p+k},\varepsilon+\omega+\Omega}
 \mathcal{L}^{(\alpha)}_{\bm{p+k},\bm{k_+}} G^K_{\bm{k_+},\varepsilon+\Omega} 
 - G^K_{\bm{p_+},\varepsilon+\omega} \overline{\mathcal{L}}^{(\beta)}_{\bm{p_+},\bm{p+k}} G^K_{\bm{p+k},\varepsilon+\omega+\Omega}
 \mathcal{L}^{(\alpha)}_{\bm{p+k},\bm{k_+}} G^K_{\bm{k_+},\varepsilon+\Omega} 
 \notag \\
 - G^R_{\bm{p_+},\varepsilon+\omega} \overline{\mathcal{L}}^{(\beta)}_{\bm{p_+},\bm{p+k}} G^A_{\bm{p+k},\varepsilon+\omega+\Omega}
 \mathcal{L}^{(\alpha)}_{\bm{p+k},\bm{k_+}} G^K_{\bm{k_+},\varepsilon+\Omega} 
 +2 G^R_{\bm{p_+},\varepsilon+\omega} \overline{\mathcal{L}}^{(\beta)}_{\bm{p_+},\bm{p+k}} G^K_{\bm{p+k},\varepsilon+\omega+\Omega}
 \mathcal{L}^{(\alpha)}_{\bm{p+k},\bm{k_+}} G^K_{\bm{k_+},\varepsilon+\Omega} 
 \notag \\
- G^K_{\bm{p_+},\varepsilon+\omega} \overline{\mathcal{L}}^{(\beta)}_{\bm{p_+},\bm{p+k}} G^R_{\bm{p+k},\varepsilon+\omega+\Omega}
 \mathcal{L}^{(\alpha)}_{\bm{p+k},\bm{k_+}} G^A_{\bm{k_+},\varepsilon+\Omega} 
 + G^A_{\bm{p_+},\varepsilon+\omega} \overline{\mathcal{L}}^{(\beta)}_{\bm{p_+},\bm{p+k}} G^R_{\bm{p+k},\varepsilon+\omega+\Omega}
 \mathcal{L}^{(\alpha)}_{\bm{p+k},\bm{k_+}} G^K_{\bm{k_+},\varepsilon+\Omega} 
 \notag\\
 -2 G^K_{\bm{p_+},\varepsilon+\omega} \overline{\mathcal{L}}^{(\beta)}_{\bm{p_+},\bm{p+k}} G^R_{\bm{p+k},\varepsilon+\omega+\Omega}
 \mathcal{L}^{(\alpha)}_{\bm{p+k},\bm{k_+}} G^K_{\bm{k_+},\varepsilon+\Omega} 
 +G^K_{\bm{p_+},\varepsilon+\omega} \overline{\mathcal{L}}^{(\beta)}_{\bm{p_+},\bm{p+k}} G^A_{\bm{p+k},\varepsilon+\omega+\Omega}
 \mathcal{L}^{(\alpha)}_{\bm{p+k},\bm{k_+}} G^R_{\bm{k_+},\varepsilon+\Omega} 
\notag \\
+ G^R_{\bm{p_+},\varepsilon+\omega} \overline{\mathcal{L}}^{(\beta)}_{\bm{p_+},\bm{p+k}} G^K_{\bm{p+k},\varepsilon+\omega+\Omega}
 \mathcal{L}^{(\alpha)}_{\bm{p+k},\bm{k_+}} G^R_{\bm{k_+},\varepsilon+\Omega} 
\Biggr \} \mathcal{L}^{(\beta)}_{\bm{k_+},\bm{q}} .
\end{gather}
Substituting the self-consistent Green function $\underline{\mathcal{G}}$ for $G$ and using the relations $[\mathcal{L}^{(\alpha)}_{\bm{qp}}]^{\textsf{(ud)}}{=}[\overline{\mathcal{L}}^{(\alpha)}_{\bm{qp}}]^{\textsf{(du)}}{=}0$, we find
\begin{gather}
[\Sigma_{\bm{q},\varepsilon}^{a,(2),R/A}]^{\textsf{(uu)}} \simeq 
  \int\limits_{\bm{p},\bm{k}} 
 \frac{\gamma^2 m^2 (d_{\bm{q}}/d_{\bm{p+k}})(\bm{k_-}\bm{p_-} +i[\bm{k_-}{\times}\bm{p_-}])}{\varepsilon-\xi_{\bm{p_+}}-\xi_{\bm{k_+}}-\xi_{\bm{p+k}}\pm i\bar{\gamma}(d_{\bm{p_+}}+d_{\bm{k_+}}+d_{\bm{p+k}})},\quad
 [\Sigma_{\bm{q},\varepsilon}^{a,(2),K}]^{\textsf{(uu)}}=[\Sigma_{\bm{q},\varepsilon}^{a,(2),R}]^{\textsf{(uu)}}- [\Sigma_{\bm{q}\omega}^{a,(2),A}]^{\textsf{(uu)}} ,
 \notag \\
 [\Sigma_{\bm{q},\varepsilon}^{a,(2),R/A/K}]^{\textsf{(ud)}}=[\Sigma_{\bm{q},\varepsilon}^{a,(2),R/A/K}]^{\textsf{(du)}}=[\Sigma_{\bm{q},\varepsilon}^{a,(2),R/A/K}]^{\textsf{(dd)}}=0.
\end{gather}

\subsection{Diagram \ref{fig:app:1}(b)}

The diagram on Fig. \ref{fig:app:1}(b) reads
\begin{gather}
\Sigma_{\bm{q},\varepsilon}^{b,(2),R} = \frac{\gamma^2}{4}\!\!\! \int\limits_{\bm{p},\bm{k};\omega,\Omega}\!\!\!
\overline{\mathcal{L}}^{(\alpha)}_{\bm{q},\bm{p_+}} \Biggl \{ G^R_{\bm{p_+},\varepsilon+\omega} 
\Bigl\{ \tr\Bigl [ \Bigl (G^A_{\bm{p+k},\varepsilon+\omega+\Omega}
\mathcal{L}^{(\alpha)}_{\bm{p+k},\bm{k_+}} G^R_{\bm{k_+},\varepsilon+\Omega} 
%\overline{\mathcal{L}}^{(\beta)}_{\bm{k_+},\bm{p+k}} \Bigr ]
+ 
 %\tr\Bigl [ 
 G^K_{\bm{p+k},\varepsilon+\omega+\Omega}
\mathcal{L}^{(\alpha)}_{\bm{p+k},\bm{k_+}} G^K_{\bm{k_+},\varepsilon+\Omega}
\Bigr ) \overline{\mathcal{L}}^{(\beta)}_{\bm{k_+},\bm{p+k}} \Bigr ]
\notag \\
-2 \tr\Bigl [ G^K_{\bm{p+k},\varepsilon+\omega+\Omega}
\mathcal{L}^{(\alpha)}_{\bm{p+k},\bm{k_+}} G^R_{\bm{k_+},\varepsilon+\Omega} \overline{\mathcal{L}}^{(\beta)}_{\bm{k_+},\bm{p+k}} \Bigr ]
-2 \tr\Bigl [ G^A_{\bm{p+k},\varepsilon+\omega+\Omega}
\mathcal{L}^{(\alpha)}_{\bm{p+k},\bm{k_+}} G^K_{\bm{k_+},\varepsilon+\Omega} \overline{\mathcal{L}}^{(\beta)}_{\bm{k_+},\bm{p+k}} \Bigr ] \Bigr \}
\notag\\
- G^K_{\bm{p_+},\varepsilon+\omega} 
\Bigl\{ \tr\Bigl [ G^K_{\bm{p+k},\varepsilon+\omega+\Omega}
\mathcal{L}^{(\alpha)}_{\bm{p+k},\bm{k_+}} G^R_{\bm{k_+},\varepsilon+\Omega} \overline{\mathcal{L}}^{(\beta)}_{\bm{k_+},\bm{p+k}} \Bigr ]
+  \tr\Bigl [ G^A_{\bm{p+k},\varepsilon+\omega+\Omega}
\mathcal{L}^{(\alpha)}_{\bm{p+k},\bm{k_+}} G^K_{\bm{k_+},\varepsilon+\Omega} \overline{\mathcal{L}}^{(\beta)}_{\bm{k_+},\bm{p+k}} \Bigr ]\Bigl \}
\Biggr \} {\mathcal{L}}^{(\beta)}_{\bm{p_+},\bm{q}} ,
\end{gather}
\begin{gather}
\Sigma_{\bm{q},\varepsilon}^{b,(2),A} = \frac{\gamma^2}{4}\!\!\! \int\limits_{\bm{p},\bm{k};\omega,\Omega}\!\!\!
\overline{\mathcal{L}}^{(\alpha)}_{\bm{q},\bm{p_+}} \Biggl \{ G^A_{\bm{p_+},\varepsilon+\omega} 
\Bigl\{ \tr\Bigl [ \Bigl ( G^R_{\bm{p+k},\varepsilon+\omega+\Omega}
\mathcal{L}^{(\alpha)}_{\bm{p+k},\bm{k_+}} G^A_{\bm{k_+},\varepsilon+\Omega} + 
G^K_{\bm{p+k},\varepsilon+\omega+\Omega}
\mathcal{L}^{(\alpha)}_{\bm{p+k},\bm{k_+}} G^K_{\bm{k_+},\varepsilon+\Omega} 
\Bigr )
\overline{\mathcal{L}}^{(\beta)}_{\bm{k_+},\bm{p+k}} \Bigr ]
\notag \\
+ 2 \tr\Bigl [ G^K_{\bm{p+k},\varepsilon+\omega+\Omega}
\mathcal{L}^{(\alpha)}_{\bm{p+k},\bm{k_+}} G^A_{\bm{k_+},\varepsilon+\Omega} \overline{\mathcal{L}}^{(\beta)}_{\bm{k_+},\bm{p+k}} \Bigr ]
+ 2 \tr\Bigl [ G^R_{\bm{p+k},\varepsilon+\omega+\Omega}
\mathcal{L}^{(\alpha)}_{\bm{p+k},\bm{k_+}} G^K_{\bm{k_+},\varepsilon+\Omega} \overline{\mathcal{L}}^{(\beta)}_{\bm{k_+},\bm{p+k}} \Bigr ] \Bigr \}
\notag\\
- G^K_{\bm{p_+},\varepsilon+\omega} 
\Bigl\{ \tr\Bigl [ G^K_{\bm{p+k},\varepsilon+\omega+\Omega}
\mathcal{L}^{(\alpha)}_{\bm{p+k},\bm{k_+}} G^A_{\bm{k_+},\varepsilon+\Omega} \overline{\mathcal{L}}^{(\beta)}_{\bm{k_+},\bm{p+k}} \Bigr ]
+  \tr\Bigl [ G^R_{\bm{p+k},\varepsilon+\omega+\Omega}
\mathcal{L}^{(\alpha)}_{\bm{p+k},\bm{k_+}} G^K_{\bm{k_+},\varepsilon+\Omega} \overline{\mathcal{L}}^{(\beta)}_{\bm{k_+},\bm{p+k}} \Bigr ]\Bigl \}
\Biggr \} {\mathcal{L}}^{(\beta)}_{\bm{p_+},\bm{q}} ,
\end{gather}
and
\begin{gather}
\Sigma_{\bm{q},\varepsilon}^{b,(2),K} = \frac{\gamma^2}{4}\!\!\! \int\limits_{\bm{p},\bm{k};\omega,\Omega}\!\!\!
\overline{\mathcal{L}}^{(\alpha)}_{\bm{q},\bm{p_+}} \Biggl \{ 
G^K_{\bm{p_+},\varepsilon+\omega} 
\Bigl\{ \tr\Bigl [ G^K_{\bm{p+k},\varepsilon+\omega+\Omega}
\mathcal{L}^{(\alpha)}_{\bm{p+k},\bm{k_+}} G^K_{\bm{k_+},\varepsilon+\Omega} 
+ 
G^R_{\bm{p+k},\varepsilon+\omega+\Omega}
\mathcal{L}^{(\alpha)}_{\bm{p+k},\bm{k_+}} G^A_{\bm{k_+},\varepsilon+\Omega} 
\Bigr )
\overline{\mathcal{L}}^{(\beta)}_{\bm{k_+},\bm{p+k}} \Bigr ]
\notag \\
+  2\tr\Bigl [ G^K_{\bm{p+k},\varepsilon+\omega+\Omega}
\mathcal{L}^{(\alpha)}_{\bm{p+k},\bm{k_+}} G^A_{\bm{k_+},\varepsilon+\Omega} \overline{\mathcal{L}}^{(\beta)}_{\bm{k_+},\bm{p+k}} \Bigr ]
-2 \tr\Bigl [ G^K_{\bm{p+k},\varepsilon+\omega+\Omega}
\mathcal{L}^{(\alpha)}_{\bm{p+k},\bm{k_+}} G^R_{\bm{k_+},\varepsilon+\Omega} \overline{\mathcal{L}}^{(\beta)}_{\bm{k_+},\bm{p+k}} \Bigr ]
\notag \\
+  2\tr\Bigl [ G^R_{\bm{p+k},\varepsilon+\omega+\Omega}
\mathcal{L}^{(\alpha)}_{\bm{p+k},\bm{k_+}} G^K_{\bm{k_+},\varepsilon+\Omega} \overline{\mathcal{L}}^{(\beta)}_{\bm{k_+},\bm{p+k}} \Bigr ]
-2 \tr\Bigl [ G^A_{\bm{p+k},\varepsilon+\omega+\Omega}
\mathcal{L}^{(\alpha)}_{\bm{p+k},\bm{k_+}} G^K_{\bm{k_+},\varepsilon+\Omega} \overline{\mathcal{L}}^{(\beta)}_{\bm{k_+},\bm{p+k}} \Bigr ]
\notag \\
+  \tr\Bigl [ G^A_{\bm{p+k},\varepsilon+\omega+\Omega}
\mathcal{L}^{(\alpha)}_{\bm{p+k},\bm{k_+}} G^R_{\bm{k_+},\varepsilon+\Omega} \overline{\mathcal{L}}^{(\beta)}_{\bm{k_+},\bm{p+k}} \Bigr ]
\Bigl \}
\notag \\
-G^A_{\bm{p_+},\varepsilon+\omega} 
\Bigl\{ \tr\Bigl [ G^K_{\bm{p+k},\varepsilon+\omega+\Omega}
\mathcal{L}^{(\alpha)}_{\bm{p+k},\bm{k_+}} G^A_{\bm{k_+},\varepsilon+\Omega} \overline{\mathcal{L}}^{(\beta)}_{\bm{k_+},\bm{p+k}} \Bigr ]
+ 
 \tr\Bigl [ G^R_{\bm{p+k},\varepsilon+\omega+\Omega}
\mathcal{L}^{(\alpha)}_{\bm{p+k},\bm{k_+}} G^K_{\bm{k_+},\varepsilon+\Omega} \overline{\mathcal{L}}^{(\beta)}_{\bm{k_+},\bm{p+k}} \Bigr ]\Bigr\} 
\notag \\
-G^R_{\bm{p_+},\varepsilon+\omega} 
\Bigl\{ \tr\Bigl [ G^K_{\bm{p+k},\varepsilon+\omega+\Omega}
\mathcal{L}^{(\alpha)}_{\bm{p+k},\bm{k_+}} G^R_{\bm{k_+},\varepsilon+\Omega} \overline{\mathcal{L}}^{(\beta)}_{\bm{k_+},\bm{p+k}} \Bigr ]
+ 
 \tr\Bigl [ G^A_{\bm{p+k},\varepsilon+\omega+\Omega}
\mathcal{L}^{(\alpha)}_{\bm{p+k},\bm{k_+}} G^K_{\bm{k_+},\varepsilon+\Omega} \overline{\mathcal{L}}^{(\beta)}_{\bm{k_+},\bm{p+k}} \Bigr ]\Bigr\} 
\Biggr \} {\mathcal{L}}^{(\beta)}_{\bm{p_+},\bm{q}} .
\end{gather}

Again, after substituting the self-consistent Green function $\underline{\mathcal{G}}$ for $G$, we obtain
\begin{gather}
[\Sigma_{\bm{q},\varepsilon}^{b,(2),R/A}]^{\textsf{(uu)}} = 
- \int\limits_{\bm{p},\bm{k}} 
 \frac{\gamma^2 m^2 \bm{k_-}^2\, d_{\bm{q}} d_{\bm{k_+}}/(d_{\bm{p+k}}d_{\bm{p_+}})}{\varepsilon-\xi_{\bm{p_+}}-\xi_{\bm{k_+}}-\xi_{\bm{p+k}}\pm i\bar{\gamma}(d_{\bm{p_+}}+d_{\bm{k_+}}+d_{\bm{p+k}})} , \quad 
 [\Sigma_{\bm{q},\varepsilon}^{b,(2),K}]^{\textsf{(uu)}} = [\Sigma_{\bm{q},\varepsilon}^{b,(2),R}]^{\textsf{(uu)}}- [\Sigma_{\bm{q},\varepsilon}^{b,(2),A}]^{\textsf{(uu)}} ,\notag \\
 [\Sigma_{\bm{q},\varepsilon}^{b,(2),R/A/K}]^{\textsf{(ud)}}=
 [\Sigma_{\bm{q},\varepsilon}^{b,(2),R/A/K}]^{\textsf{(du)}}
 =[\Sigma_{\bm{q},\varepsilon}^{b,(2),R/A/K}]^{\textsf{(dd)}}
 =0 .
\end{gather}

\subsection{Diagram \ref{fig:app:1}(c)}

The next contribution, diagram \ref{fig:app:1}(c), corresponds to diagram \ref{fig:app:1}(a) with all matrices $\overline{\mathcal{L}}_{\bm{qp}}^{(\alpha)}$ and $\overline{\mathcal{L}}_{\bm{qp}}^{(\beta)}$ interchanged with $\mathcal{L}_{\bm{qp}}^{(\alpha)}$ and ${\mathcal{L}}_{\bm{qp}}^{(\beta)}$, respectively. We obtain
\begin{gather}
\Sigma_{\bm{q},\varepsilon}^{c,(2),R}= \frac{\gamma^2}{4}\!\!\! \int\limits_{\bm{p},\bm{k};\omega,\Omega} \!\!\! {\mathcal{L}}^{(\alpha)}_{\bm{q},\bm{p_+}}
\Biggl \{  G^R_{\bm{p_+},\varepsilon+\omega} {\mathcal{L}}^{(\beta)}_{\bm{p_+},\bm{p+k}} G^K_{\bm{p+k},\varepsilon+\omega+\Omega}
 \overline{\mathcal{L}}^{(\alpha)}_{\bm{p+k},\bm{k_+}} G^K_{\bm{k_+},\varepsilon+\Omega}  -G^K_{\bm{p_+},\varepsilon+\omega} {\mathcal{L}}^{(\beta)}_{\bm{p_+},\bm{p+k}} G^K_{\bm{p+k},\varepsilon+\omega+\Omega}
 \overline{\mathcal{L}}^{(\alpha)}_{\bm{p+k},\bm{k_+}} 
 \notag \\
 \times
 G^R_{\bm{k_+},\varepsilon+\Omega} 
 - 2 G^R_{\bm{p_+},\varepsilon+\omega} {\mathcal{L}}^{(\beta)}_{\bm{p_+},\bm{p+k}} G^K_{\bm{p+k},\varepsilon+\omega+\Omega}
 \overline{\mathcal{L}}^{(\alpha)}_{\bm{p+k},\bm{k_+}} G^R_{\bm{k_+},\varepsilon+\Omega} 
 + G^K_{\bm{p_+},\varepsilon+\omega} {\mathcal{L}}^{(\beta)}_{\bm{p_+},\bm{p+k}} G^A_{\bm{p+k},\varepsilon+\omega+\Omega}
 \overline{\mathcal{L}}^{(\alpha)}_{\bm{p+k},\bm{k_+}} G^K_{\bm{k_+},\varepsilon+\Omega} 
 \notag \\
 - 2 G^K_{\bm{p_+},\varepsilon+\omega} {\mathcal{L}}^{(\beta)}_{\bm{p_+},\bm{p+k}} G^A_{\bm{p+k},\varepsilon+\omega+\Omega}
 \overline{\mathcal{L}}^{(\alpha)}_{\bm{p+k},\bm{k_+}} G^R_{\bm{k_+},\varepsilon+\Omega} 
 - G^R_{\bm{p_+},\varepsilon+\omega} {\mathcal{L}}^{(\beta)}_{\bm{p_+},\bm{p+k}} G^A_{\bm{p+k},\varepsilon+\omega+\Omega}
 \overline{\mathcal{L}}^{(\alpha)}_{\bm{p+k},\bm{k_+}} G^R_{\bm{k_+},\varepsilon+\Omega} 
\Biggr \} \overline{\mathcal{L}}^{(\beta)}_{\bm{k_+},\bm{q}} ,
\end{gather}
\begin{gather}
\Sigma_{\bm{q},\varepsilon}^{c,(2),A}= -\frac{\gamma^2}{4}\!\!\! \int\limits_{\bm{p},\bm{k};\omega,\Omega} \!\!\! {\mathcal{L}}^{(\alpha)}_{\bm{q},\bm{p_+}}
\Biggl \{  G^A_{\bm{p_+},\varepsilon+\omega} {\mathcal{L}}^{(\beta)}_{\bm{p_+},\bm{p+k}} G^K_{\bm{p+k},\varepsilon+\omega+\Omega}
 \overline{\mathcal{L}}^{(\alpha)}_{\bm{p+k},\bm{k_+}} G^K_{\bm{k_+},\varepsilon+\Omega}  -G^K_{\bm{p_+},\varepsilon+\omega} {\mathcal{L}}^{(\beta)}_{\bm{p_+},\bm{p+k}} G^K_{\bm{p+k},\varepsilon+\omega+\Omega}
 \overline{\mathcal{L}}^{(\alpha)}_{\bm{p+k},\bm{k_+}}  \notag \\
 \times G^A_{\bm{k_+},\varepsilon+\Omega} 
 - 2 G^A_{\bm{p_+},\varepsilon+\omega} {\mathcal{L}}^{(\beta)}_{\bm{p_+},\bm{p+k}} G^K_{\bm{p+k},\varepsilon+\omega+\Omega}
 \overline{\mathcal{L}}^{(\alpha)}_{\bm{p+k},\bm{k_+}} G^A_{\bm{k_+},\varepsilon+\Omega} 
 - G^K_{\bm{p_+},\varepsilon+\omega} {\mathcal{L}}^{(\beta)}_{\bm{p_+},\bm{p+k}} G^R_{\bm{p+k},\varepsilon+\omega+\Omega}
 \overline{\mathcal{L}}^{(\alpha)}_{\bm{p+k},\bm{k_+}} G^K_{\bm{k_+},\varepsilon+\Omega} 
 \notag \\
 - 2 G^A_{\bm{p_+},\varepsilon+\omega} {\mathcal{L}}^{(\beta)}_{\bm{p_+},\bm{p+k}} G^R_{\bm{p+k},\varepsilon+\omega+\Omega}
 \overline{\mathcal{L}}^{(\alpha)}_{\bm{p+k},\bm{k_+}} G^K_{\bm{k_+},\varepsilon+\Omega} 
+ G^A_{\bm{p_+},\varepsilon+\omega} {\mathcal{L}}^{(\beta)}_{\bm{p_+},\bm{p+k}} G^R_{\bm{p+k},\varepsilon+\omega+\Omega}
 \overline{\mathcal{L}}^{(\alpha)}_{\bm{p+k},\bm{k_+}} G^A_{\bm{k_+},\varepsilon+\Omega} 
\Biggr \} \overline{\mathcal{L}}^{(\beta)}_{\bm{k_+},\bm{q}} ,
\end{gather}
and
\begin{gather}
\Sigma_{\bm{q},\varepsilon}^{c,(2),K}= \frac{\gamma^2}{4}\!\!\! \int\limits_{\bm{p},\bm{k};\omega,\Omega} \!\!\! {\mathcal{L}}^{(\alpha)}_{\bm{q},\bm{p_+}}
\Biggl \{  \Bigl [ G^A_{\bm{p_+},\varepsilon+\omega} {\mathcal{L}}^{(\beta)}_{\bm{p_+},\bm{p+k}} G^K_{\bm{p+k},\varepsilon+\omega+\Omega}
 \overline{\mathcal{L}}^{(\alpha)}_{\bm{p+k},\bm{k_+}}
   +2G^K_{\bm{p_+},\varepsilon+\omega} {\mathcal{L}}^{(\beta)}_{\bm{p_+},\bm{p+k}} G^K_{\bm{p+k},\varepsilon+\omega+\Omega}
 \overline{\mathcal{L}}^{(\alpha)}_{\bm{p+k},\bm{k_+}} \Bigr ]
 G^A_{\bm{k_+},\varepsilon+\Omega} 
 \notag \\
 - 2 G^K_{\bm{p_+},\varepsilon+\omega} {\mathcal{L}}^{(\beta)}_{\bm{p_+},\bm{p+k}} G^A_{\bm{p+k},\varepsilon+\omega+\Omega}
 \overline{\mathcal{L}}^{(\alpha)}_{\bm{p+k},\bm{k_+}} G^K_{\bm{k_+},\varepsilon+\Omega} 
 - G^K_{\bm{p_+},\varepsilon+\omega} {\mathcal{L}}^{(\beta)}_{\bm{p_+},\bm{p+k}} G^K_{\bm{p+k},\varepsilon+\omega+\Omega}
 \overline{\mathcal{L}}^{(\alpha)}_{\bm{p+k},\bm{k_+}} G^K_{\bm{k_+},\varepsilon+\Omega} 
 \notag \\
 - G^R_{\bm{p_+},\varepsilon+\omega} {\mathcal{L}}^{(\beta)}_{\bm{p_+},\bm{p+k}} G^A_{\bm{p+k},\varepsilon+\omega+\Omega}
 \overline{\mathcal{L}}^{(\alpha)}_{\bm{p+k},\bm{k_+}} G^K_{\bm{k_+},\varepsilon+\Omega} 
 - 2 G^R_{\bm{p_+},\varepsilon+\omega} {\mathcal{L}}^{(\beta)}_{\bm{p_+},\bm{p+k}} G^K_{\bm{p+k},\varepsilon+\omega+\Omega}
 \overline{\mathcal{L}}^{(\alpha)}_{\bm{p+k},\bm{k_+}} G^K_{\bm{k_+},\varepsilon+\Omega}
 \notag \\
 - G^K_{\bm{p_+},\varepsilon+\omega} {\mathcal{L}}^{(\beta)}_{\bm{p_+},\bm{p+k}} G^R_{\bm{p+k},\varepsilon+\omega+\Omega}
 \overline{\mathcal{L}}^{(\alpha)}_{\bm{p+k},\bm{k_+}} G^A_{\bm{k_+},\varepsilon+\Omega} 
 + G^A_{\bm{p_+},\varepsilon+\omega} {\mathcal{L}}^{(\beta)}_{\bm{p_+},\bm{p+k}} G^R_{\bm{p+k},\varepsilon+\omega+\Omega}
 \overline{\mathcal{L}}^{(\alpha)}_{\bm{p+k},\bm{k_+}} G^K_{\bm{k_+},\varepsilon+\Omega}
 \notag \\
 + 2 G^K_{\bm{p_+},\varepsilon+\omega} {\mathcal{L}}^{(\beta)}_{\bm{p_+},\bm{p+k}} G^R_{\bm{p+k},\varepsilon+\omega+\Omega}
 \overline{\mathcal{L}}^{(\alpha)}_{\bm{p+k},\bm{k_+}} G^K_{\bm{k_+},\varepsilon+\Omega}
 + G^K_{\bm{p_+},\varepsilon+\omega} {\mathcal{L}}^{(\beta)}_{\bm{p_+},\bm{p+k}} G^A_{\bm{p+k},\varepsilon+\omega+\Omega}
 \overline{\mathcal{L}}^{(\alpha)}_{\bm{p+k},\bm{k_+}} G^R_{\bm{k_+},\varepsilon+\Omega}
 \notag \\
 + G^R_{\bm{p_+},\varepsilon+\omega} {\mathcal{L}}^{(\beta)}_{\bm{p_+}\bm{p+k}} G^K_{\bm{p+k},\varepsilon+\omega+\Omega}
 \overline{\mathcal{L}}^{(\alpha)}_{\bm{p+k},\bm{k_+}} G^R_{\bm{k_+},\varepsilon+\Omega}
\Biggr \} \overline{\mathcal{L}}^{(\beta)}_{\bm{k_+},\bm{q}} .
\end{gather}

Substituting the self-consistent Green function $\underline{\mathcal{G}}$ instead of $G$, we find
\begin{gather}
[\Sigma_{\bm{q},\varepsilon}^{c,(2),R/A}]^{\textsf{(dd)}} = 
 \int\limits_{\bm{p},\bm{k}} 
 \frac{\gamma^2 m^2 (d_{\bm{q}}/d_{\bm{p+k}})(\bm{k_-}\bm{p_-} -i[\bm{k_-}{\times}\bm{p_-}])}{\varepsilon+\xi_{\bm{p_+}}+\xi_{\bm{k_+}}+\xi_{\bm{p+k}}\pm i\bar{\gamma}(d_{\bm{p_+}}+d_{\bm{k_+}}+d_{\bm{p+k}})} ,
 \quad 
 [\Sigma_{\bm{q},\varepsilon}^{c,(2),K}]^{\textsf{(dd)}}  = - [\Sigma_{\bm{q},\varepsilon}^{c,(2),R}]^{\textsf{(dd)}} + [\Sigma_{\bm{q},\varepsilon}^{c,(2),A}]^{\textsf{(dd)}} , 
  \notag \\
 [\Sigma_{\bm{q},\varepsilon}^{c,(2),R/A/K}]^{\textsf{(uu)}}=[\Sigma_{\bm{q},\varepsilon}^{c,(2),R/A/K}]^{\textsf{(ud)}}=[\Sigma_{\bm{q},\varepsilon}^{c,(2),R/A/K}]^{\textsf{(du)}}=0 .
\end{gather}

\subsection{Diagram \ref{fig:app:1}(d)}

The last contribution, diagram \ref{fig:app:1}(d), is obtained from diagram \ref{fig:app:1}(b) upon interchanging
all matrices $\overline{\mathcal{L}}_{\bm{qp}}^{(\alpha)}$ and $\overline{\mathcal{L}}_{\bm{qp}}^{(\beta)}$ with $\mathcal{L}_{\bm{qp}}^{(\alpha)}$ and $\mathcal{L}_{\bm{qp}}^{(\beta)}$, respectively. We obtain
\begin{gather}
\Sigma_{\bm{q},\varepsilon}^{d,(2),R} = \frac{\gamma^2}{4}\!\!\! \int\limits_{\bm{p},\bm{k};\omega,\Omega}\!\!\!
{\mathcal{L}}^{(\alpha)}_{\bm{q},\bm{p_+}} \Biggl \{ G^R_{\bm{p_+},\varepsilon+\omega} 
\Bigl\{ \tr\Bigl [ \Bigl ( G^A_{\bm{p+k},\varepsilon+\omega+\Omega}
\overline{\mathcal{L}}^{(\alpha)}_{\bm{p+k},\bm{k_+}} G^R_{\bm{k_+},\varepsilon+\Omega} 
+  
 G^K_{\bm{p+k},\varepsilon+\omega+\Omega}
\overline{\mathcal{L}}^{(\alpha)}_{\bm{p+k},\bm{k_+}} G^K_{\bm{k_+},\varepsilon+\Omega} 
\Bigr )
\mathcal{L}^{(\beta)}_{\bm{k_+},\bm{p+k}} \Bigr ]
\notag \\
+2 \tr\Bigl [ G^K_{\bm{p+k},\varepsilon+\omega+\Omega}
\overline{\mathcal{L}}^{(\alpha)}_{\bm{p+k},\bm{k_+}} G^R_{\bm{k_+},\varepsilon+\Omega} \mathcal{L}^{(\beta)}_{\bm{k_+},\bm{p+k}} \Bigr ]
+2 \tr\Bigl [ G^A_{\bm{p+k},\varepsilon+\omega+\Omega}
\overline{\mathcal{L}}^{(\alpha)}_{\bm{p+k},\bm{k_+}} G^K_{\bm{k_+},\varepsilon+\Omega} \mathcal{L}^{(\beta)}_{\bm{k_+},\bm{p+k}} \Bigr ] \Bigr \}
\notag\\
- G^K_{\bm{p_+},\varepsilon+\omega} 
\Bigl\{ \tr\Bigl [ G^K_{\bm{p+k},\varepsilon+\omega+\Omega}
\overline{\mathcal{L}}^{(\alpha)}_{\bm{p+k},\bm{k_+}} G^R_{\bm{k_+},\varepsilon+\Omega} \mathcal{L}^{(\beta)}_{\bm{k_+},\bm{p+k}} \Bigr ]
+  \tr\Bigl [ G^A_{\bm{p+k},\varepsilon+\omega+\Omega}
\overline{\mathcal{L}}^{(\alpha)}_{\bm{p+k},\bm{k_+}} G^K_{\bm{k_+},\varepsilon+\Omega} \mathcal{L}^{(\beta)}_{\bm{k_+},\bm{p+k}} \Bigr ]\Bigl \}
\Biggr \} \overline{\mathcal{L}}^{(\beta)}_{\bm{p_+},\bm{q}} ,
\end{gather}
\begin{gather}
\Sigma_{\bm{q},\varepsilon}^{d,(2),A} = \frac{\gamma^2}{4}\!\!\! \int\limits_{\bm{p},\bm{k};\omega,\Omega}\!\!\!
{\mathcal{L}}^{(\alpha)}_{\bm{q},\bm{p_+}} \Biggl \{ G^A_{\bm{p_+},\varepsilon+\omega} 
\Bigl\{ \tr\Bigl [ \Bigr ( G^R_{\bm{p+k},\varepsilon+\omega+\Omega}
\overline{\mathcal{L}}^{(\alpha)}_{\bm{p+k},\bm{k_+}} G^A_{\bm{k_+},\varepsilon+\Omega}
+ 
G^K_{\bm{p+k},\varepsilon+\omega+\Omega}
\overline{\mathcal{L}}^{(\alpha)}_{\bm{p+k},\bm{k_+}} G^K_{\bm{k_+},\varepsilon+\Omega} 
\Bigr )
\mathcal{L}^{(\beta)}_{\bm{k_+},\bm{p+k}} \Bigr ]
\notag \\
-2 \tr\Bigl [ G^K_{\bm{p+k},\varepsilon+\omega+\Omega}
\overline{\mathcal{L}}^{(\alpha)}_{\bm{p+k},\bm{k_+}} G^A_{\bm{k_+},\varepsilon+\Omega} \mathcal{L}^{(\beta)}_{\bm{k_+},\bm{p+k}} \Bigr ]
-2 \tr\Bigl [ G^R_{\bm{p+k},\varepsilon+\omega+\Omega}
\overline{\mathcal{L}}^{(\alpha)}_{\bm{p+k},\bm{k_+}} G^K_{\bm{k_+},\varepsilon+\Omega} \mathcal{L}^{(\beta)}_{\bm{k_+},\bm{p+k}} \Bigr ] \Bigr \}
\notag\\
- G^K_{\bm{p_+},\varepsilon+\omega} 
\Bigl\{ \tr\Bigl [ G^K_{\bm{p+k},\varepsilon+\omega+\Omega}
\overline{\mathcal{L}}^{(\alpha)}_{\bm{p+k},\bm{k_+}} G^A_{\bm{k_+},\varepsilon+\Omega} \mathcal{L}^{(\beta)}_{\bm{k_+},\bm{p+k}} \Bigr ]
+  \tr\Bigl [ G^R_{\bm{p+k},\varepsilon+\omega+\Omega}
\overline{\mathcal{L}}^{(\alpha)}_{\bm{p+k},\bm{k_+}} G^K_{\bm{k_+},\varepsilon+\Omega} \mathcal{L}^{(\beta)}_{\bm{k_+},\bm{p+k}} \Bigr ]\Bigl \}
\Biggr \} \overline{\mathcal{L}}^{(\beta)}_{\bm{p_+},\bm{q}} ,
\end{gather}
and
\begin{gather}
\Sigma_{\bm{q},\varepsilon}^{d,(2),K} = \frac{\gamma^2}{4}\!\!\! \int\limits_{\bm{p},\bm{k};\omega,\Omega}\!\!\!
{\mathcal{L}}^{(\alpha)}_{\bm{q},\bm{p_+}} \Biggl \{ G^K_{\bm{p_+},\varepsilon+\omega} 
\Bigl\{ \tr\Bigl [ \Bigl ( G^K_{\bm{p+k},\varepsilon+\omega+\Omega}
\overline{\mathcal{L}}^{(\alpha)}_{\bm{p+k},\bm{k_+}} G^K_{\bm{k_+},\varepsilon+\Omega} 
+ 
G^R_{\bm{p+k},\varepsilon+\omega+\Omega}
\overline{\mathcal{L}}^{(\alpha)}_{\bm{p+k},\bm{k_+}} G^A_{\bm{k_+},\varepsilon+\Omega} 
\Bigr ) 
\mathcal{L}^{(\beta)}_{\bm{k_+},\bm{p+k}} \Bigr ]
\notag \\
-2 \tr\Bigl [ G^K_{\bm{p+k},\varepsilon+\omega+\Omega}
\overline{\mathcal{L}}^{(\alpha)}_{\bm{p+k},\bm{k_+}} G^A_{\bm{k_+},\varepsilon+\Omega} \mathcal{L}^{(\beta)}_{\bm{k_+},\bm{p+k}}\Bigr ]
+2 \tr\Bigl [ G^K_{\bm{p+k},\varepsilon+\omega+\Omega}
\overline{\mathcal{L}}^{(\alpha)}_{\bm{p+k},\bm{k_+}} G^R_{\bm{k_+},\varepsilon+\Omega} \mathcal{L}^{(\beta)}_{\bm{k_+},\bm{p+k}}\Bigr ]
\notag \\
- 2 \tr\Bigl [ G^R_{\bm{p+k},\varepsilon+\omega+\Omega}
\overline{\mathcal{L}}^{(\alpha)}_{\bm{p+k},\bm{k_+}} G^K_{\bm{k_+},\varepsilon+\Omega} \mathcal{L}^{(\beta)}_{\bm{k_+},\bm{p+k}}\Bigr ]
+ 2 \tr\Bigl [ G^A_{\bm{p+k},\varepsilon+\omega+\Omega}
\overline{\mathcal{L}}^{(\alpha)}_{\bm{p+k},\bm{k_+}} G^K_{\bm{k_+},\varepsilon+\Omega} \mathcal{L}^{(\beta)}_{\bm{k_+},\bm{p+k}}\Bigr ]
\notag \\
+ \tr\Bigl [ G^A_{\bm{p+k},\varepsilon+\omega+\Omega}
\overline{\mathcal{L}}^{(\alpha)}_{\bm{p+k},\bm{k_+}} G^R_{\bm{k_+},\varepsilon+\Omega} \mathcal{L}^{(\beta)}_{\bm{k_+},\bm{p+k}}\Bigr ]
\Bigl \}
\notag \\
-G^R_{\bm{p_+},\varepsilon+\omega} 
\Bigl\{ \tr\Bigl [ G^A_{\bm{p+k},\varepsilon+\omega+\Omega}
\overline{\mathcal{L}}^{(\alpha)}_{\bm{p+k},\bm{k_+}} G^K_{\bm{k_+},\varepsilon+\Omega} \mathcal{L}^{(\beta)}_{\bm{k_+},\bm{p+k}} \Bigr ]
+ 
 \tr\Bigl [ G^K_{\bm{p+k},\varepsilon+\omega+\Omega}
\overline{\mathcal{L}}^{(\alpha)}_{\bm{p+k},\bm{k_+}} G^R_{\bm{k_+},\varepsilon+\Omega} \mathcal{L}^{(\beta)}_{\bm{k_+},\bm{p+k}} \Bigr ]\Bigr \}
\notag \\
-G^A_{\bm{p_+},\varepsilon+\omega} 
\Bigl\{ \tr\Bigl [ G^R_{\bm{p+k},\varepsilon+\omega+\Omega}
\overline{\mathcal{L}}^{(\alpha)}_{\bm{p+k},\bm{k_+}} G^K_{\bm{k_+},\varepsilon+\Omega} \mathcal{L}^{(\beta)}_{\bm{k_+},\bm{p+k}} \Bigr ]
+ 
 \tr\Bigl [ G^K_{\bm{p+k},\varepsilon+\omega+\Omega}
\overline{\mathcal{L}}^{(\alpha)}_{\bm{p+k},\bm{k_+}} G^A_{\bm{k_+},\varepsilon+\Omega} \mathcal{L}^{(\beta)}_{\bm{k_+},\bm{p+k}} \Bigr ]\Bigr \}
\Biggr \} \overline{\mathcal{L}}^{(\beta)}_{\bm{p_+},\bm{q}} .
\end{gather}

Substitution of the self-consistent Green function $\underline{\mathcal{G}}$ for $G$ yields
\begin{gather}
[\Sigma_{\bm{q},\varepsilon}^{d,(2),R/A}]^{\textsf{(dd)}} {=} 
{-} \int\limits_{\bm{p},\bm{k}}
 \frac{\gamma^2 m^2 \bm{k_-}^2 d_{\bm{q}} d_{\bm{k_+}}/(d_{\bm{p+k}}d_{\bm{p_+}})}{\varepsilon+\xi_{\bm{p_+}}+\xi_{\bm{k_+}}+\xi_{\bm{p+k}}\pm i\bar{\gamma}(d_{\bm{p_+}}+d_{\bm{k_+}}+d_{\bm{p+k}})}
  ,
  \quad 
  [\Sigma_{\bm{q},\varepsilon}^{d,(2),K}]^{\textsf{(dd)}} = - [\Sigma_{\bm{q},\varepsilon}^{d,(2),R}]^{\textsf{(dd)}}+ [\Sigma_{\bm{q},\varepsilon}^{d,(2),A}]^{\textsf{(dd)}} ,
  \notag \\
 [\Sigma_{\bm{q},\varepsilon}^{d,(2),R/A/K}]^{\textsf{(uu)}} =[\Sigma_{\bm{q},\varepsilon}^{d,(2),R/A/K}]^{\textsf{(ud)}} =[\Sigma_{\bm{q},\varepsilon}^{d,(2),R/A/K}]^{\textsf{(du)}} =0 .
\end{gather}

\subsection{The total result}

There are also contributions which are similar to diagrams \ref{fig:app:1}(a) and (b) with $\overline{\mathcal{L}}_{\bm{qp}}^{(\alpha)}$ interchanged with $\mathcal{L}_{\bm{qp}}^{(\alpha)}$, or, alternatively, $\overline{\mathcal{L}}_{\bm{qp}}^{(\beta)}$ interchanged with $\mathcal{L}_{\bm{qp}}^{(\beta)}$. However, once the self-consistent Green functions are substituted, these diagrams vanish. Therefore, in total, we find the following non-zero components
\begin{gather}
[\Sigma_{\bm{q},\varepsilon}^{(2),R/A}]^{\textsf{(uu)}} = \bar{\gamma}^2 d_q \Upsilon_{\pm}(\bm{q},\varepsilon), 
\quad [\Sigma_{\bm{q},\varepsilon}^{(2),R/A}]^{\textsf{(dd)}}
= - \bar{\gamma}^2 d_q \Upsilon_{\mp}(\bm{q},-\varepsilon) ,\quad 
\Sigma_{\bm{q},\varepsilon}^{(2),K} = \Sigma_{\bm{q},\varepsilon}^{(2),R} \sigma_z - \sigma_z \Sigma_{\bm{q},\varepsilon}^{(2),A} ,
\notag\\
\Upsilon_{\pm}(\bm{q},\varepsilon) = -\frac{m^2}{2n^2} \int\limits_{\bm{p},\bm{k}} \frac{1}{d_{\bm{p+k}}}
 \frac{\left (\bm{k_-}\sqrt{d_{\bm{k_+}}/d_{\bm{p_+}}} - \bm{p_-} \sqrt{d_{\bm{p_+}}/d_{\bm{k_+}}}\right )^2}{\varepsilon-\xi_{\bm{p_+}}-\xi_{\bm{k_+}}-\xi_{\bm{p+k}}\pm i\bar{\gamma}(d_{\bm{p_+}}+d_{\bm{k_+}}+d_{\bm{p+k}})} .
 \label{eq:Sigma:final:res:0}
 \end{gather}
Evaluating $\Upsilon_\pm(\bm{q},\varepsilon)$ at $q{=}\varepsilon{=}0$, we obtain 
\begin{equation}
\Upsilon_\pm(0,0)=\frac{1}{4\pi} \frac{1}{1\mp i\bar{\gamma}} 
\frac{m^d}{n} 
\begin{cases}
3 , & d=1 ,\\
\ln (n/m^2), & d=2 .
\end{cases}
\label{eq:app:Tau:Phi}
\end{equation}
Here we assumed that $n{\gg} m^d$ for $d{=}1,2$. This result suggests that 
deviations from the self-consistent Born approximation (due to crossing diagrams) is fully controlled by the small parameter $m^d/n {\ll} 1$ (even for $\bar{\gamma}$ of the order of unity).

If the unitary dynamics is absent, $\xi_{\bm{p}}{=}0$, we find that 
\begin{equation}
\Upsilon_\pm(0,0)=\pm \frac{i}{4\pi\bar{\gamma}} 
\frac{m^d}{n} 
\begin{cases}
3 , & d=1 ,\\
\ln (n/m^2), & d=2 .
\end{cases}
\label{eq:app:Tau:Phi:2}
\end{equation}
Again, the above result suggests that 
deviations from the self-consistent Born approximation are fully controlled by the small parameter $m^d/n {\ll} 1$.

In the absence of unitary dynamics, $\xi_q=0$, the real part of the self-energy \eqref{eq:Sigma:final:res:0} could in principle have resulted in the appearance of an effective spectrum for the particles. Let us examine this. Writing $\varepsilon{=}\varepsilon^\prime{+}i\varepsilon^{\prime\prime}$, we obtain the following equation for the real part of the spectrum, $\varepsilon^\prime$:
\begin{equation}
\varepsilon^\prime  = - \bar{\gamma}^2 d_q  \frac{m^2}{2n^2} \int\limits_{\bm{p},\bm{k}} \frac{\varepsilon^\prime}{d_{\bm{p+k}}}
 \frac{\bm{k_-}^2d_{\bm{k_+}}/d_{\bm{p_+}}+ \bm{p_-}^2d_{\bm{p_+}}/d_{\bm{k_+}}
 - 2 \bm{k_-}\bm{p_-}}{\varepsilon^{\prime 2}+[\bar{\gamma}(d_{\bm{p_+}}+d_{\bm{k_+}}+d_{\bm{p+k}})+\varepsilon^{\prime\prime}]^2} \, .
 \label{epsilon:prime:A1}
\end{equation}
However, since $\bm{k_-}^2d_{\bm{k_+}}/d_{\bm{p_+}}{+}\bm{p_-}^2d_{\bm{p_+}}/d_{\bm{k_+}}{-}2 \bm{k_-}\bm{p_-}{\geqslant} 0$, the only solution of Eq. \eqref{epsilon:prime:A1} is $\varepsilon^\prime{=}0$.

\section{Hybridization of the `up' and `down' bands by an external scalar potential\label{App:ToyModel}}

Due to the presence of the off-diagonal matrix elements  $\Phi^{\textsf{(ud)}}$ and $\Phi^{\textsf{(du)}}$, an external scalar potential results in a non-zero density of the particles in the upper band. This effect was illustrated by a toy model at the end of Sec. \ref{Sec:Nonlinear}. In this Appendix, we estimate the corresponding density $\delta n^{\textsf{(u)}}_{\bm{q},\omega}$ induced by  transitions caused by the external potential. Since such hybridization occurs even in the absence of dissipation we set $\gamma{=}0$ for the sake of simplicity.

Using Eqs. \eqref{eq:T:1} and \eqref{eq:T:2} for $\gamma{=}0$, we write 
\begin{gather}
T^{\textsf{(uu)}}_{\bm{p},\bm{q};\omega}     
= 2\pi i \int\limits_{\varepsilon} \Phi^{\textsf{(ud)}}_{\bm{p_++Q_-/2,-Q_-};\Omega_+}
\Phi^{\textsf{(du)}}_{\bm{p_-+Q_+/2,Q_+};-\Omega_-}
\Biggl [\frac{F_{\varepsilon_-}^{\textsf{(u)}} \delta(\varepsilon_--\xi_{\bm{p_-}}) }{(\varepsilon_+-\xi_{\bm{p_+}}+i0^+)(\varepsilon-\Omega+\xi_{\bm{p+Q}}+i 0^+)} \notag\\
+  
\frac{F_{\varepsilon_+}^{\textsf{(u)}} \delta(\varepsilon_+-\xi_{\bm{p_+}})}{(\varepsilon-\Omega+\xi_{\bm{p+Q}}-i 0^+)(\varepsilon_--\xi_{\bm{p_-}}-i0^+)} 
+ \frac{F_{\varepsilon-\Omega}^{\textsf{(d)}}\delta(\varepsilon-\Omega+\xi_{\bm{p+Q}})}{(\varepsilon_+-\xi_{\bm{p_+}}+i0^+)(\varepsilon_--\xi_{\bm{p_-}}-i0^+)} \Biggr] .
\label{eq:comment:1}
\end{gather}
Here $F_{\varepsilon}^{\textsf{(u/d)}}$ stands for the distribution function of the  `up'/`down' particles, respectively. For the dark state we have $F_{\varepsilon}^{\textsf{(u)}}{=}{-}F_{\varepsilon}^{\textsf{(d)}}{=}1$. Integrating over $\varepsilon$ and using Eq. \eqref{eq:eq:du:2} strictly at $\gamma{=}0$, 
we obtain the following result,
\begin{gather}
\delta n^{\textsf{(u)}}_{\bm{q},\omega}
= \frac{i}{2} \int_{\bm{p}} T^{\textsf{(uu)}}_{\bm{p},\bm{q};\omega} = 
\int_{\bm{p,Q},\Omega} 
\frac{\Phi^{\textsf{(ud)}}_{\bm{p_++Q_-/2,-Q_-};\Omega_+}
\Phi^{\textsf{(du)}}_{\bm{p_-+Q_+/2,Q_+};-\Omega_-}}
{(-\Omega_-+\xi_{\bm{p_-}}+\xi_{\bm{p+Q}}+i 0^+)(-\Omega_++\xi_{\bm{p_+}}+\xi_{\bm{p+Q}}-i 0^+)}  .
\label{eq:comment:3}
\end{gather}
The above equation suggests the following interpretation. We set $q{=}\omega{=}0$ on the right-hand side for simplicity. The matrix element $\Phi^{\textsf{(ud)}}$ corresponds to the transition from a state with the energy $\xi_{\bm{p_+}}$ in the upper band to a virtual state with the energy $\xi_{\bm{p}}{+}\xi_{\bm{p+Q}}{-}\Omega$ in the down band. The matrix element $\Phi^{\textsf{(du)}}$ corresponds to the transition back from the virtual state with the energy  $\xi_{\bm{p}}{+}\xi_{\bm{p+Q}}{-}\Omega$ in the down band to a state with the energy $\xi_{\bm{p_-}}$ in the upper band.

As expected, the imaginary part of the response function \eqref{eq:comment:3} is nonzero for $\Omega{>}2m^2$ only. For the real part, we obtain at low frequencies
\begin{equation}
 \delta n^{\textsf{(u)}}_{\textsf{0}}(\bm{x},t)  \simeq  \bm{E}^2(\bm{x},t) \int_{\bm{p}} \frac{m^2}{4\xi^4_{\bm{p}}} \propto \frac{\chi}{m^2}
 \bm{E}^2(\bm{x},t) .
 \label{eq:comment:4}
\end{equation}
In the derivation we took into account that a typical $p{\sim} m$ is much larger than both $Q$ and $q$.

\section{Cartoon example of the dark state instability\label{App:ToyModel:2}}
In this section, we consider a simple discrete 1D two-band model which mimics some aspects of the full model investigated in the main text.
Specifically, we start with the following Hamiltonian
\begin{equation} \label{eq:H_toy_appendix}
H_0=t\sum_{j}\left(\psi_{\uparrow,j+1}^\dagger \psi_{\downarrow,j}+h.c.\right).
\end{equation}
Note that $H_0$ is block-diagonal as only the pairs of states $(\uparrow, j+1)$ and $(\downarrow, j)$ for all $j$ are coupled. Formally, this Hamiltonian is equivalent to the Su–Schrieffer–Heeger model~\cite{Su1979} in the maximally-dimerized limit $t^\prime/t=0$.

One can easily check that Eq.~\eqref{eq:H_toy_appendix} can be brought to the diagonal form
\begin{equation}
H_0= t\sum\limits_{j}\left(l_{{\sf{u}},j}^\dagger l_{{\sf{u}},j}-l_{{\sf{d}},j}^\dagger l_{{\sf{d}},j} \right)
\end{equation}
by means of the following transformation
\begin{gather}
\psi_{\uparrow,j}=\frac{1}{\sqrt{2}}\left(l_{{\sf{u}},j-1}-l_{{\sf{d}},j-1}\right)\;,\quad  \psi_{\downarrow,j}=\frac{1}{\sqrt{2}}\left(l_{{\sf{u}},j}+l_{{\sf{d}},j}\right)\;,\notag\\ \label{eq:transformation_l_psi}
l_{{\sf{u}},j}= \frac{1}{\sqrt{2}}\left(\psi_{\downarrow,j}+\psi_{\uparrow,j+1}\right)\;,\quad l_{{\sf{d}},j}= \frac{1}{\sqrt{2}}\left(\psi_{\downarrow,j}-\psi_{\uparrow,j+1}\right).
\end{gather}
The operators $l_{{\sf{u}}/{\sf{d}},j}$ correspond to the hybridized orbitals belonging to a single dimer.
Note that, similarly to the operators $l_{{\sf{u}}/{\sf{d}}}({\bm x})$ introduced in the main text (cf. Eq.~\eqref{eq:l_def_op}), our $l_{{\sf{u}}/{\sf{d}},j}$ can be expressed through linear combination of the lattice derivatives $\delta_j\equiv \psi_{\uparrow,j+1}-\psi_{\uparrow,j}$ as well as the sum and difference of the two onsite annihilation operators, $n_j\equiv \psi_{\uparrow,j}+\psi_{\downarrow,j}$ and $m_j\equiv \psi_{\uparrow,j}-\psi_{\downarrow,j}$,
\begin{equation} 
l_{{\sf{u}},j}= \frac{1}{\sqrt{2}}\left(n_j+\delta_{j}\right),\quad l_{{\sf{d}},j}= -\frac{1}{\sqrt{2}}\left(m_j+\delta_{j}\right).
\end{equation}
At half-filling, the ground state of $H_0$ consists of all the ${\sf{d}}$-orbitals being occupied on all sites, while all ${\sf{u}}$-orbitals are empty. Following the notation introduced in the main text, we will refer to this state as a `dark state' and denote it by $|D\rangle$.

The dynamics is governed by the GKSL equation, taking the form
\begin{equation}
\frac{d\rho}{dt} = 
 -i [H_{\rm 0},\rho] +\sum_{\sigma=\uparrow/\downarrow} \gamma_\sigma  \sum\limits_{j} 
\bigl (2 L_{j}^{(\sigma)} \rho (L_j^{(\sigma)})^\dagger - \{(L_j^{(\sigma)})^\dagger  L_{j}^{(\sigma)}, \rho\}\bigr ).
\label{eq:GKSLtoy}
\end{equation}
The dissipative part of dynamics is specified by the following jump operators
\begin{equation} 
L^{(\sigma)}_j=\psi_{\sigma,j}^\dagger l_{{\sf{u}},j},\quad \sigma =\uparrow,\downarrow,
\end{equation}
with the associated coupling constants $\gamma_{\uparrow}$ and $\gamma_{\downarrow}$, respectively. We emphasize that this particular choice of $L^{(\sigma)}_j$ is very similar to $L_{1/2}(\bm{x})$ introduced in the main text (cf. Eq.~\eqref{eq:Jump:Operators1:def}). It is convenient to make use of Eq.~\eqref{eq:transformation_l_psi} and re-write these operators in terms of $l_{{\sf{u}}/{\sf{d}},j}$ only. As a result, we obtain
\begin{equation}
L^{(\uparrow)}_j=\frac{1}{\sqrt{2}}\left(l_{{\sf{u}},j-1}^\dagger-l_{{\sf{d}},j-1}^\dagger\right) l_{{\sf{u}},j},\quad L^{(\downarrow)}_j=\frac{1}{\sqrt{2}}\left(l_{{\sf{u}},j}^\dagger+l_{{\sf{d}},j}^\dagger\right) l_{{\sf{u}},j}.
\end{equation}
The physical process represented by these operators is simple: They take a particle from the `up' band on a given site, and either dump it into the `down' band, or keep it in the `up' band. In both cases, $L^{(\uparrow)}_j$ also slightly shifts the particle in real space, while $L^{(\downarrow)}_j$ keeps it on the same site. Let us now highlight some of the important features of these jump operators (which they, in part, share with their more involved relatives $L_{1/2}(\bm{x})$). First of all, $L^{(\sigma)}_j$ are not products of eigen-operators of $H_0$, but rather consist of a linear combination of such. In simple terms, $L^{(\sigma)}_j$ can be written as $L^{(\sigma)}_j= A^{(\sigma)}_j+B^{(\sigma)}_j$, where $A^{(\sigma)}_j$ and $B^{(\sigma)}_j$ correspond to only one of particular physical processes just described (interband and intraband transitions). This is different from the case when $A^{(\sigma)}_j$ and $B^{(\sigma)}_j$ act as separate jump operators, since the GKSL equation contains terms quadratic in $L$. Therefore, our choice of $L^{(\sigma)}_j$ allows for mixed operators of the form $(A^{(\sigma)}_j)^\dagger B^{(\sigma)}_j$, etc. Second, while $L^{(\downarrow)}_j$ is purely local (i.e. it only acts on a single site $j$), $L^{(\uparrow)}_j$ actually couples nearest neighbors $j$ and $j-1$. As we will see, this condition enables the mixed operators to act non-trivially and produce additional excitations. Finally, we have $L^{(\sigma)}_j|D\rangle=0$, implying that the density matrix $\rho_D=|D\rangle \langle D|$ is a possible steady state solution of the corresponding GKSL equation. Since both $L^{(\sigma)}_j$ contain terms that move particles from the `up' band to the `down' band (and not the other way around), one could na\"{\i}vely expect that at sufficiently long times local perturbations around the dark state $\rho_D$ should relax towards it. We shall now see that this is \textit{not} the case.

Our strategy will be to contrast the two limiting cases: (a) $t=\gamma_{\uparrow}=0$, and (b) $t=\gamma_{\downarrow}=0$. We will show that $\rho_D$ is a {\it unique} (and `attractive') steady state only in the case (a), whereas (b) features a continuous family of attractive steady state solutions with a finite occupation number in the `up' band.

As a warm-up, let us start with case (a), $t=\gamma_{\uparrow}=0$. Since the dynamics is purely local, it is sufficient to consider only the following two states
\begin{equation}\label{eq:basis_one_site}
   |..{\sf{u}}..\rangle=l_{{\sf{u}},j}^\dagger |..\circ..\rangle,\quad   |..{\sf{d}}..\rangle=l_{{\sf{d}},j}^\dagger |..\circ..\rangle,
\end{equation}
where the symbol `$\circ$' denotes a completely empty site $j$, while the rest of the sites can have arbitrary occupation (the density matrix is factorized). Note that the states of the form $l_{{\sf{u}},j}^\dagger l_{{\sf{d}},j}^\dagger |..\circ..\rangle$ are allowed, but they cannot relax under the dynamics introduced by $L^{(\downarrow)}_j$ (clearly, $L^{(\downarrow)}_j$ acting on these states gives zero), so we ignore them. One can make use of the operator identity
\begin{equation}
 \left(L^{(\downarrow)}_j\right)^\dagger L^{(\downarrow)}_j = \frac{1}{2} l_{{\sf{u}},j}^\dagger l_{{\sf{u}},j}\left(1-l_{{\sf{d}},j}^\dagger l_{{\sf{d}},j}\right),
\end{equation}
which leads to the following matrix elements
\begin{equation}
\label{eq:L_down_matrix}
L^{(\downarrow)}_j = \frac{1}{\sqrt{2}}\begin{pmatrix}
1 & 0\\
1 & 0
\end{pmatrix}_j,\quad \left(L^{(\downarrow)}_j\right)^\dagger L^{(\downarrow)}_j = \begin{pmatrix}
1 & 0\\
0 & 0
\end{pmatrix}_j.
\end{equation}
The index $j$ refers to the local basis \eqref{eq:basis_one_site}. The most general ansatz for the site-$j$ reduced density matrix is given by
\begin{equation}
\rho_j(t)=\begin{pmatrix}
1-\rho_D(t)& \rho_m(t)\\
\rho_m(t) & \rho_D(t)
\end{pmatrix}_j.
\end{equation}
After substituting this expression into the corresponding GKSL equation, and making use of the matrix representation \eqref{eq:L_down_matrix}, we obtain
\begin{equation}
\dot{\rho}_D(t)=1-\rho_D(t),\quad \dot{\rho}_m(t)=1-\rho_D(t)-\rho_m(t).
\end{equation}
The initial condition $\rho_D(0)=\rho_m(0)=0$, leads to $\rho_D(t)= 1-\exp(-t)$, $\rho_m(t)= t\exp(-t)$, with the attractive steady state solution $\rho_D(t=\infty)=1$, $\rho_m(t=\infty)=0$. This result indeed confirms our na\"{\i}ve expectations regarding the dynamics of this system.

Case (b), $t=\gamma_{\downarrow}=0$, is much more involved. To see how the complications arise, one can use the identity 
\begin{gather}
\left(L^{(\uparrow)}_j\right)^\dagger L^{(\uparrow)}_j = \frac{1}{2} l_{{\sf{u}},j}^\dagger l_{{\sf{u}},j} \left(2- l_{{\sf{u}},j-1}^\dagger l_{{\sf{u}},j-1}-l_{{\sf{d}},j-1}^\dagger l_{{\sf{d}},j-1}+ l_{{\sf{u}},j-1}^\dagger l_{{\sf{d}},j-1}+l_{{\sf{d}},j-1}^\dagger l_{{\sf{u}},j-1}\right),
\end{gather}
and act by this operator on the initial state $l_{{\sf{u}},j}^\dagger l_{{\sf{d}},j}|D\rangle$ corresponding to a simple one-particle excitation around the dark state (i.e., only one particle in the `up' band). The result reads
\begin{equation}
    \left(L^{(\uparrow)}_j\right)^\dagger L^{(\uparrow)}_jl_{{\sf{u}},j}^\dagger l_{{\sf{d}},j}|D\rangle=\frac{1}{2}l_{{\sf{u}},j}^\dagger l_{{\sf{d}},j}|D\rangle +\frac{1}{2}l_{{\sf{u}},j-1}^\dagger l_{{\sf{d}},j-1}l_{{\sf{u}},j}^\dagger l_{{\sf{d}},j} |D\rangle, \label{eq:appendix_pair}
\end{equation}
where the last term stems from the interference between the two terms in $L^{(\uparrow)}_j$. Crucially, this extra contribution represents a state with two particles in the `up' band. The corresponding physical process is the following: We first use the operator $l_{{\sf{u}},j-1}^\dagger l_{{\sf{u}},j}$ appearing $L^{(\uparrow)}_j$ to move an `up' particle from the site $j$ to $j-1$, and then use $l_{{\sf{u}},j}^\dagger l_{{\sf{d}},j-1}$ from $(L^{(\uparrow)}_j)^\dagger$ to move a `down' particle to the `up' band. Thus, this process provides a mechanism for `pumping' of particles to the `up' band. Importantly, this process competes with the recombination of `up' particles and `down' holes (described by the remaining terms in the GKSL equation), which tends to relax the system towards the dark state. In order to analyze this competition in more detail, we will solve a two-site version of this problem only (which is sufficient to illustrate the general situation).

%%%%%%%%%%%%%%%%%%%
\begin{figure*}[t!]
\centerline{\includegraphics[width=0.9\textwidth]{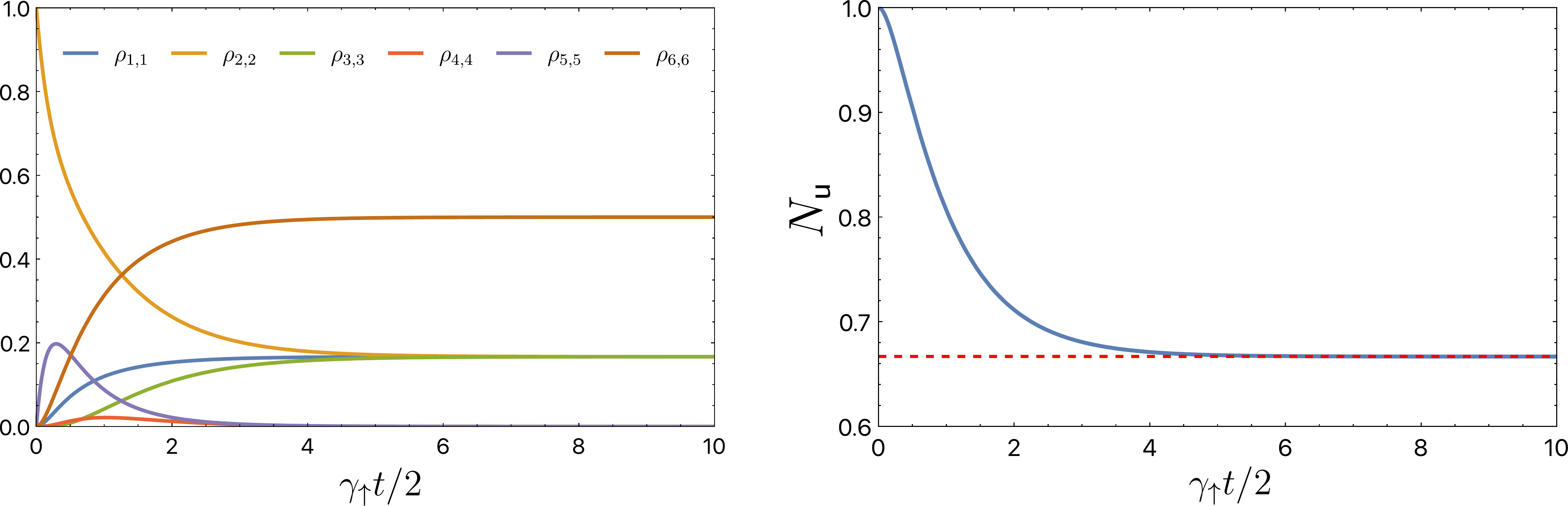}
}
\caption{Numerical solution of the GKSL equation for the toy two-band model on two sites (for$t=\gamma_{\downarrow}=0$) prepared in the initial state $|{\sf{u}},{\sf{d}}\rangle$ [see Eq.~\eqref{eq:basis} for the definition]. Left panel: The time dependence of the diagonal elements $\rho_{k,k}$ of the density matrix $\rho(t)$. Right panel: The expectation value of the total number of particles in the `up' band. The red dashed line corresponds to the steady state limit $N_{\sf{u}}=2/3$. }
\label{fig:twosite}
\end{figure*}
%%%%%%%%%%%%%%%%%%%

The full basis of a two-site model at half-filling consists of six states
\begin{gather}
   |{\sf{u}},{\sf{u}}\rangle = l_{{\sf{u}},1}^\dagger l_{{\sf{u}},2}^\dagger |\circ,\circ\rangle,\quad |{\sf{u}},{\sf{d}}\rangle = l_{{\sf{u}},1}^\dagger l_{{\sf{d}},2}^\dagger |\circ,\circ\rangle,\quad |{\sf{d}},{\sf{u}}\rangle = l_{{\sf{d}},1}^\dagger l_{{\sf{u}},2}^\dagger |\circ,\circ\rangle, \notag \\ \label{eq:basis}
  |{\sf{u}}/{\sf{d}}, \circ \rangle = l_{{\sf{u}},1}^\dagger l_{{\sf{d}},1}^\dagger |\circ,\circ\rangle,\quad  |\circ, {\sf{u}}/{\sf{d}} \rangle=l_{{\sf{u}},2}^\dagger l_{{\sf{d}},2}^\dagger |\circ,\circ\rangle,\quad |{\sf{d}}, {\sf{d}} \rangle\equiv|D\rangle = l_{{\sf{d}},1}^\dagger l_{{\sf{d}},2}^\dagger |\circ,\circ\rangle.
\end{gather}
The matrix elements of the jump operators read as follows
 \begin{equation}
L_1^{(\uparrow)}=\frac{1}{\sqrt{2}}\begin{pmatrix}
0 & 0 & 0 & 0 & 0 & 0\\
0 & 0 & 0 & 0 & 0 & 0\\
0 & 0 & 0 & -1 & 0 & 0\\
0 & 0 & 0 & 0 & 0 & 0\\
1 & 1 & 0 & 0 & 0 & 0\\
0 & 0 & 0 & 1 & 0 & 0
\end{pmatrix},\quad \quad L_2^{(\uparrow)}=\frac{1}{\sqrt{2}}\begin{pmatrix}
0 & 0 & 0 & 0 & 0 & 0\\
0 & 0 & 0 & 0 & 1 & 0\\
0 & 0 & 0 & 0 & 0 & 0\\
-1 & 0 & -1 & 0 & 0 & 0\\
0 & 0 & 0 & 0 & 0 & 0\\
0 & 0 & 0 & 0 & -1 & 0
\end{pmatrix}. \end{equation}
Here we assumed periodic boundary conditions. One can easily check that the right hand side of the corresponding GKSL equation vanishes if evaluated with the following two-parameter density matrix
\begin{equation}\label{eq:steady_state_twosite}
\rho_{*}=\left(
\begin{array}{cccccc}
 \rho _{{\sf{u}}} & -\rho _{{\sf{u}}} & -\rho _{{\sf{u}}} & 0 & 0 & \rho _{m} \\
 -\rho _{{\sf{u}}} & \rho _{{\sf{u}}} & \rho _{{\sf{u}}} & 0 & 0 & -\rho _{m} \\
 -\rho _{{\sf{u}}} & \rho _{{\sf{u}}} & \rho _{{\sf{u}}} & 0 & 0 & -\rho _{m} \\
 0 & 0 & 0 & 0 & 0 & 0 \\
 0 & 0 & 0 & 0 & 0 & 0 \\
 \rho _{m} & -\rho _{m} & -\rho _{m} & 0 & 0 & 1-3 \rho _{{\sf{u}}} \\
\end{array}
\right),
\end{equation}
with the condition $ (1-6\rho_{{\sf{u}}})^2+12 \rho_m^2\leq 1$, required for the eigenvalues of $\rho_{*}$ to remain nonnegative (note that $\operatorname{Tr} \rho_{*} =1$ as it should be). In fact, Eq.~\eqref{eq:steady_state_twosite} is the most general form of the steady state solution in this model. In particular, the choice $\rho_m=\rho_{\sf{u}}=0$ gives the dark state $\rho_{*}=\rho_{D}$. The expectation value of the total number of particles in the `up' band can be easily computed as
\begin{equation}
N_{{\sf{u}}} = \sum\limits_j\operatorname{tr}( l_{{\sf{u}},j}^\dagger l_{{\sf{u}},j} \rho_{*}) =  4\rho_{{\sf{u}}}.
\end{equation}
The resulting steady state (with particular $\rho_{{\sf{u}}}$ and $\rho_m$) depends on the initial conditions. Let us consider $\rho(t=0)=|{\sf{u}},{\sf{d}}\rangle \langle {\sf{u}},{\sf{d}}|$ (thus, $N_{\sf{u}}(t=0)=1$), and study how the system evolves with time. In Fig.~\ref{fig:twosite}, we demonstrate a numerical solution of the corresponding GKSL equation. One can see that the diagonal matrix elements corresponding to the states $ |{\sf{u}},{\sf{u}}\rangle$, $|D\rangle$ and $ |{\sf{d}},{\sf{u}}\rangle$ grow with time monotonically (although with different rates), while the amplitude for $ |{\sf{u}},{\sf{d}}\rangle$ decays, and eventually saturates at a finite value. At the same time, the amplitudes for $ |\circ,{\sf{u}}/{\sf{d}}\rangle$ and $ |{\sf{u}}/{\sf{d}},\circ\rangle$ initially increase, but then quickly approach zero, in full agreement with the steady state structure of Eq.~\eqref{eq:steady_state_twosite}. We emphasize that the competition between recombination and `pumping' results in a finite expectation value for the total number of particles in the `up' band (see the right panel in Fig.~\ref{fig:twosite}). This aspect of the toy model resembles our results for the full model studied in the main text. As a final remark, we note that in addition to the jump operators $L_j^{(\uparrow)}$ containing $l_{{\sf{u}},j}$, one could include operators with $l_{{\sf{d}},j}^\dagger$ since they also annihilate the dark state. In particular, it is easy to verify that the following choice $\tilde{L}_j^{(\downarrow)}=\psi_{\downarrow,j-1}l_{{\sf{d}},j}^\dagger = \left(l_{{\sf{u}},j-1}+l_{{\sf{d}},j-1}\right)l_{{\sf{d}},j}^\dagger/\sqrt{2}$ leads to the same steady density matrix as in Eq.~\eqref{eq:steady_state_twosite}.

\end{widetext}

\bibliography{Lindblad-ref}

\end{document}